\documentclass[journal,10pt,twocolumn,twoside]{IEEEtran}

\usepackage{ifpdf}
\usepackage{cite}
\usepackage{comment}
\usepackage{amsmath,amssymb,amsfonts,bm}
\usepackage{algorithmic}
\usepackage{graphicx, adjustbox}
\usepackage{array}
\usepackage[shortlabels]{enumitem}
\usepackage{breqn}
\usepackage{bm}
\usepackage{dsfont}
\usepackage{upgreek}
\usepackage{textcomp}
\usepackage{multirow}
\usepackage{algorithm}
\usepackage{algorithmic}
\usepackage{pgfplots}
\usepackage{tkz-euclide}
\usepackage{subcaption}
\usepackage{tikz}{\tiny }
\usepackage{hyperref}
\usepackage{balance}
\usepackage{ tipa }
\usepackage{dirtytalk}
\usepackage{subfiles}

\usepackage{fancyhdr}
\usepackage{upgreek}
\usepackage{soul}
\usepackage{breqn}
\usepackage[english]{babel}

\usetikzlibrary{shapes, arrows, positioning, shadows}

\addto\captionsenglish{}
\allowdisplaybreaks

\usepackage{amsthm}

\usepackage{fancyhdr}
\usepackage{ upgreek }
\usepackage{soul}
\makeatletter
\newcommand{\vast}{\bBigg@{4}}

\newcommand{\major}[1]{{\leavevmode{#1}}}
\newcommand{\Vast}{\bBigg@{5}}
\makeatother
\pgfplotsset{compat=1.18}
\begin{document}

\title{\huge A Survey on Intelligent Internet of Things: \\Applications, Security, Privacy, and Future Directions
}
	
\author{Ons Aouedi, Thai-Hoc Vu, Alessio Sacco, Dinh C. Nguyen, \\Kandaraj Piamrat, Guido Marchetto, and Quoc-Viet Pham

\thanks{Ons Aouedi is with SnT, SIGCOM, University of Luxembourg, Luxembourg (e-mail: ons.aouedi@uni.lu).}

\thanks{Kandaraj Piamrat is with Nantes University, École Centrale Nantes, CNRS, INRIA, LS2N, UMR 6004, France  (e-mail: kandaraj.piamrat@ls2n.fr).}

\thanks{Thai-Hoc Vu is with the Department of Electrical, Electronic and Computer Engineering, University of Ulsan, Republic of Korea (e-mail: vuthaihoc1995@gmail.com).}

\thanks{Alessio Sacco and Guido Marchetto are with DAUIN, Politecnico di Torino, 10129 Turin, Italy (e-mail: alessio\_sacco@polito.it; guido.marchetto@polito.it).
}

\thanks{Dinh C. Nguyen is with the Department of Electrical and Computer Engineering, University of Alabama in Huntsville, USA (e-mail: Dinh.Nguyen@uah.edu). }

\thanks{Quoc-Viet Pham (\textit{corresponding author}) is with the School of Computer Science and Statistics, Trinity College Dublin, The University of Dublin, Dublin 2, D02 PN40, Ireland (e-mail: viet.pham@tcd.ie).}
}


\maketitle

\begin{abstract}
The rapid advances in the Internet of Things (IoT) have promoted a revolution in communication technology and offered various customer services. Artificial intelligence (AI) techniques have been exploited to facilitate IoT operations and maximize their potential in modern application scenarios. In particular, the convergence of IoT and AI has led to a new networking paradigm called Intelligent IoT (IIoT), which has the potential to significantly transform businesses and industrial domains. This paper presents a comprehensive survey of IIoT by investigating its significant applications in mobile networks, as well as its associated security and privacy issues. Specifically, we explore and discuss the roles of IIoT in a wide range of key application domains, from smart healthcare and smart cities to smart transportation and smart industries. Through such extensive discussions, we investigate important security issues in IIoT networks, where network attacks, confidentiality, integrity, and intrusion are analyzed, along with a discussion of potential countermeasures. Privacy issues in IIoT networks were also surveyed and discussed, including data, location, and model privacy leakage. Finally, we outline several key challenges and highlight potential research directions in this important area. \end{abstract}
	
	
\IEEEpeerreviewmaketitle

\section{Introduction}
\label{introduce}
The Internet of Things (IoT) has evolved in recent years as a paradigm driving the development of modern companies and smart cities~\cite{broring2022intelliot}. 
IoT enables connections between distributed devices, such as mobile phones, tablets, and computers, which sense and transfer data from external environments to serve end users. The concept of IoT relies mostly on communication between devices for local services, for example, collaborative data collection, and interconnections between devices and the server, for example, cloud servers, edge servers, or data centers, for high-level services, such as data management and network monitoring. 

Modern IoT applications require new service standards and high quality of service, where machine learning (ML) and artificial intelligence (AI) techniques have been integrated across the complete data lifecycle, beginning at the point of data generation or collection by IoT devices, all the way to its utilization by end-users. The integration of intelligent solutions with ML/AI in IoT networks forms a new networking paradigm called {Intelligent IoT (IIoT)}\footnote{The acronym ``IIoT'' can refer to both ``Industrial Internet of Things'' and ``Intelligent Internet of Things''. In the context of this paper, ``Industrial IoT'' refers to the application of IoT technologies in industrial settings for purposes such as improving productivity and efficiency, while ``IIoT'' refers to Intelligent IoT that signifies the incorporation of AI and ML methodologies to make devices and IoT systems smarter and more autonomous}, which has transformed IoT applications such as intelligent healthcare, intelligent transportation, and smart industries~\cite{zhang2021intelligent}. Specifically, IIoT opens up a multitude of opportunities for device advancement, such as equipping local IoT devices with on-device intelligence powered by integrated AI models, and service provision, which includes intelligent data transmission and AI-assisted data management~\cite{zhou2019edge}. Leveraging AI in IoT networks paves the way for economical network operations and increases user experience. This empowers consumers in the IoT to harness the benefits of smart functions such as traffic prediction in intelligent vehicular IoT networks and automated disease detection in healthcare-focused IoT systems~\cite{elbir2022federated}. {Moreover, AI integration into IoT applications presents a broad spectrum of commercial benefits. These include enhanced operational efficiency through intelligent automation and pattern recognition, improved precision cost by eliminating human errors, predictive maintenance to save costs and avoid business disruptions, enhanced customer service for increased satisfaction, and increased scalability to handle extensive IoT ecosystems~\cite{nagaty2023iot}. Commercial applications of AI in IoT are now prevalent across various sectors. For example, manufacturing robots equipped with AI and sensors improve processes and efficiency. Autonomous vehicles use AI to predict behaviors and optimize driving conditions. In retail, AI-driven analytics and smart devices like intelligent shopping carts and security cameras improve customer service and operational efficiency. In healthcare, wearable devices use AI to monitor and provide insights into health metrics, promoting proactive healthcare. Finally, smart agriculture sensors apply AI to optimize resource use and crop yields, demonstrating the transformative potential of AI in IoT across diverse industries.}

\subsection{State-of-the-Art and Our Contributions}
\label{Subsec:Comparison_and_Contributions}

\begin{table*}[t!]
    \renewcommand{\arraystretch}{1.15}
	\caption{Summary of related reviews on IIoT}
	\label{Table:Summary_ExistingSurveys}
	\centering
	\resizebox{\textwidth}{!}{%
	\begin{tabular}{|p{1.30cm}|p{7.25cm}|p{6.25cm}|}
		\hline 
	\multirow{1}{*}{\textbf{References}} & \multirow{1}{*}{\textbf{Contributions}} & \multirow{1}{*}{\textbf{Limitations}} \\
	\hline
	\hline
		
	\multirow{1}{*}{\cite{lin2017survey}} & A survey of IoT with respect to system architecture, enabling technologies, security, and privacy issues, along with the integration of fog/edge computing and its applications. & The application of ML in IoT networks has not been explored.\\ \hline
	\multirow{1}{*}{\cite{al2020survey}, \cite{samie2019cloud}} & A survey on ML/DL in mitigating the security threats in IoT including the discussion of using ML/DL to secure IoT systems.  &  The use of FL in IoT networks and privacy issues have not been discussed.\\ \hline
    \multirow{1}{*}{\cite{amiri2020survey}} & A survey on the use of ML to address privacy issues of IoT, including scalability, interoperability and resource limitations, such as computation and energy. & Security and FL-based solutions have not been presented.\\ \hline
    \multirow{1}{*}{\cite{jamalipour2021taxonomy}} & A survey on ML-based IDS for IoT systems. & The privacy issues have not been explored and discussed. \\ \hline
    \multirow{1}{*}{\cite{chen2021deep}, \cite{lei2020deep}}  & A review of the recent studies on using DRL algorithms to address communication, computing, caching, and control problems so as to enable a wide variety of IoT applications. & The paper only focuses on deep reinforcement learning in IoT systems. \\ \hline
    \multirow{1}{*}{\cite{wu2020research}} & A systematic review of the use of AI in solving the IoT security problems. & The paper only focuses on the use of AI with security aspects in IoT, while its roles with privacy issues have not been presented. \\ \hline
    \multirow{1}{*}{\cite{koroniotis2019forensics}} & A review of DL mechanisms used to investigate botnets and their applicability in the IoT systems. & The paper only focuses on DL mechanisms for botnets in IoT.\\ \hline
     \multirow{1}{*}{\cite{khalil2021deep}} & A survey on DL techniques and their use cases for IIoT systems, including smart manufacturing, smart metering, and smart agriculture.  & The IIoT security aspects were partially covered and the privacy issues have not been discussed.  \\ \hline
    \multirow{1}{*}{\cite{amin2020edge}} & A review of edge intelligence in healthcare IoT, including architecture and frameworks. & The security, threats, and privacy issues have not been explored and discussed.\\ \hline 
    \multirow{1}{*}{\cite{ferrag2023edgelearning}} & A survey on edge learning vulnerabilities and defenses for 6G-enabled IoT. & The IIoT applications and privacy issues have not been explored. \\ \hline
    \multirow{1}{*}{\textbf{This paper}} & An extensive survey on IIoT. Particularly, \newline
    - We extensively discuss the role of ML, DL, FL, and DRL in IoT applications including smart healthcare, smart city, smart transportation, and smart industry. \newline
    - We identify and discuss the security and privacy issues in IIoT systems. \newline
    - We provide a taxonomy table and key lessons learned. The research challenges and directions are also highlighted. 
    & - \\ \hline
	\end{tabular}
	}
\end{table*}

Recent concerns about the security and privacy of the IoT have inspired numerous reviews and analyses. For example, in their work referenced as~\cite{lin2017survey}, the authors presented a comprehensive overview of the IoT, including aspects such as system architecture, enabling technologies, and challenges related to security and privacy. This study also delved into the integration of fog and edge computing and its real-world applications. In addition to this, there has been a marked interest in the potential of ML and Deep Learning (DL)-based models within the IoT domain. These models are particularly notable for their unique ability to tackle complex problems, and various surveys have emerged that explore ML/DL-based models for IoT systems from different perspectives. {For instance, the authors in~\cite{al2020survey} and~\cite{samie2019cloud} examined the challenges involved in utilizing ML/DL models for enhancing IoT security. Furthermore, the work in~\cite{amiri2020survey} provided information on the use of these models to address privacy issues within IoT, focusing on concerns such as scalability, interoperability, and resource constraints such as computational power and energy efficiency. On the other hand,~\cite{jamalipour2021taxonomy} offered an extensive review of ML-based Intrusion Detection Systems (IDS), including the principles, benefits, and drawbacks of DL-based and RL-based IDS, with particular attention to platform-specific IDS applications. The role of deep RL (DRL) within IoT was highlighted in~\cite{chen2021deep} and~\cite{lei2020deep}, with an emphasis on the application of DRL algorithms to manage communication, computing, caching, and control in various IoT scenarios.} 

The complexities related to IoT security protection and the potential of ML in addressing these challenges were further explored in~\cite{wu2020research}. The study also noted the scalability of IoT-enabled botnets, pinpointing this as a significant cybersecurity challenge. In a pioneering survey,~\cite{koroniotis2019forensics} detailed DL techniques and forensic methods specific to botnet detection within IoT environments. Furthermore, the use of DL in industrial IoT was discussed in~\cite{khalil2021deep}, encompassing areas such as smart manufacturing, smart meters, and smart agriculture. Lastly,~\cite{amin2020edge} looked at the deployment of ML and DL models in various layers and nodes of IoT healthcare architectures. A comprehensive comparison of these works and our research can be found in Table~\ref{Table:Summary_ExistingSurveys}, providing a summary of existing surveys and contributions in the field.

Despite the extensive study of IIoT in the existing literature, there appears to be a notable absence of comprehensive reviews focused on the application of ML, DL, {Federated Learning (FL)}, and DRL to address security and privacy concerns within IoT systems. To fill this research gap, we provide a review of IIoT systems. Specifically, we provide a state-of-the-art survey on IIoT applications in smart healthcare, smart cities, smart transportation, and smart industries. In addition, we focus on highlighting security vulnerabilities and privacy threats in the IIoT. The contributions of our work can be summarized as follows.
\begin{itemize}
    \item We present a review of the state-of-the-art IIoT applications, where the use of AI techniques in IoT is investigated and analyzed within different domains, including smart healthcare, smart cities, smart transportation, and smart industry. 
    \item We identify and discuss key security issues in IIoT networks and systems, with a focus on network attacks, confidentiality, integrity, and intrusion. Potential countermeasures are also provided to address these security issues. 
    \item We study privacy concerns in current IIoT networks and applications in three main domains: data privacy leakage, location privacy leakage, and model privacy leakage.
    \item We highlight several key technical challenges and discuss promising directions in IIoT to provide more insight into future research in this important area. 
\end{itemize}

\subsection{Structure of the Survey}

\begin{figure}[t]
    \centering    \includegraphics[width=0.85\linewidth,keepaspectratio]{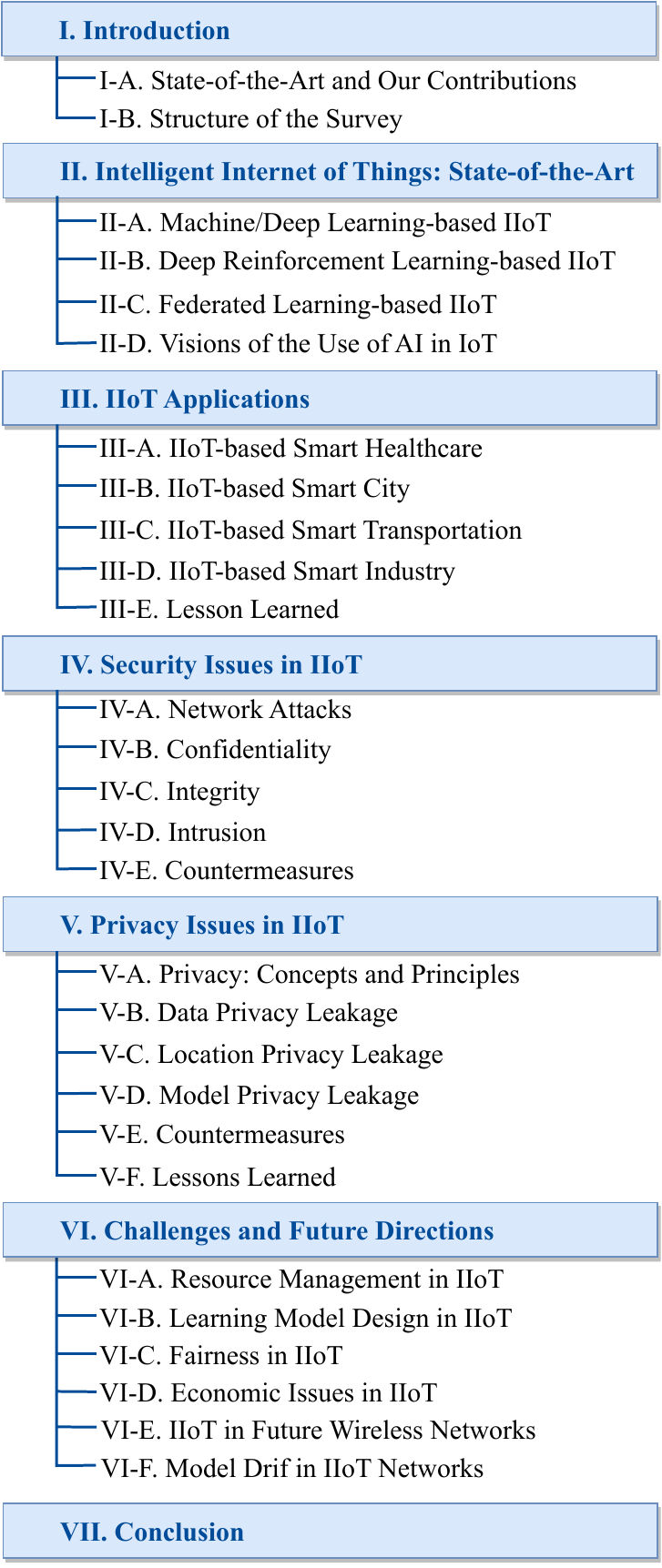}
    \caption{The outline of this survey paper.}
    \label{fig:structure}
\end{figure}

We summarize the structure of this survey in Fig.~\ref{fig:structure}. We start by examining current studies on IIoT in Section~\ref{sec:iiot} and the applications and usages of IIoT in Section~\ref{sec:iiot_app}. Then, we focus on two of the main issues when dealing with such a large amount of data: security (Section~\ref{sec:security}) and privacy (Section~\ref{sec:privacy}). Finally, we summarize the challenges of current solutions in Section~\ref{sec:challenges}, and conclude the paper in Section~\ref{sec:conclusion}.
A list of key acronyms and abbreviations used throughout the paper is provided in Table~\ref{Tab:acronym}.

\begin{table}[!th]
    \renewcommand{\arraystretch}{1.05}
    \caption{{List of key acronyms.}}
    \centering
    \begin{tabular}{|p{1.5cm}|p{5cm}|}
        \hline
        \textbf{Acronyms} & \textbf{Description}\\
        \hline
        {IoT} & Internet of Things \\
        \hline
        {IIoT} & Intelligent IoT \\
        \hline
        {FL} & Federated Learning\\
        \hline
        {ML} & Machine Learning\\
        \hline
        {DL} & Deep Learning\\
        \hline
        {FTL} & Federated Transfer Learning\\
        \hline
        {CNN} & Convolutional Neural Networks\\
        \hline
        {RNN} & Recurrent Neural Network\\
        \hline
        {GRU} & Gated Recurrent Unit \\
        \hline
        {BiGRU} & Bidirectional GRU \\
        \hline
        {LSTM} & Long Short Term Memory \\
        \hline
        {BiLSTM} & Bidirectional LSTM \\
        \hline
        {DNN} & Deep Neural Network\\
        \hline
        {GAN} & Generative Adversarial Network\\
        \hline
        {TL} & Transfer Learning \\
        \hline
        {GNN} & Graph Neural Network \\
        \hline
        {HFL} & Horizontal FL \\
        \hline
        {VFL} & Vertical FL \\
        \hline
        {DRL} & Deep RL\\
        \hline
        {MLP} & Multi-layer Perceptron \\
        \hline
        {COVID-19} &  CoronaVirus Infection Disease\\
        \hline
        {EHR} &  Electronic Health Records\\
        \hline
        {RL} & Reinforcement Learning\\
        \hline
        {DQN} & Deep Q-network \\
        \hline
        {MARL} & Multi-Agent Reinforcement Learning \\
        \hline
        {AE} & AutoEncoder \\
        \hline
        {VAE} & Variational AE \\
        \hline
        {SVM} & Support Vector Machines \\
        \hline
        {KNN} & K-Nearest Neighbors \\
        \hline
        {DT} & Decision Tree \\
        \hline
        {RF} & Random Forest \\
        \hline
        {PCA} & Principal Component Analysis \\
        \hline
        {HAR} & Human Activity Recognition \\
        \hline
        {IoMT} & Internet of Medical Things \\
        \hline
        {CPS} & Cyber-Physical System \\
        \hline
        {IID} & Independent and Identically Distributed \\
        \hline
        {AAL} & Ambient Assisted Living \\
        \hline
        {AR} & Augmented Reality \\
        \hline
        {IDS} & Intrusion Detection Systems \\
        \hline
        {IPS} & Intrusion Preventive Systems \\
        \hline
        {WSN} & Wireless Sensor Network \\
        \hline
        {HE} & Homomorphic Encryption\\
        \hline
        {DP} & Differential Privacy \\
        \hline
        {SMPC} & Secure Multi-party Computation \\
        \hline
        {RFID} & Radio Frequency Identification \\
        \hline
        {OBU} & OnBoard Unit \\
        \hline
        {GDPR} & General Data Protection Regulation \\
        \hline
        {DoS} & Denial-of-Service \\
        \hline
        {DDoS} & Distributed DoS \\
        \hline
        {DBN} & Deep Belief Networks \\
        \hline
        {NN} & Neural Network \\
        \hline
        {{UAV}} & {Unmanned Aerial Vehicle}\\
        \hline
        {{RSU}} & {Road Side Unit}\\
        \hline
        {{SNN}} & {Spiking Neural Networks}\\
        \hline
        {{EV}} & {Electric Vehicle} \\
        \hline
        {{BS}} & {Base Station} \\ \hline
        {{VANET}} & {Vehicular ad hoc Network} \\ \hline
        {{IBE}} & { Identity-Based Encryption} \\
        \hline
    \end{tabular}
    \label{Tab:acronym}
\end{table}
\section{Intelligent Internet of Things: State-of-the-Art}
\label{sec:iiot}

\begin{figure*}[!htbp]
\centering
\includegraphics[width=0.85\linewidth]{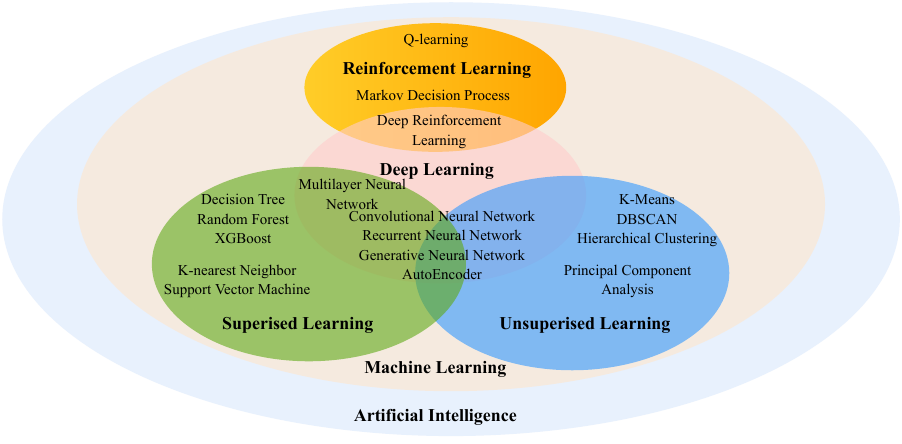}
\caption{AI, ML, DL, and DRL relationships and their main models.}
\label{fig:AIML}
\end{figure*}

In the following section, we provide an in-depth look at the cutting-edge developments in ML, DL, FL, and DRL within the context of IoT systems. Additionally, we explore and discuss the concepts and prospects of integrating these technologies. Fig.~\ref{fig:AIML} shows the classification and the relationship of the main AI techniques.

\subsection{Machine/{Deep} Learning-based IIoT}
\label{subsec:ml_iiot}

AI is a pioneering science that has been in development since 1956. Drawing on knowledge from a diverse array of disciplines such as philosophy, mathematics, economics, neuroscience, psychology, computer engineering, control theory, cybernetics, and linguistics, AI has made significant strides over the years~\cite{russell2010artificial}. ML, a branch of AI, focuses on the development of computer systems that can learn and improve from experience, as outlined in \cite{jordan2015machine}. {Rather than following explicitly programmed instructions, ML strives to generalize the examples provided in a training set. This approach can be described as ``programming by example''~\cite{aouedi2022intelligent}.} ML has emerged as an area of significant interest for both academia and industry, with major players such as Google, Apple, Facebook, Netflix, and Amazon investing in its development. In contrast to traditional programming, the objective of ML in general is to find ``rules'' or build a model using the training data. The model is then applied to unseen data for prediction/classification and automated decision-making processes. {Mathematically, the general formula for a predictive model in ML can be written as follows:
\[y = f(x; \theta),\]
where $y$ is the prediction, $x$ is the input data, $f(\cdot )$ is the model, and $\theta$ represents the model parameters.}



ML can be classified as either \textit{shallow learning} or \textit{deep learning}. The performance of shallow models is heavily dependent on the representation of features, while DL encompasses deep structured learning, hierarchical learning, and deep feature learning~\cite{b137}.
{In recent years, DL has gained significant popularity in IoT systems, mainly due to its exceptional performance when handling large datasets. The improvement in DL-based models can be attributed to their inherent non-linearity. This allows for an increase in model size and consequently the ability to scale efficiently with larger datasets. Furthermore, the vast amounts of data generated by IoT networks have been a driving force behind the success of DL. DL-based models employ intricate combinations of linear and non-linear functions to effectively learn pertinent features~\cite{goodfellow2016deep}. These models consist of two main components, which are activation functions and layers. Activation functions (e.g., ReLU, sigmoid) are responsible for determining the output of neurons given input data. Herein, a neural network function can be represented as \vspace{-0.1 cm}\[y = \sigma(Wx + b),\] where $\sigma$ is a non-linear activation function, $W$ is the weight matrix, $x$ is the input vector and $b$ is the bias vector.}

{Layers in DL refer to organized sets of neurons, which can be fully connected, convolutional, pooling, or recurrent, depending on the specific architecture. For instance, multilayer perceptron (MLP) networks and autoencoder (AE) are built using fully connected layers, while convolutional neural networks (CNNs) are characterized by alternating groups of convolutional and pooling layers. Recurrent neural networks (RNNs), such as vanilla RNNs, gated recurrent units (GRUs), and long-short-term memory (LSTMs), are based on recurrent layers. For a comprehensive introduction to these neural network architectures, please refer to reference~\cite{goodfellow2016deep}. Depending on the specific problem, these models can be trained using supervised, unsupervised, or semi-supervised approaches. In a deep-supervised learning approach, DNN, CNN, RNN, LSTM, and GRU are the different DL techniques mostly used in IoT applications. With deep semi-/unsupervised learning, generative adversarial networks (GAN), and AE are also one of the most popular models. 
}


To achieve ML-based intelligence in IoT systems, a wide range of techniques are used, such as supervised learning, unsupervised learning, semi-supervised learning, and RL ({Fig.~\ref{fig:AIML}}).  \major{\textit{Supervised learning} is the most used approach of ML process in IoT systems. With this learning approach, the models operate by assigning an observation to a predefined class based on a set of observed features related to that observation. Instead, \textit{unsupervised learning} aims to discover relationships between input data without prior knowledge of the output. In particular, unsupervised learning seeks to categorize input data into distinct clusters (i.e., groups) by analyzing the similarities between the input data. Finally, \textit{semi-supervised learning} combines supervised and unsupervised learning and incorporates labeled and unlabeled data to train the model. As labeling the data can be challenging and time-consuming, semi-supervised learning aims to mitigate these issues by utilizing a few labeled examples alongside a large collection of unlabeled data. Such a learning approach is particularly suitable when large amounts of unlabeled data are available, as is often the case with network traffic. Consequently, in recent years, there has been a growing interest in semi-supervised learning within IoT systems. On the other hand, the main idea of RL was inspired by biological learning systems. It is different from supervised and unsupervised learning, where, instead of trying to find a pattern or learning from a training set of labeled data, the only data source for RL is the feedback that a software agent receives from its environment after performing an action~\cite{b136}.}

\subsection{Reinforcement Learning-based IIoT}
\label{subsec:rl_iiot}

RL is a subfield of ML that emphasizes sequential decision-making under uncertain conditions. 
Unlike other learning approaches, RL does not require predefined target outcomes; its primary objective is to enable an agent to maximize its cumulative reward by taking a series of actions in response to a dynamic environment. RL has been successfully applied to a wide range of IoT problems, including robotics, autonomous vehicles, recommendation systems, and optimization of network resources~\cite{chen2021deep, sacco2022partially}. In RL, an agent interacts with its environment by performing actions, receiving feedback in the form of rewards or penalties, and updating its knowledge based on these experiences~\cite{b136}. The agent's goal is to learn an optimal strategy for selecting actions that maximize the expected cumulative reward over time. This learning process typically involves exploring the environment to gather new information and taking advantage of the acquired knowledge to make better decisions. {The key formula in RL is the update rule for Q-learning, given by \[Q(s,a) \leftarrow  Q(s,a) + \alpha [r + \gamma \max_{a'} Q(s',a') - Q(s,a)],\] where $ Q(s,a)$ is the value of taking action $a$ in state $s$, $r$ is the reward, $\gamma $ is the discount factor and $\alpha$ is the learning rate. 
}

In the context of highly dynamic IoT networks, the development of decision-making strategies learned by agents is a promising research direction. The adaptive capabilities of these agents allow them to respond to changes effectively. Moreover, in an IIoT ecosystem, various devices often need to work together to achieve a common goal, and RL provides a framework for these multi-agent systems. It allows devices to learn, cooperate, and perform tasks more efficiently. Additionally, RL can also play a role in edge computing, a critical aspect of IIoT, where computations are performed closer to where the data are generated. RL can help optimize data processing and management tasks on the edge, reducing latency and bandwidth requirements. As a result, within the IIoT ecosystem, RL can have essential capabilities by enabling more efficient, adaptable, and autonomous systems, improving the overall effectiveness and utility of IIoT applications~\cite{kokkonen2022autonomy}. 


\begin{table*}[!htbp]
\centering
\caption{Usage of AI Paradigms in IIoT Applications.}

\scalebox{1.0}{

\begin{tabular}{|l|p{8.75cm}|p{5.75cm}|}
\hline
 \textbf{Paradigm} &  \textbf{Characteristics} &
\textbf{Usage within IIoT} \\
\hline
Shallow ML & - {Fast training.} \newline
- {Perform well on small datasets.} \newline
- {Tend to be more interpretable.} \newline
- {Process data in real-time.} & - {Detect faults or anomalies in manufacturing processes.} \newline
- {Predict energy consumption patterns of IoT devices.} \\
\hline
DL models & - {Learn hierarchical representations of data.} \newline
- {Capture intricate patterns and relationships in different types of data.} \newline
- {Process vast amounts of data.} \newline
- Outperform shallow ML-based models in many complex IoT applications. & - {Analyze visual data from surveillance cameras or drones.} \newline
- {Capture temporal and spatial dependencies for predictive maintenance.} \newline
- Integrate data from multiple sensors in IoT systems such as vision, audio, and sensor data. \\
\hline
DRL & - {Learns directly from interactions with the environment.} \newline
- {Focuses on sequential decision-making tasks.} \newline
- Learn efficiently from limited interaction samples with the environment. \newline
- {Leverages transfer learning.} & - {Empowers autonomous systems in IoT.}
\newline 
- {Optimizes resource allocation in IoT networks.} \newline
- {Optimizes irrigation schedules.}
\\
\hline
Federated Learning & - {Ensures that sensitive or private data does not communicated.} \newline 
- {Operates on decentralized data sources.} \newline 
- Promotes collaborative learning among the decentralized devices or servers. \newline 
- {Accommodates a large number of devices participating in the training process.} & - Allows IoT devices to collaboratively train models for anomaly detection. \newline
- {Enables edge devices to collaboratively train models.} \newline 
- Supports traffic prediction in IoT-enabled systems.\\
\hline
\end{tabular}
}
\label{Modelusage}
\end{table*}

\subsection{Federated Learning-based IIoT}
\label{subsec:fl_iiot}

Innovative AI solutions that can withstand both data delays and data sensitivity are required for IoT applications so that they can operate locally without sending data to a centralized organization~\cite{nguyen2021federated, le2023applications}. Due to the distributed and privacy-enhancing features of FL, some IIoT applications have undergone significant changes since their introduction ~\cite{mcmahan2017communication}, FL is a decentralized learning strategy in which training data that cannot be shared with third parties are distributed across devices~\cite{konevcny2016federated}. In particular, FL aims to build a global model that minimizes the average loss function on local datasets. FL is one of the most effective solutions for creating distributed IoT systems due to recent improvements in mobile technology and growing concerns about privacy leakage. To ensure that user data remain at the point of origin and are not shared directly with third parties, FL delegated AI functions, such as data training, to the edge of the network where IoT devices are located. In terms of network resource conservation and privacy enhancement, this strategy enables cooperative training of a shared global model, offering benefits to both network operators and IoT consumers. 

As a result, FL becomes a solid substitute for traditional centralized models, accelerating the implementation of IoT services and applications on a larger scale. IoT devices and the aggregated server are the two key parts of the FL concept in IoT networks, and training can take place in a centralized or decentralized manner. One of the most commonly utilized FL topologies in FL-IoT systems is centralized FL. A weighted averaging approach, such as Federated Averaging (FedAvg)~\cite{lim2020federated, le2023applications}, is used to aggregate the models learned from each IoT device, which trains the model locally using its private data. {The formula for FedAvg can be mathematically represented as \[\theta = \sum_{k=1}^{K} \frac{n_k}{n} \theta_k,\] where $ \theta$ is the global model parameters, $K$ is the number of local models, $n_k$ is the number of data points in the $k$-th local dataset, $n $ is the total number of data points, and $ \theta_k$ are the parameters of the $k$-th local model.}

Using the received global model, the training is performed on the consensus model that the server has returned. In a decentralized or peer-to-peer FL system, the process of model aggregation does not rely on a central server. As an alternative, each device shares its trained models with some or all of its peers, and each one is responsible for its aggregate.

Additionally, FL may be separated into Horizontal FL (HFL), Vertical FL (VFL), and Federated Transfer Learning (FTL), depending on how the client data are dispersed. IoT devices use HFL, which uses multiple sample spaces while maintaining the same feature space. Customers may use the same model for in-person training due to the characteristics of constant data. An example of HFL is intrusion detection~\cite{aouedi2022federated} in the intelligent industry. Representative and low-dimensional features are learned by training an AE model on each device (using unlabeled local data). Then, a cloud server uses FedAvg to combine these models into a global AE. The cloud server then builds a supervised neural network by including fully connected layers on top of the global encoder and trains the resultant model using data that have been tagged and are readily available online. In contrast to HFL, the learning of shared models is addressed by VFL on devices with the same sample space but several data feature spaces ~\cite{feng2020multi}. A smart city's shared learning model, which includes e-commerce businesses and financial institutions, is an illustration of VFL in IIoT applications. Different data characteristics that serve city clients inside a smart city, an e-commerce business, and a bank can participate in a VFL process to collaboratively train a model utilizing their datasets. These datasets might contain past user payments from online retailers and bank user account balances. This approach enables VFL to predict the best-tailored loans for each consumer based on their online buying habits. Last but not least, the FTL combines the FL and transfers learning ideas to~\cite{liu2020secure}. By adding additional learning clients who have datasets with various sample and feature spaces, it seeks to increase the sample space in the VFL architecture. FTL can help detect disease in IoT networks by promoting cooperation between various nations with numerous hospitals that have unique patients (sample space) and a variety of diagnostic procedures (feature space). By doing this, the FTL may enhance the output of the shared model, increasing diagnostic precision.


\begin{figure*}[!htbp]
\centering
\includegraphics[width=0.75\linewidth]{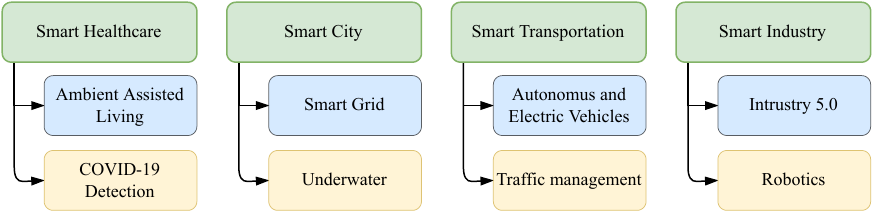}
\caption{IIoT application domains.}
\label{IIoT_app}
\end{figure*}

\subsection{Visions of the Use of AI in IoT}

IIoT represents the fusion of advanced technologies, such as ML, DL, RL, FL, data analytics, cloud, and edge computing, to create smart devices that communicate with each other through the Internet. By integrating these technologies, vast amounts of data can be harnessed for big data analytics, leading to valuable insights and scientific knowledge. In particular, ML and DL models can be used to create data analytics tools that can handle information collected from a variety of IoT devices, including sensors, smartphones, computers, and radio frequency identification (RFID) devices. These models are capable of handling a wide range of IoT data types, including enormous data volumes (data streaming from millions of sensing devices) and many modalities (images, time series data, video and text)~\cite{mohammadi2018deep, le2023wirelessly}. In an unmanned aerial vehicle (UAV)-assisted cellular offloading situation, for instance, the study~\cite{zhong2021multi} investigated the combined optimization of the UAV trajectory design and power allocation. By coordinating many devices to execute model training in IoT devices, FL enables IIoT to achieve better scalability and privacy protection while guaranteeing that data stay in the location where it was created. Table~\ref{Modelusage} provides an overview of how the different ML, DL, DRL, and FL paradigms can be used in IIoT applications.

Although the integration of AI and IoT allows IoT devices not only to collect and communicate data, but also to learn from it, and to make decisions based on those data, security and privacy remain critical concerns in the IIoT landscape. With devices collecting, processing, and storing enormous volumes of data, a significant portion of which can be sensitive or personal, there is a mandatory need to develop efficient privacy preservation methods. In IIoT, devices are continuously communicating with each other and with the cloud, requiring secure communication channels to guarantee confidentiality and integrity. Moreover, AI models deployed in IIoT can become targets for adversarial attacks, aimed at manipulating the model's behavior. Addressing these challenges requires multidisciplinary research, including areas such as cryptography, ML/DL-based models~\cite{aouedi2023f}, hardware design, and network security. It is also worth noting that any solution needs to take into account the specific constraints and requirements of IIoT, such as resource limitations, the need for real-time operation, or the scalability to large numbers of devices~\cite{aouedi2024towards}.
 

\section{IIoT Applications}
\label{sec:iiot_app}

As stated in the previous sections, the integration of AI/ML into the IoT has favored a new plethora of novel applications.
This section provides an extensive discussion of such representative IIoT applications, including smart healthcare, smart cities, smart transportation, and smart industries, along with applied use cases, as shown in Fig.~\ref{IIoT_app}.

\subsection{IIoT-based Smart Healthcare}

Recently, patient monitoring has been a crucial concern; therefore, healthcare is projected to be the dominant application of IIoT systems~\cite{al2015internet}. Advances in IIoT-enabled systems may lead to significant benefits in healthcare, allowing medical personnel to identify anomalies earlier, monitor and prognosis critical diseases, and reduce potential risks for critically ill patients as much as possible. The ubiquity of wearable, lifestyle and assistive IoT devices helps to collect massive amounts of data to train and improve the performance of ML-based models and, in turn, provide better healthcare services. For instance, the study in~\cite{walinjkar2017personalized} used heart rate variability for the prediction of arrhythmias with the KNN classifier, which led to a high accuracy of 97\% in detecting cardiac arrhythmias in real time. Here, we focus on examining the functions of IIoT in healthcare care using two use scenarios, namely COVID-19 detection and ambient assisted living (AAL).

\subsubsection{Ambient Assisted Living (AAL)}

AAL technologies are designed to support people without obstruction in their daily activities, enhancing their quality of life, safety, and independence. IIoT can play an important role in this area by monitoring health parameters, creating a more comfortable and manageable living environment, and reducing feelings of isolation and loneliness. Sleep monitoring is one of the services associated with AAL as appropriate sleep is an integral part of a healthy lifestyle. The recent development of smartphone applications and sleep sensing devices, such as electroencephalograms (EEG) and electrooculograms (EOG), has increased the analysis of sleep patterns. To detect changes in sleep patterns, the work in~\cite{hong2017multivariate} proposed an accurate deep hybrid model followed by a K-Mediod algorithm. It combines both Deep Belief Networks (DBNs) for unsupervised feature learning and {LSTM} to improve classification performance and learn long-range dependencies in time series data. The features used to train these models were total sleep duration, sleep efficiency, hypnagogic time, number of awakenings, self-evaluations of sleep, and physical condition. Longitudinal sleep sequences consisting of daily sleep types were clustered using a K-Medoid algorithm to find the patterns of each sleep sequence cluster. 
Another approach to monitoring posture during sleep by using a pressure sensor mattress that records pressure in different areas of the mattress has been proposed in~\cite{matar2016internet}. The proposition consists of two phases: in the first phase, the sensor data are pre-processed using Principal Component Analysis (PCA) for feature extraction and dimensionality reduction, whereas the second phase uses the extracted features to train and then the SVM model to classify the data into prone/supine, left, and correct lateral positions. 

Although PCA can learn and extract relevant features, its linear combination of features does not have sufficient capability to model complex non-linear dependencies. To overcome these issues, CNN models have represented important research on AAL trends. This is because of their capability to extract relevant features and classify large-scale images. It has certain advantages over traditional shallow learning techniques for image treatment and analysis, particularly in AAL and Human Activity Recognition (HAR). The first CNN-based approach for HAR was proposed in~\cite{zeng2014convolutional}. It automatically learns and extracts human activity features without any domain knowledge (such as activities in the kitchen, jogging, walking, etc.). The experimental results showed that the CNN-based approach can outperform the PCA method. Similarly, the authors of~\cite{yang2015deep} investigated the efficiency of a CNN by comparing its performance with state-of-the-art methods using benchmark datasets. This study shows that it serves as a competitive tool for feature learning and classification for HAR applications. Furthermore, an end-to-end CNN model was proposed that showed good results in predicting three movements of the arm performed in daily activities in~\cite{panwar2017cnn}. Despite the performance of the CNN model, its complex architecture due to the growing number of hyperparameters may increase computational costs and make it unsuitable for AAL applications with strict latency constraints. Therefore, a model that can balance the trade-off between accuracy and computational cost is required. To do so, the authors of~\cite{cheng2022real} proposed, for the first time, a computation-efficient CNN by replacing conventional convolutions with the conditionally parameterized convolution algorithm (CondConv). Using WISDM, PAMAP2, UNIMIB-SHAR, and OPPORTUNITY datasets, the proposed solution always performed better than its counterpart without CondConv without compromising inference speed.

However, CNN-based models sometimes fail to extract temporal features, and capturing these temporal dynamics is fundamental for successful AAL applications. To overcome these limitations, CNN-LSTMs have been proposed as a combination of CNN and LSTM layers. In this context, the authors of~\cite{ordonez2016deep} proposed a DL framework composed of convolutional and LSTM models, called \texttt{DeepConvLSTM}. The convolutional layers provide abstract representations of the input sensor data, the LSTM models, and the temporal dynamics of the activation of the extracted features. Performance evaluation against CNN baseline shows that \texttt{ DeepConvLSTM} achieves a higher F1 score with the OPPORTUNITY dataset.

Processing of data collected from all users by a central body is often necessary for these systems to operate well. Data in healthcare systems are extremely sensitive and governed by laws such as the United States Health Insurance Portability and Accountability Act (HIPPA) compared to other IIoT applications~\cite {drolet2017electronic}. As a result, operators have trouble collecting a lot of data, especially if patients and hospitals are reluctant to provide their sensitive information. The latency and storage resources that arise when patient-sensitive data need to be collected in the cloud for model training are the next challenge for standard ML models. As a consequence, FL has been used as a promising alternative to deploy IIoT, as it provides intelligence with privacy awareness and without the need to share end-user data. Several research efforts have been devoted to the use of FL for AAL, enabling intelligent solutions in healthcare systems and protecting privacy-sensitive medical data~\cite{aouedi2022handling}. For instance, the authors of~\cite{zehtabian2021privacy} proposed an FL-based approach to the predictor of HAR in a smart environment in a realistic scenario. The designed approach uses the LSTM model and trains it in three scenarios: local, centralized, and federated. Using the CASAS dataset, the evaluation results show that the FL approach can provide competitive results compared to centralized learning.

Labeling data for FL-based models is frequently challenging and time-consuming because FL model training typically takes place on equipment that is out of reach for humans. Therefore, to scale up an HAR system over a large number of devices, researchers have begun to reformulate FL as a semi-supervised issue~\cite{presotto2021semi}. According to the semi-supervised FL for HAR proposed in the work in~\cite{zhao2020semi}, clients would locally train an AE model using their unlabeled data, and the server would then include the local AE models that were created into the pipeline of the supervised learning process. Their experimental findings demonstrate that HAR with semi-supervised FL may reach accuracy on par with supervised FL without being unaffected by non-independent and identically distributed (non-IID) data. Similarly, the work in~\cite{bettini2021personalized} combines Active Learning (AL) with label propagation, dubbed textttFedHAR, to federated semi-automatically annotate the end-user data. The simulation results using the two well-known datasets MobiAct and WISDM show that the labeled data produced by AL offer pertinent informative data and, consequently, allow \texttt{FedHAR} to achieve competitive performance with a fully supervised FL model (FedAvg).

Furthermore, to reduce the impact of non-IID data, transfer learning techniques have been introduced on the client side to improve personalization. For example,~\cite{wu2020personalized} investigates the emerging personalized FL method by proposing  \texttt{PerFit} framework that can mitigate the negative effects caused by statistical heterogeneities. The authors investigated the performance of two customized FL-based solutions: FTL and federated distillation (FD) using the \texttt{PerFit} framework. With FTL, each client adjusts the model they downloaded from the cloud server using their unique data, but with FD, each client may create their model based on their unique needs. The experimental findings demonstrate that customized FL can reduce the performance loss caused by non-IID data and that both FTL and FD are capable of collecting user data.

Additionally, using extremely small Internet of Medical Things (IoMT) devices, also known as nanodevices, may be necessary to distribute healthcare services across larger regions. These units, called straggler clients, may respond very slowly or may not be able to complete the anticipated number of local iterations. There are a few ways that can prevent choosing these customers throughout the training process. However, dropping stragglers can diminish the number of active clients and, consequently, the performance of the model~\cite{aouedi2022handling}. To solve this problem, the authors of~\cite{imteaj2021fedparl} introduced a lightweight Federated Proximal, Activity, and Resource-Aware model for a resource-constrained IoT environment, called \texttt{FedPARL}. To allow the client to accommodate the reduced model sizes in the first layer, \texttt{FedPARL} conducts sample-based model pruning on the server. To choose competent and reliable clients for training, the second layer of \texttt{FedPARL} additionally looks at previous actions and resource availability (CPU, memory, battery life, and data volume). By allocating local epochs in the third layer following the client's resource availability, partial work was made possible. The results show that \texttt{FedPARL} achieves an improved stable accuracy compared to the FedAvg and FedProx approaches. On the other hand, the authors of~\cite{shaik2022fedstack} proposed a heterogeneous federated stacking model to monitor patient activities based on classification models, called \texttt{FedStack}. This model can process a variety of client-model architectures and group the local models into a global robust model, which in turn can reduce the impact of the stragglers' clients on the classification performance. To do so, three different DL-based models (MLP, CNN, and LSTM) were used and individually trained on each data set. Then, the model predictions were stacked homogeneously and heterogeneously and then passed to the global model. The results show that the proposed \texttt{FedStack} achieves better accuracy than several baseline models for HAR.

\subsubsection{COVID-19 Detection}

Recently, the COVID-19 pandemic has reached 214 countries and regions across the world, profoundly affecting daily life in those areas. The potential of IIoT to support the fight against COVID-19 has been investigated in~\cite{aouedi2022handling, pham2020artificial}. The IIoT can help with timely quarantine and medical treatment. The use of ML/DL models with computed tomography (CT) images for COVID-19 detection was considered in several works~\cite{shi2020review}. For example, in~\cite{barstugan2020coronavirus} an early detection of COVID-19 based on SVM has been proposed. Features were extracted through several features extraction methods (Grey Level Co-occurrence Matrix (GLCM), Local Directional Pattern (LDP), Grey Level Run Length Matrix (GLRLM), Grey-Level Size Zone Matrix (GLSZM), and Discrete Wavelet Transform (DWT)) followed by SVM as a classifier. Despite the performance of the SVM, it cannot scale very well. To overcome the scalability problem of the shallow model CNN-based models have been used for chest CT image treatment and processing~\cite{zheng2020deep}. In this context, to classify COVID-positive and COVID-negative, the work in~\cite{narin2021automatic} used different CNN models, which are pre-trained models ResNet50, ResNet101, ResNet152, InceptionV3, and InceptionResNetV2 with three binary data sets that include X-ray images of patients with normal (healthy), COVID-19, bacterial, and viral pneumonia. The result is promising, especially since ResNet50 and ResNet101 have achieved performance with 96.1\%. It should be noted that the COVID-19 dataset~\cite{cohen2020covid} and Kaggle’s Chest X-ray Images (Pneumonia) are also used to form the data set in this study. 

Although the efficiency of the above models was observed, the similarity of different datasets was not considered. At the same time, making knowledge transfer inter-domains possibly leads to a more precise and personalized model, and one way to achieve this is via TL. TL, which is widely used in the field of DL, may be a promising solution for COVID-19 detection. It enables the detection of various COVID-19 using small medical image datasets, helps to avoid learning from scratch, and, in turn, reduces energy consumption. For instance, the study in~\cite{apostolopoulos2020covid} suggested that transfer learning can extract relevant features related to COVID-19 disease and, in turn, improve detection performance. In a similar work~\cite{pathak2020deep}, a deep TL model is used to classify COVID-19 infected patients and showed that DTL is a useful approach to the classification of COVID-19. In addition, FL has also been used to detect positive cases by training models from isolated medical institutions. Each medical facility that uses FL participates in training using its own local COVID-19 images, such as X-ray and CT scans, and only model parameters are exchanged-sensitive user data are not necessary. In this regard,~\cite{feki2021federated} proposes a multi-institutional collaborative FL framework for COVID-19 identification. Similarly, \cite{wang2021auxiliary} has suggested a 5G-enabled architecture for COVID-19 diagnostics that communicates with several hospitals through FL while protecting privacy. The authors of~\cite{zhang2021dynamic}also suggested a dynamic fusion-based FL technique to increase the performance of the model and the effectiveness of communication to identify patients infected with COVID-19. In addition to the aforementioned applications, employ DL, such as the LSTM model in real-time to anticipate the dynamics of the COVID-19 epidemic to improve health and policy initiatives~\cite{sarumi2022potential, alazab2020covid}.

\subsection{IIoT-based Smart City}

IIoT plays a crucial role in smart city development, as it can facilitate data collection, exchange, and analysis, leading to improved services, efficiency, and quality of life. There are several key domains in which IIoT can offer useful services for smart cities, including smart grid and water management.

\subsubsection{IIoT for Smart Grids}

With the rapid growth of the population and the increase in the number of industries, traditional grids for reducing power consumption and forecasting demand have become inefficient. To address these issues, an intelligent infrastructure is needed for the successful processing of large amounts of data. Over the past decade, we have seen a shift toward smart grids, which has replaced traditional power grids largely~\cite{syed2020smart}. Due to recent advances in IIoT, smart grids can intelligently learn patterns and ensure intelligent and automated power grids. Smart grids are among the most prominent applications of the IIoT, with their convergence unveiling considerable potential for modernizing electric power systems. They established a fully automated and intelligent energy delivery network by integrating all users connected to them~\cite{baumeister2010literature}. Beyond their role in energy networks, smart grids play a crucial role in the future of renewable energy systems. They improve the reliability of power supplies, help reduce $CO_{2}$ emissions, and promote the adoption of green energy sources~\cite{reka2018future}. Therefore, with the integration of smart grids and IIoT infrastructure, end users can analyze and monitor their daily energy consumption patterns. This enables greater control and management of energy use, offering the potential to schedule energy use through a simple mobile application.

Recently, ML has also been applied to energy management in smart grids. An example is presented in~\cite{ahmed2020machine} where an ML-based energy management framework was proposed to mitigate energy management problems on the demand side. This approach integrated Gaussian process regression (GPR) with ML to develop an efficient energy management model (EMM). Performance parameters calculated from an optimized model for Prosumer Energy Surplus (PES), Prosumer Energy Cost (PEC), and Grid Revenue (GR) were fed into the ML-based GPR system for training. This adaptive framework of service level agreements (SLAs) between energy consumers and the grid benefits all stakeholders involved. Furthering the integration of ML in smart grid systems, \cite{alazab2020multidirectional} introduced a smart grid system augmented with a Cyber-Physical System (CPS) model. They proposed a multidirectional LSTM, known as the (\texttt{MLSTM}) model, which is specifically designed to predict the stability of a smart grid network. The experimental results validate that the \texttt{MLSTM} approach outperforms other conventional ML methods in terms of performance, emphasizing the power of advanced ML techniques in this domain.

Furthermore, the integration of FL with smart grids presents a novel approach that benefits from the power of decentralized machine learning. However, during FL training rounds, communication between clients and the server may fail due to the time-varying reliability properties of links in a wireless network of smart grids. Such communication failures not only hinder the convergence rate of the model, but also lead to resource wastage, including energy expended on unsuccessful local training. Recognizing these challenges, \cite{zhai2021dynamic} introduced a dynamic FL problem in a grid mobile edge computing (GMEC) environment, considering the high dynamism of link reliability. To mitigate communication failures between industrial clients and the server, the authors proposed a delay-deadline-constrained FL framework. This structure aims to prevent excessively long training delays by formulating a dynamic client selection problem, maximizing the computing utility while also minimizing the communication latency during the FL process. At the same time, several other FL algorithms have been formulated to address similar challenges. For example, \cite{cao2020ifed} presented a framework for power learning in IoT networks comprising electricity providers and users. The distinctive communication model they developed was based on the FL process. This model seeks to balance the trade-offs between resource consumption, user utility, and local differential privacy, thereby presenting a comprehensive and holistic approach to managing smart grid networks using FL. In this context, the study in~\cite{su2021secure} proposed an FL approach for the secure and efficient sharing of personal energy data in smart grids with edge-cloud collaboration. This hierarchical FL framework promotes the analysis of energy data that preserves privacy and is communication efficient within smart grids. This approach considered non-IID data from heterogeneous users, which inspired the development of a local data evaluation model for cost modeling and two optimization problems specifically tailored for energy service providers and energy data owners. In the final phase, the study introduced a DRL-based incentive algorithm designed to manage multidimensional user private information and vast state spaces. This strategy facilitates the determination of optimal pricing strategies for energy data providers and optimal training strategies for energy data owners. Similarly, the research conducted by \cite{tun2021federated} proposed a cloud-edge FL-based strategy that involved clustering of the FL-based energy demand predictor system. This innovative system groups clients with similar attributes, which, in turn, allows the aggregation of model updates from clients within the same cluster.  This novel approach leverages the synergy between advanced ML techniques and the capabilities of smart grids, indicating promising new directions for research and application in this field.

\subsubsection{IIoT for Underwater}

The demand for water from industries, factories and mining continues to grow with increasing urbanization. Simultaneously, there is an increase in wastewater disposal without appropriate treatment from natural sources, which also pollutes unpolluted water~\cite{zhao2020application}. The implementation of smart water management mechanisms for effective distribution, conservation, and maintenance of water quality is now more crucial than ever. For example, in agriculture, significant challenges are associated with access to water, efficient use, and the incorporation of sustainable practices for water conservation and harvesting.

\begin{figure}[t]
\centering
\includegraphics[width=0.98\linewidth]{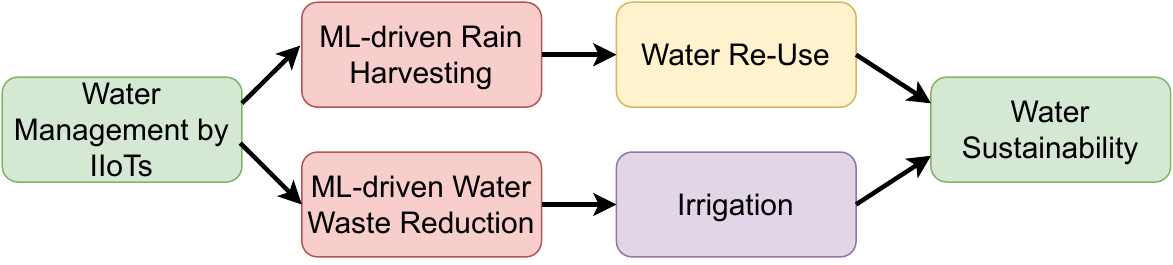}
\caption{IIoT for water management parameters.}
\label{water_manegm}
\end{figure}

Technological progress in the field of industrial IoT presents an exciting prospect of significant improvements in underwater exploration, environmental monitoring, and industrial processes~\cite{hou2021machine}. These technologies can play an instrumental role in creating an intelligent water management system (illustrated in Fig.~\ref{water_manegm}), which comprehensively covers aspects such as wastewater management, irrigation, rainwater harvesting, water reuse, and sustainable sourcing from natural resources. The study carried out in Tunisia as a joint project with the Water Production and Management of Water capitalized on the integration of AI and IoT technologies to improve productivity by reducing wasteful consumption and improving user access to accurate and timely information~\cite{ktari2022lightweight}. The study introduced a novel technique to monitor water consumption using an optical character recognition device (OCR) in conjunction with the YoLo 4 ML model, focusing on smart cities' paradigms. Meanwhile, other researchers have focused on the simulation of water quality and levels. For instance, a study in the Nakdong River basin of South Korea conducted such simulations, taking into account parameters such as organic carbon, phosphorus, and nitrogen content~\cite{baek2020prediction}. Another study explored the application of DL, proposing integrated automatic detection models that used U-net and CNN to identify temporal resolution and high spatial imagery systems that are the key to mapping pivotal irrigation systems in the center~\cite{saraiva2020automatic}. These advancements underscore the transformative power of IIoT in redefining the proposed approach toward sustainable and intelligent water management.

On the other hand, increasing water pollution around the world is an endangering factor for water quality. To predict water quality, the work in~\cite{wang2017water} used a BiLSTM with time-series data. The model used monthly data collections from quality reports for six years (2013-2019) on the Yamuna River, New Delhi. The BiLSTM model was unique in its approach, as it focused not only on training, but also on missing value imputation, thereby ensuring a comprehensive analysis of the available data. In the same context, the authors of~\cite{prasad2022analysis} demonstrated that the usage of auto DL techniques for the determination of water quality gives better results. Auto DL, one of the most recent and promising technologies, enables straightforward interpretation and model creation with possibly minimal coding requirements. Its application in this domain yielded favorable results, further validating the potential of Auto DL in environmental analysis. Conventional Internet of Underwater Things (IoUT) systems frequently employ centralized learning to ensure efficiency, reliability, and timeliness. Recent research has emphasized the significance of privacy and security in mission-critical IoUT frameworks~\cite{victor2022federated}. FL as a secure, decentralized ML system can be used to address the IoUT challenges. The study outlines FL's potential applicability in IoUT, highlighting its issues, unresolved topics, and future research directions. In this context, the authors of~\cite{zhao2021federated} proposed a federated meta-learning enhanced acoustic radio cooperative framework for the Ocean of Things, which takes advantage of the data distributed on the surface nodes. This continuous exploration of advanced FL techniques in IoUT projects a promising future for this field.


\subsection{IIoT-based Smart Transportation}

The recent development of IIoT-based smart transportation systems improves the safety and sustainability of urban mobility and therefore makes daily life more convenient and efficient. It can facilitate smart transportation by treating and analyzing the data collected from a wide range of sensors to make decisions in real time. Existing researchers have made great efforts to investigate different aspects of applying IIoT in smart transportation development, including autonomous and electric vehicles, as well as traffic management systems.

\subsubsection{Autonomous and Electric Vehicles}

Intelligent and autonomous driving has become possible with the help of IIoT systems. Traffic sign recognition is a crucial task in advanced driver assistance and autonomous systems. Manual localization and recognition of traffic signs can be extremely time-consuming when applied to thousands of kilometers of roads. Therefore, the automatic and smart detection and recognition of traffic signs play a pivotal role in smart transportation. In this context, IIoT is used to support vehicles in autonomously obeying traffic rules by correctly recognizing traffic signs~\cite{balali2015detection}.

The following aspects highlight IIoT's impact on traffic sign detection and recognition:
\begin{itemize}
    \item \textit{IIoT for Traffic Sign Detection and Recognition}: Shallow models, such as decision trees and SVM classifiers, are widely used for the recognition of traffic signs due to their ease of training and updating. These models achieved good results~\cite{zaklouta2012real}~\cite{chen2019fire}. Using feature learning based on real examples can help easily adapt and capture a high degree of variability in the appearance of a large number of traffic signs. The authors of~\cite{tabernik2019deep} use the Mask Region-CNN (R-CNN) model for large-scale detection and recognition of traffic signs through automated end-to-end learning. The simulation results showed excellent performance in the localization and recognition of traffic-sign instances. Based on~\cite{tabernik2019deep}, the authors of~\cite{yang2018deep} added an attention network to the fast R-CNN model to extract more relevant features from the input data. Another traffic sign recognition system is presented in~\cite{arcos2018deep}. The authors used multiple spatial transformers and extensively studied their impact on network configurations within the CNN model. 
    The use of DL-based models to identify traffic signs in real-time with limited equipment resources was proposed in~\cite{zhang2020lightweight}. The authors focused on reducing the computational cost of the model, making it suitable for IIoT scenarios. To do so, a knowledge distillation-based model is proposed, in which the shallower student model is trained through the softened output of the teacher model on the target datasets. The performance of the proposed model demonstrates that a lightweight network can reduce the number of redundant parameters while maintaining comparable accuracy. Similar to the healthcare service, collecting traffic sign data in the cloud presents serious privacy leakage risks owing to the inclusion of location privacy information. To preserve the privacy of traffic sign data with limited computing resources, the work in~\cite{xie2022efficient} proposed \texttt{FedSNN}. Without disclosing any personal information, the authors employed FL to carry out cooperative training for accurate recognition models. They also proposed Spiking Neural Networks (SNNs), a third generation of neural networks, to further save energy and processing resources. As shown in Fig.~\ref{fl_snn}, the networked vehicle trains the SNN model using the local traffic sign dataset during each communication round. The trained model is subsequently delivered to a Road Side Unit (RSU), where it is utilized for global aggregation, after data encoding.

\begin{figure}[ht!]
\centering
\includegraphics[width=\linewidth]{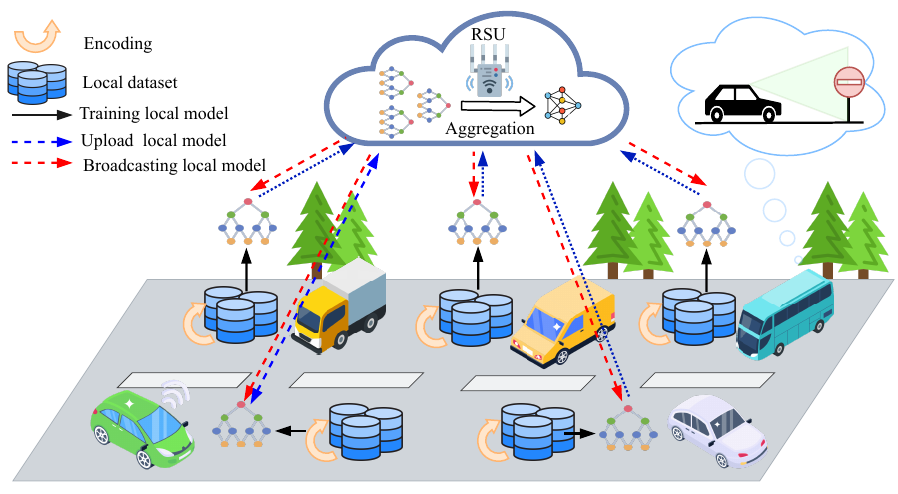}
\caption{Overview of FedSNN for road traffic sign recognition~\cite{xie2022efficient}.}
\label{fl_snn}
\end{figure}

    \item \textit{Energy consumption of electric vehicles}: According to the International Energy Agency~\cite{b35}, the number of electric vehicles (EV) on the road will increase tremendously by more than 4000\% in 2030. This trend will lead to an explosion in energy consumption, and providing convenient charging services is of critical importance to enhance the user experience. As a result, many IIoT-based solutions have been proposed to predict the energy demand for EVs. In this context, the authors of~\cite{majidpour2014fast} introduced a mobile application that can estimate energy availability and/or expected charging finishing time at the charging station using the KNN algorithm. The application collects historical data on the energy consumed and charging duration of connected electric vehicles on a remote server, which is then used to train KNN for the prediction task. In a different approach,~\cite{saputra2019energy} proposed a Federated Energy Demand Learning (\texttt{FEDL}) method for predicting energy demand. To increase the efficacy of the \texttt{FEDL} method, the authors introduced a clustering technique that groups charging stations into groups before performing the learning process. The effectiveness of the system was tested using real data from charging stations in Dundee City, the United Kingdom, collected between 2017 and 2018. The results showcased improved prediction accuracy while maintaining a good communication overhead, thus demonstrating the potential of this approach. Furthermore, the study in~\cite{chics2016reinforcement} adopted the RL-based algorithm to solve the daily energy that should be added to the batteries of electric vehicles. In particular, the authors cast the EV charging problem as a Markov decision process for choosing the amount of energy to be charged, to reduce the long-term cost of charging for an individual plug-in EV based on the current known day-ahead and following predicted electricity prices. 
    Also, a novel approach has been proposed to model the behavior of plug-in electric vehicles (PEV) in the energy market has been proposed in~\cite{jahangir2020plug}. The authors introduced a clustering technique to group PEVs that exhibit similar behavior patterns. Subsequently, an LSTM model is assigned to each cluster, which is tasked with capturing and forecasting the unique behavior of each cluster. The use of clustering techniques helps decrease the complexity of the model, improving its performance in predicting PEV behavior. Performance evaluations showcased the effectiveness of the proposed method in accurately modeling and predicting PEV behavior in the energy market, outperforming traditional forecasting methods. Furthermore, the authors of~\cite{liu2022mobile} aimed to increase the proportion of vehicles charged with insufficient energy and to reduce the charging costs of these vehicles. To achieve this, they proposed an efficient placement strategy for idle mobile charging stations, termed \texttt{ FL-PDMIM}. This innovative approach indicates the continued search for more effective solutions in the management of energy consumption for electric vehicles. 
    {On the other hand, vehicle emissions have become a major and costly problem in many countries with the growth of the population and motor vehicles. To solve this problem, eco-driving is one such technique, which aims to reduce vehicle fuel consumption and emissions by providing proper guidance. In this context, the authors of~\cite{shi2018application} used traditional (non-deep) Q learning to minimize $CO_2$ emission at signalized intersections. Similarly, this study~\cite{wegener2021automated} used a deep RL algorithm to learn eco-driving strategies in an urban environment, where the agent only received minimal connectivity data and no explicit prediction of the traffic situation.
    The work in~\cite{pozzi2020ecological} developed a Deep Deterministic Policy Gradient (DDPG) algorithm to “learn” an eco-driving velocity planner for a plug-in hybrid electric vehicle within a model-free approach.}

\end{itemize}

\subsubsection{Traffic Management Systems}

Traffic flow prediction is an important functional component of Intelligent Transportation Systems (ITS). It guarantees a more pleasant driving environment and eliminates the possibility of traffic accidents. It is one of the largest segments within smart transportation, where the adoption of IIoT technologies is considered the most prominent. An enormous amount of spatio-temporal traffic data and vehicle-related information is being produced through devices such as cameras, various sensors, and Global Positioning Systems (GPS)~\cite{yuan2021survey}. This data is then transferred to traffic management centers, allowing for timely decision making. Often, techniques such as Autoregressive (AR) and Moving Average (MA) are used to model the time series data, giving rise to models like Autoregressive Moving Average (ARMA) and Autoregressive Integrated Moving Average (ARIMA). However, these methods can prove to be insufficient for traffic prediction due to the highly dynamic and non-linear nature of spatio-temporal correlations between different positions. 
Therefore, this scenario calls for the development and adoption of more sophisticated models that can effectively capture these complex dynamics.



Compared to classical statistical models, data-driven ML-based traffic prediction models can handle high-dimensional data and obtain their nonlinear relationships well. Thus, ML models can be established as strong competitors to classical statistical models and receive tremendous attention in traffic prediction. In this context,  the study in~\cite{cai2016spatiotemporal} proposed an improved KNN model to improve the accuracy of the forecast based on spatiotemporal correlation and achieve multi-step forecasting. Rather than relying on traditional time series, this model represents the traffic state via the spatio-temporal state matrix. The model uses Gaussian weighted Euclidean distance to identify 'nearest neighbors' and then applies the Gaussian function to assign weights to each nearest neighbor. Furthermore, Random Forest, a robust ML algorithm, has also been proposed as a viable solution for traffic prediction~\cite{johansson2014regression}. These developments underscore the potential and effectiveness of ML to improve traffic prediction outcomes.

Although the feasibility and reliability of KNN and other shallow models may not perform well due to their limited parameter space in modeling complex IIoT traffic. In other words, the complexity of road traffic predicts future traffic volume a very challenging task and beyond the ability of such models. In this context, DL has been used as an alternative solution to infer information from large datasets and requires very little domain knowledge and engineering by hand~\cite{lara2021experimental}. Consequently, significant efforts have been devoted to DL for IIoT-based traffic prediction. Generally, CNN is used to extract the spatial correlation of grid-structured data~\cite{sun2020city}~\cite{zhao2019deep}, then Graph Convolutional Network (GCN)~\cite{kipf2016semi},  diffusion convolutional recurrent neural network (DCRNN)~\cite{li2017diffusion}, and graph Wavenet models~\cite{wu2019graph} have been proposed as an improved version of the convolution operation to capture spatial correlations in non-Euclidean data and show a more efficient representation of the traffic structure. Also, to reduce the prediction error, the authors of~\cite{guo2020optimized} added an attention mechanism to a multi-component graph convolutional network to form an attention-based spatial-temporal graph convolutional network. Another graph-based model called \texttt{GMAN} employed attention structures in both spatial and temporal dimensions, enabling a more accurate representation of dynamic spatio-temporal correlations~\cite{zheng2020gman}.

Moreover, to efficiently capture the temporal along the spatial features in traffic flow, the RNN model and its variants LSTM and GRU are combined with CNN or Graph neural network to extract the temporal correlations. For instance, the authors of~\cite{li2019hybrid} proposed a hybrid DL approach, called \texttt{GLA}. Within this \texttt{GLA} framework, the output from the {GCN} model is fed into an LSTM model to capture the temporal dependencies of traffic flow. In the same direction, the study presented in~\cite{wang2018crowd} successfully used LSTM to capture both spatial and temporal components of the traffic data by removing dense kernels with convolutional ones. Furthermore, FL is a practical replacement for conventional centralized ML methods in traffic prediction applications. However, developing efficient traffic prediction while preserving privacy is a promising direction. In this context,~\cite{liu2020privacy} introduced a solution that combines GRU and FL to predict traffic flow. They proposed a clustering \texttt{FedGRU} algorithm, which integrates the optimal global model and captures the spatio-temporal correlation without necessitating the collection of traffic flow data in the cloud. This strategic approach emphasizes both efficiency and data privacy. Several innovative FL-based models for traffic prediction have also been introduced, as highlighted in recent works such as~\cite{yuan2022fedstn} and~\cite{zhang2021fastgnn}. 

Currently, some researchers have turned their attention toward energy considerations for resource-constrained RSUs. These units often tend to serve vehicles in closer proximity, a practice that can lead to incomplete services and an impaired user experience. To mitigate this issue, the authors of~\cite{atallah2016reinforcement} proposed a Q-learning-based intelligent RSU scheduling system. In this architecture, the Q-learning agent chooses which cars to serve as output after taking into account the system statuses and vehicle demands as input. Additional fines for unfinished services are introduced into the incentive structure to encourage a better customer experience. The agent saves the transition of each interaction in the replay memory to further increase the efficiency of the data. The Deep Q-Network (DQN) parameters are then updated by picking a mini-batch of independent transitions at random. The suggested method can serve more cars and prevent insufficient service, according to the simulation findings. Building on~\cite{atallah2016reinforcement}, the multi-agent system (MAS) used in the work in~\cite{yan2018smart} expands the technique to include many RSUs working together to meet user needs and increase throughput. Multi-Agent Reinforcement Learning (MARL) is a technique that RSUs may use to give improved utility over a larger region. Thus, both service availability and overall system performance are enhanced by this method.

\subsection{IIoT-based Smart Industry}

The term ``intelligent industry'' refers to the integration of smart technology into production processes, and IIoT is essential in the analysis of large amounts of data produced by industrial machinery and IoT devices. In different production phases, these methodologies allow process modeling, monitoring, prediction, and control~\cite{ge2017data}. Here, we concentrate on examining the functions of IIoT in robots and Industry 5.0.

\begin{figure*}[!htbp]
\centering
\includegraphics[scale=0.65]{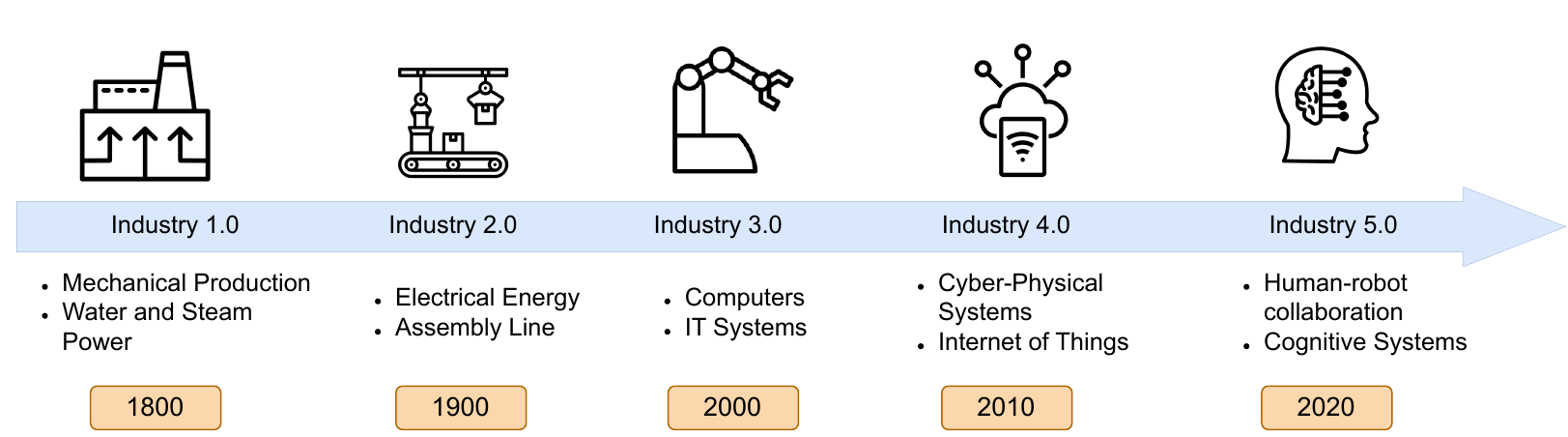}
\caption{Illustration of Industrial Evolution from Industry 1.0 to Industry 5.0.}
\label{fig_indus}
\end{figure*}

\subsubsection{Industry 5.0}

The advent of Industry 1.0 in 1874 marked a significant shift in industrial production. Fig.~\ref{fig_indus} provides an overview of the evolution of Industry X.0. The advent of Industry 5.0 in 2020 ushered in the concept of smart manufacturing for the future. While the fourth industrial revolution, Industry 4.0, focused primarily on the use of technology to optimize the means of production (with technologies such as AI, robotics, IoT, and cloud computing), Industry 5.0 is expected to place a stronger emphasis on the collaboration between humans and machines. It emphasizes the harmony between humans and machines as well and views machines as partners rather than replacements for human labor. This includes using robotics for tasks that are hazardous to humans or require extreme precision and leaving tasks that need creativity and decision-making skills to humans. As societies become increasingly aware of environmental challenges, Industry 5.0 emphasizes sustainable and environmentally friendly production. This includes optimizing resources, reducing waste, and implementing technologies that reduce the environmental impact of production processes.



Industry 5.0 embodies the integration of human subjectivity and intelligence along with the efficiency, precision, and AI capabilities of machines in industrial production~\cite{maddikunta2022industry}. This fusion highlights the importance of humanistic care and paves the way for the evolution toward a symbiotic ecosystem.
A key component of this industrial revolution is IIoT, which can be used to support intelligent systems within the context of Industry 5.0. By facilitating seamless communication and interaction between machines and devices, IIoT plays a critical role in driving the shift towards more intelligent and efficient industries. As an example, an automated material identification using machine vision is introduced in~\cite{penumuru2020identification}. In this study, multiple classification algorithms - including SVM, DT, RF, Logistic Regression, and KNN - are trained on color features extracted from various materials. The experimental results confirm the effectiveness of these algorithms, as demonstrated by their remarkable accuracy in material classification. The adoption of such advanced methodologies ushers in a new era of efficiency and precision in industrial operations. Another example in which these shallow ML models are applied can be found in~\cite{candanedo2018machine}, where logistic regression and RF classifiers are used for automatic monitoring and analysis of the performance of heating, ventilation, and air conditioning (HVAC) systems. These examples demonstrate the potential and versatility of ML in various aspects of Industry 5.0 applications. Similarly, in this work~\cite{calabrese2020sophia}, DT-based models including RF, gradient boosting (GB), and XGBoost are used to predict machine failure. Further, in~\cite{bajic2018machine}, the authors explore the performance of a variety of fundamental ML models within the purview of smart manufacturing. This evaluation provides valuable insight into the strengths and applicability of different shallow models in the industrial context, highlighting their essential role in the realization of Industry 5.0.

Similar to other IIoT applications, DL algorithms have also been extensively investigated for various applications within Industry 5.0. These applications range from predictive maintenance and quality control to system optimization and automated decision-making, demonstrating the vast potential and adaptability of DL techniques in the context of the fourth industrial revolution. In this context, a DL-based approach has been proposed in~\cite{mivskuf2016comparison}, where a unique approach that takes advantage of DL has been proposed to manage data analytics tasks. This approach uses an open-source framework, H2O AI, to implement the DL-based model. The proposed model was evaluated on a text recognition dataset sourced from the UCI repository. This exemplifies the power of DL for complex tasks such as text recognition, highlighting its relevance to the increasingly sophisticated demands of IIoT applications. Additionally, the authors of~\cite{subakti2018indoor} introduced a DL-based framework using the MobileNet model for automatic machine and machine component detection. This model is integrated into their AR-based solution aimed at visualization, analysis, and interaction with machines in indoor environments. The effectiveness of this model underlines the potential of DL to improve Industry 5.0 capabilities. In a different context, the authors of~\cite{mozaffar2018data} utilized a GRU model to predict the high-dimensional thermal history in direct energy deposition (DED) processes. Their findings show that the use of GRU for thermal history prediction is not only highly accurate, but also adaptable across a broad spectrum of processing parameters and part geometries. Furthermore, in this study~\cite{francis2019deep} an implementation of a novel CNN-based framework has been proposed to predict distortions in laser-based additive manufacturing (LBAM). This demonstrates the versatility of DL models, as they can be adapted to address complex and specific problems within the Industry 5.0 environment.

On the other hand, RL seems like a potential option in the context of Industry 5.0 when taking into account the dynamic nature of industrial operations. For example, multi-agent RL was utilized in the paper~\cite{cao2020multiagent}t for cooperative caching and workload offloading in Multi-access Edge Computing (MEC). The multi-agent deep deterministic policy gradients technique was specifically suggested by the authors for usage on actor-critic networks at each MTA (machine-type agent). The MTAs then swap their locally learned models to create a global model. The efficacy of this multi-agent MEC system is predicated on the idea that all edge nodes can interact flawlessly with one another throughout the training phase. However, due to inadequate connectivity between edge nodes and extremely dynamic network topologies in many real-world circumstances, such an assumption may not hold. Additionally, some frameworks combine RL with FL, as the one described in~\cite{messaoud2020deep}. Within industrial IoT networks, the authors of this paper presented a deep-federated reinforcement learning (DFRL) technique to help with dynamic resource allocation and network management. To provide slice-based resource allocation in Industrial IoT, the plan used Deep Federated Q-Learning (DFQL). The authors presented two strategies: Multi-Agent Deep Q-Learning (MAQL), which dynamically slices transmit power and spreading factor to maximize Quality of Service (QoS) requirements, such as throughput and delay, and FL, which is used to learn a multi-agent model to facilitate decision-making on Industrial IoT virtual network slices.

Recently, an integrated FL-blockchain architecture was developed for industrial IoT networks~\cite{qu2020blockchained}, to offer security throughout the construction of IIoT systems. FL provides an answer to Industry 5.0's privacy and productivity problems. Using sophisticated selection methods, blockchain technology (BCT) can be used to strengthen FL against the potential of poisoning attacks. This cutting-edge technology amalgam shows how safe and decentralized processes can be included in the Industry 5.0 framework. To add security to the implementation of FL, citing~\cite{arachchige2020trustworthy} also uses blockchain. The industrial sector has also seen extensive use of the GNN paradigm. For example, to simulate the spatio-temporal interdependence of heat reactions in additive manufacturing processes, the authors of~\cite{mozaffar2021geometry} used a GNN model. Their findings highlighted the potential of GNN in the context of Industry 5.0 by showcasing enhanced performance on lengthy thermal histories of unknown geometries.

\subsubsection{Robotics}

GNN also plays an important role in the robotics field, especially in understanding and interpreting the environment as well as the interactions between different entities. A GNN works by considering entities as nodes and the interactions or relationships between these entities as edges, forming a graph~\cite{dong2023graph}. For instance, the research conducted in~\cite{zhou2022graph} used robots to communicate, share information with neighboring entities, and select actions through a GNN-based learning framework. To capture the complex topological properties and node features of these graphs, a Variational Autoencoder (VAE) and GNN, known as \texttt{NeatNet}, have been proposed~\cite{kapelyukh2022my}. This innovative architecture extracts a low-dimensional latent preference vector from users based on their arrangement of scenes, offering a powerful approach to learning latent representations within graph-based systems.

In addition to GNN-based models, in the context of IIoT, RL also plays a significant role in the field of robotics, primarily because it offers a way for robots to autonomously learn optimal behavior from their own experiences. Unlike other forms of ML, which require a fixed training phase, RL allows robots to continuously learn and improve their performance over time. In this study~\cite{ruan2019mobile}, an end-to-end approach has been proposed for mobile robot navigation in unknown environments. By leveraging DRL, a mobile robot progressively acquires knowledge about its surroundings by exploring the environment. Consequently, it learns to navigate autonomously toward a target destination relying solely on an RGB-D camera for sensory input. This approach underscores the potential of DRL to facilitate self-guided navigation systems in the realm of robotics.

Similar to other IIoT applications, FL plays a crucial role in robotics in several aspects, such as collaborative learning, data privacy/security, communication overhead, and network efficiency. Numerous scholarly contributions have been made in this field. For instance, the authors of~\cite{liu2020federated} and~\cite{liu2019lifelong} explored a federated imitation learning approach for cloud robotics. Under this scheme, each robot contributes to the training of an imitation Neural Network (NN) using its unique sensor image dataset. The updated parameters are subsequently offloaded to the cloud for knowledge fusion, where the server aggregates the knowledge from various robots to construct a robust learning model. The aggregated knowledge is then disseminated back to the robots for the subsequent round of learning, enabling them to gain insights from the shared knowledge. Using this FL approach, the efficiency and accuracy of local robots in imitation learning can be significantly enhanced compared to traditional centralized learning methods. This distributed learning strategy allows robots to draw on the knowledge and experiences of other robots, thereby augmenting their learning capabilities.

The importance of fog/edge computing in the context of vertical domains in 5G communications and beyond has been underlined in~\cite{pham2020survey}. Fog/edge computing essentially decentralizes the cloud computing concept by relocating computational resources closer to the network edge, thereby facilitating lower-latency communication. In the realm of FL, fog/edge computing offers the computational, network, and storage resources required for robots to share, collaborate, and learn tasks collectively. Edge computing was also used with FL in~\cite{lim2020federated} to facilitate cooperative learning between robotic devices. Due to the inherent complexity and dynamic nature of the system, each device operates its own unique RL model to formulate its control policy. After developing a mature policy model, the devices shared their parameters with a central cloud server for aggregation. The experimental results confirmed that the proposed method significantly enhances learning performance, especially in terms of speed, across different learning clients. Consequently, we can conclude that the combination of cloud computing and fog/edge computing can play a crucial role in enabling federated robotics, where robots can collaborate, exchange knowledge, and collectively improve their learning and performance through FL techniques.

\subsection{Lesson Learned}

In this sub-section, we discuss the key lessons acquired from using AI in different IoT applications.

\subsubsection{IIoT for smart Healthcare}

IIoT has the potential to revolutionize the realm of smart healthcare by infusing advanced AI functionalities. This transformation can improve healthcare services, improve predictive analyzes for patients, and reduce latency, all by fostering cooperation among various entities, including patients, healthcare providers, and medical institutions. For example, AI can provide different solutions to combat the COVID-19 pandemic in several ways, such as by supporting the prediction of outbreaks. We also find that shallow models can be easier to interpret and explain and may require fewer computational resources. However, their performance may not be that strong when dealing with complex patterns and large datasets. Additionally, CNN-based models have been a key component of image processing due to their ability to extract complex ideas from visual data. Before we can use the CNN-based model on a large scale, the computational cost must be addressed~\cite{cheng2022real}. Modern CNN models in particular have great inference accuracy, but are becoming increasingly complicated with millions of parameters. The IoT is moving toward edge computing, which performs processing at or close to the data source, emphasizing the need for effective and lightweight models. This is due to the limited computational resources and storage capacity of edge devices. We also find that FL can provide flexible and privacy-preserving healthcare applications. However, 
according to~\cite{wu2020personalized}, to address the issue of data heterogeneity in distributed health IoT networks, personalized FL is essential as it facilitates the development of high quality personalized models.

\subsubsection{IIoT for Smart City}
AI techniques have been extensively utilized to infuse intelligence into the realization of smart cities. Their ability to manage large-scale data generated from various sensors, devices, and human interactions in real time gives them the ability to provide cities with smart features. Most of the proposed AI-based smart city solutions are based on DL models. These models are important for extracting patterns from the data generated through ubiquitous IoT devices. For example, DL helps improve the accuracy and efficiency of surveillance systems, providing safer environments by detecting anomalies or tracking objects across camera networks. Also, DL can predict energy demand and adjust supply accordingly, optimizing the use of renewable energy sources and reducing energy waste. In addition, it processes large volumes of environmental data from various sensors in the city to predict water quality. Similarly, FL plays a critical role in smart cities, particularly in addressing the challenges related to data privacy and network efficiency. In general, IIoT can enable smart cities to be more efficient, intelligent, and privacy-preserving, making it a key technology for their development and growth.

\subsubsection{IIoT for Smart Transportation}

Several possible applications of AI in smart transportation include autonomous vehicles and traffic management. AI is the foundation for self-driving cars and trucks, as it allows these vehicles to interpret sensory data, make decisions in real time, and navigate the environment safely without human intervention. It can power advanced driver assistance systems (ADAS) to improve road safety, offering features such as automatic braking, collision avoidance, and lane departure warnings. Specifically, RL can continually learn and adapt to new situations, which can be particularly useful in complex and ever-changing traffic scenarios. Moreover, a new generation of neural networks, known as SNN, has been used for autonomous driving. These models, known for their energy and computational efficiency, can bring significant advances to the field of autonomous vehicles, furthering the goal of full automation. Furthermore, in the realm of traffic management, DL models, especially GNNs, are revolutionizing the way we interpret and respond to traffic data. These models can analyze real-time traffic data, forecast congestion, and recommend the most efficient routes for drivers. This is largely due to their ability to extract significant information from complex, high-dimensional, and unstructured datasets. Therefore, AI is not only influencing, but actively shaping the future of smart transportation.

\subsubsection{IIoT for Smart Industry}
Recently, IIoT has been introduced to bring AI functions to the IoT to empower the smart industry, including the advent of Industry 5.0 and advances in robotics. Industry 5.0 often referred to as the \say{cognitive industry}, marks a significant shift from automation to collaboration. It uses AI to enhance human-machine interaction. The use of AI algorithms enables real-time decision-making, predictive maintenance, and optimization of production processes. In effect, AI contributes to more flexible, efficient, and personalized production lines, while upholding the value of human creativity and innovation. In particular, RL can optimize process control and production planning, leading to increased efficiency and reduced costs. RL can also be used to train robots to perform complex tasks by learning optimal strategies through trial and error. For example, an RL-trained robot could learn the most efficient path to move parts across a factory floor. However, GNNs offer valuable insight and opportunities to improve various aspects of Industry 5.0, thereby improving productivity and efficiency in the new age of industrial operations. Furthermore, we find that RNN-based models (e.g., GRU and LSTM) hold significant potential for smart industry applications because industrial machinery generally generates time-series data that can be analyzed to predict failures.

In summary, ML, DL, RL, and FL solutions have great potential in IIoT applications, their successful deployment requires careful consideration of several factors, including data quality, computational requirements, privacy, robustness, and security.

\begin{table*}[ht!]
\renewcommand{\arraystretch}{1.00}
\centering
\caption{Summary of IIoT Applications}
\label{tab:IoT Services}
\scalebox{.9}{
\begin{tabular}{|p{2cm}|l|p{2cm}|p{1.5cm}|p{6.5cm}|p{4.75cm}|}
\hline
\multicolumn{1}{|c|}{\textbf{Application}} &
\multicolumn{1}{c|}{\textbf{Ref.}} &
\multicolumn{1}{c|}{\textbf{Use case}} & 
{\textbf{Model type}} &
\multicolumn{1}{c|}{\textbf{Contributions}} &
{\textbf{Limitations and Challenges}} \\ \hline
\multirow{4}{*}{\begin{tabular}[c]{@{}c@{}}\\  Smart Healthcare \end{tabular}} &
\cite{hong2017multivariate} & Ambient Assisted Living & K-mediod, DBN, LSTM & A DL-model for remote sleep monitoring. &
The computation cost has not been evaluated.\\ \cline{2-6} 
 &
 \cite{matar2016internet} & 
 Ambient Assisted Living & PCA, SVM & A shallow ML-based model for remote body pressure monitoring. &
 The proposed scheme is simple and not scalable. \\ \cline{2-6} 
 &
 \cite{zeng2014convolutional} & Ambient Assisted Living & CNN & A DL-based model to learn and extract human activity features without any domain knowledge. &
 The comparison with other state-of-art DL-based schemes has been missed.\\ \cline{2-6} 
 &
 \cite{yang2015deep} & Ambient Assisted Living & CNN & An investigation of DL model efficiency for features extraction for HAR. &
 The complexity of the proposed scheme has not been discussed. \\ \cline{2-6}
 &
 \cite{cheng2022real} & Ambient Assisted Living & CNN & A computation-efficient DL-based model for AAL. &
Privacy of data has not been taken into account.\\ \cline{2-6} 
  &
 \cite{ordonez2016deep} & Ambient Assisted Living & CNN, LSTM & A spatio-atemporal fearures extraction in AAL. &
 The model was trained in a fully supervised way. \\ \cline{2-6} 
  &
 \cite{zehtabian2021privacy} & Ambient Assisted Living & LSTM & An FL-based approach for the HAR predictor in a smart environment. &
 The scalability of the FL scheme has not been discussed.\\ \cline{2-6} 
  &
 \cite{zhao2020semi} & Ambient Assisted Living & MLP, AE &A semi-supervised FL scheme for HAR. &
 The communication cost has not been evaluated. \\ \cline{2-6} 
  &
 \cite{bettini2021personalized} & Ambient Assisted Living & DNN & A semi-supervised federated Active learning framework for HAR. &
 The impact of the non-IID data has not been investigated. \\ \cline{2-6} 
  &
 \cite{wu2020personalized} & Ambient Assisted Living & CNN & Personalised FL model to mitigate the negative effects caused by statistical heterogeneity. &
 The system heterogeneity has not been taken into account.\\ \cline{2-6} 
  &
 \cite{imteaj2021fedparl} & Ambient Assisted Living & - & A resource-Aware Lightweight FL model for AAL. &
 The energy consumption of the proposed scheme has not been discussed. \\ \cline{2-6}
 &
 \cite{shaik2022fedstack} & Ambient Assisted Living & Ensemble model & An heterogeneous federated stacking model to monitor patient activities. & Privacy-related issues in the proposed scheme have not been investigated. \\ \cline{2-6}
 &
  \cite{barstugan2020coronavirus} & COVID-19 & SVM & An early detection of COVID-19 based on shallow model. &
 The scalability issue of SVM has been ignored. \\ \cline{2-6} 
  &
  \cite{narin2021automatic} & COVID-19 & CNN & A CNN-based models for COVID-19 images classification. & A small dataset has been used for the performance evaluation. \\ \cline{2-6} 
  &
  \cite{apostolopoulos2020covid} & COVID-19 & CNN & A Transfer Learning scheme for COVID-19 image classification. &
 The model was trained in a fully supervised way. \\ \cline{2-6} 
  &
  \cite{wang2021auxiliary} & COVID-19 & - & A 5G-enabled architecture for the diagnosis of COVID-19 based on FL. &
 One dataset has been used for performance evaluation. \\ \cline{2-6}
  &
  \cite{zhang2021dynamic} & COVID-19 & CNN & A dynamic fusion-based FL method to improve communication efficiency and model performance to detect COVID-19-infected patients. & The effect of the non-IID data has not been investigated.\\ \cline{2-6}
  &
  \cite{sarumi2022potential} & COVID-19 & LSTM & An FL framework to predict the dynamics of the COVID-19 outbreak. & The scalability of the model has not been discussed. \\
 \hline
 
\multirow{3}{*}{\begin{tabular}[c]{@{}c@{}}  Smart City \end{tabular}} &
\cite{ahmed2020machine} & Smart Grids & - & An ML-based energy management framework to mitigate demand-side energy management issues. &
A comparative analysis of different MLs has been ignored.\\ \cline{2-6} 
 &
 \cite{alazab2020multidirectional} & Smart Grids & LSTM & 
A Multidirectional LSTM model to predict the stability of a smart grid network. &
Convergence latency should be included.\\ \cline{2-6} 
 & 
 \cite{zhai2021dynamic} & Smart Grids & RNN & A delay-deadline constrained FL framework to predict link reliability in mobile edge computing of a grid. & The impact of the straggler nodes on local training has not been considered.
\\ \cline{2-6}

 & 
 \cite{cao2020ifed} & Smart Grids & NN & An FL framework for power learning in IoT networks comprising electric providers and users. & Comparison of ML models in FL simulation has been missed.
\\ \cline{2-6}
 & 
 \cite{su2021secure} & Smart Grids & DRL & An FL approach for secure and efficient personal energy data sharing using edge-cloud collaboration. & Model security has not been investigated. 
\\ \cline{2-6}
 & 
 \cite{tun2021federated} & Smart Grids & RNN & A cloud-edge FL-based scheme for the prediction of energy demand. & One dataset has been used for the performance evaluation.
\\ \cline{2-6}
 & 
 \cite{ktari2022lightweight} & Underwater & YoLo 4 & A novel technique to monitor water consumption using an optical character recognition device. & No details simulations are given.
\\ \cline{2-6}
 & 
 \cite{baek2020prediction} & Underwater & CNN & An integrated automatic detection model to identify temporal resolution and high spatial imagery systems. & Convergence of the model has not been verified.
\\ \cline{2-6}
 & 
 \cite{wang2017water} & Underwater & BiLSTM & A DL-based model for water quality prediction. & The computation cost of the proposed scheme has not been discussed.
\\ \cline{2-6}
 & 
 \cite{prasad2022analysis} & Underwater & Auto DL & Auto DL model for water quality prediction. & Data privacy has not been investigated.
\\ \cline{2-6}
 & 
 \cite{zhao2021federated} & Underwater & NN & A federated Meta-Learning Enhanced Acoustic Radio Cooperative Framework. & The effect of the non-IID data has not been evaluated.
\\ \hline
\end{tabular}%
}
\end{table*}

\begin{table*}[ht!]
\renewcommand{\arraystretch}{1.00}
\centering
\caption{Summary of IIoT Applications (Continued)}
\label{tab:IoT Services_2}
\scalebox{.9}{
\begin{tabular}{|p{2.5cm}|l|p{1.5cm}|p{1.5cm}|p{6.5cm}|p{4.5cm}|}
\hline
\multicolumn{1}{|c|}{\textbf{Application}} &
\multicolumn{1}{c|}{\textbf{Ref.}} &
\multicolumn{1}{c|}{\textbf{Use case}} & 
{\textbf{Model type}} &
\multicolumn{1}{c|}{\textbf{Contributions}} &
{\textbf{Limitations and Challenges}} \\ \hline
\multirow{4}{*}{\begin{tabular}[c]{@{}c@{}}\\  Smart Transportation \end{tabular}} &
\cite{tabernik2019deep} & Autonomous Vehicles & CNN & An automatic end-to-end learning for large-scale detection and recognition of traffic signs through.  &
  The computation cost has not been discussed. \\ \cline{2-6} 
 &
 \cite{yang2018deep} & Autonomous Vehicles & CNN & An attention network to extract more relevant features from the input data.  &
 The convergence of the proposed scheme has not been verified. \\ \cline{2-6} 
 &
 \cite{arcos2018deep} & Autonomous Vehicles & Transformers & A Multiple spatial transformers for traffic. &
 The generalization capability of the proposed scheme has not been evaluated. \\ \cline{2-6} 
 &
 \cite{zhang2020lightweight} & Autonomous Vehicles & CNN & A knowledge distillation-based model to reduce the computation cost. & 
 A small dataset has been used for the performance evaluation.\\ \cline{2-6}

& \cite{xie2022efficient} & Autonomous Vehicles & SNN & An energy FL-based scheme for accurate recognition models. & 
The communication cost has not been verified. \\
 
 \cline{2-6}%
 &
 \cite{majidpour2014fast} & Autonomous Vehicles & KNN & A mobile application that can estimate energy availability and/or expected charging finishing time. &
 The scalability of the proposed scheme has not been explored. \\ \cline{2-6} 
  &
 \cite{saputra2019energy} & Autonomous Vehicles & NN & A Federated Energy Demand Learning Scheme for Energy Demand Prediction. &
 Training latency has not been analyzed.\\ \cline{2-6} 
  &
 \cite{chics2016reinforcement} & Autonomous Vehicles & RL & A Markov decision process to choose the amount of energy to be charged. & 
 The choice of the model hyper-parameters has not been explored. \\ \cline{2-6} 
  &
 \cite{jahangir2020plug} & Autonomous Vehicles & LSTM & A novel approach to model the behavior of plug-in electric vehicles in the energy market. &
 The privacy of the collected data has not been taken into account.\\ \cline{2-6} 
  &
 \cite{liu2022mobile} & Autonomous Vehicles & LSTM & An FL-based placement strategy for idle mobile charging stations. &
 The communication cost of the proposed scheme has not been evaluated. \\ \cline{2-6} 
  &
 \cite{cai2016spatiotemporal} & Traffic management & KNN & an improved KNN model to improve forecast accuracy based on spatiotemporal correlation and achieve multistep forecasting. &
 The generalization capability of the proposed scheme has not been verified. \\ \cline{2-6} 
  &
 \cite{johansson2014regression} & Traffic management & RF & A shallow sample model for traffic prediction. &
 The proposed model is simple. \\ \cline{2-6}
 &
  \cite{li2019hybrid} & Traffic management & GCN, LSTM & An hybrid GNN and LSTM to better capture the temporal dependencies of traffic flow. & 
 One dataset has been used for the performance evaluation. \\ \cline{2-6} 
  &
  \cite{liu2020privacy} & Traffic management & GRU & A clustering-based FL scheme to efficiently capture spatio-temporal correlation without privacy concerns. &
 Privacy and security performance have not been evaluated.\\ \cline{2-6} 
  &
  \cite{atallah2016reinforcement} & Traffic management & RL & An intelligent RSU scheduling algorithm based on Q-learning. &
 The proposed scheme is not scalable to multi-agent systems. \\ \cline{2-6} 
 & \cite{yan2018smart} & Traffic management & Multi-agent RL & Enhancing the utility and cover a more extensive area. &
 Communication privacy has not been taken into account. \\ \hline
 
\multirow{3}{*}{\begin{tabular}[c]{@{}c@{}} Smart Industry \end{tabular}} &
\cite{penumuru2020identification} & Industry Intelligence & SVM, DT, RF, KNN & A shallow model to automate material identification.
 &
  The scalability of the proposed schemes has not been evaluated. \\ \cline{2-6} 
 &
 \cite{candanedo2018machine} & Industry intelligence & RF & An automatic monitoring and analysis of the performance of heating, ventilation, and air conditioning systems.
 &
 Numerical results for performance evaluation are lacking. \\ \cline{2-6} 
 & 
 \cite{calabrese2020sophia} & Industry Intelligence & RF, XGBoost & A DT-based models to predict machine failure. & The proposed work is simple with a lack of detail analysis.
\\ \cline{2-6}
 & 
 \cite{subakti2018indoor} & Industry Intelligence & CNN & A DL-based framework for automatic machine and machine component detection. & The computation cost of the proposed scheme has not been evaluated.
\\ \cline{2-6}
 & 
 \cite{mozaffar2018data} & Industry Intelligence & GRU & A DL scheme to predict the high-dimensional thermal history in directed energy deposition processes. & The comparison with other state-of-the-art DL-based schemes has been missed.
\\ \cline{2-6}
 & 
 \cite{francis2019deep} & Industry Intelligence & CNN & A novel DL-based framework to predict distortions in laser-based additive manufacturing. & The generalization capability of the proposed scheme has not been verified.
\\ \cline{2-6}
 & 
 \cite{cao2020multiagent} & Industry Intelligence & Multi-agent RL & A multi-agent
RL scheme to reduce the computation delay and increase the channel access success rate in industry IoT without knowing the system parameters in advance. & The security risk of the IIoT devices has been ignored.
\\ \cline{2-6}
 & 
 \cite{messaoud2020deep} & Industry Intelligence & RL & A federated RL to facilitate dynamic resource allocation and network management within industrial IoT networks. & The communication cost of the proposed scheme has not been evaluated.
\\ \cline{2-6}
 & 
 \cite{arachchige2020trustworthy} & Industry Intelligence & GNN & A FL-based blockchain scheme to provide security for FL implementation in industry IoT networks. & Training latency has not been analyzed.
\\ \cline{2-6}
 & 
 \cite{mozaffar2021geometry} & Industry Intelligence & GNN & A GNN model to capture the spatiotemporal dependencies of thermal responses in additive manufacturing processes. & The computation cost of the proposed schemes has not been verified.
\\ \cline{2-6}
 & 
 \cite{zhou2022graph} & Robotics & GNN & GNN models to capture local interactions of robots and learn decentralized decision-making for robots. & The communication cost has not been explored.
\\ \cline{2-6}
 & 
 \cite{kapelyukh2022my} & Robotics & VAE, GNN & VAE architecture using GNN to extract a low-dimensional latent preference vector from a user. & The models' complexity has not been discussed.
\\ \cline{2-6}
 & 
 \cite{ruan2019mobile} & Robotics & DRL & An end-to-end DRL-based approach for mobile robot navigation in unknown environments. & The proposed model has not been evaluated with dynamic environments.
\\ \cline{2-6}
 & 
 \cite{liu2020federated} & Robotics & NN & A federated imitation learning approach for cloud robotics with heterogeneous sensor data. & The scalability of the proposed scheme has not been investigated.
\\ \cline{2-6}
&
 \cite{liu2019lifelong} & Robotics & RL & An FL-based scheme for robotic management & Data loss caused by communication has not been considered.
\\ \hline
\end{tabular}%
}
\end{table*}

\section{Security Issues in IIoT}
\label{sec:security}

As IIoT networks have become an integral part of our lives, the use of learning algorithms in a variety of domains (e.g., home, industry, and healthcare) can help communication processes and analytics with real-time responses. However, the proliferation of IoT devices makes IoT mining easy, and the lack of security standards and policies leaves IIoT networks open to unauthorized attacks. Against this background, understanding network attacks, confidentiality, integrity, and intrusion is necessary to develop countermeasures.

\subsection{Network Attacks}

Network attacks, often known as cyberattacks, attempt to obtain unauthorized access to or alter the regular operation of a network or its components. These attacks can target any aspect of a network. In what follows, we first present four typical types of IIoT networks: denial-of-service attacks, poisoning attacks, adversarial attacks, and membership inference attacks.  Then, we introduce some potential ways to evaluate the security levels, along with some security standards.

\subsubsection{Denial-of-service Attacks}

A denial-of-service (DoS) attack is a kind of malicious attack that exploits the availability of single or multiple interconnected IoT devices, sensors, or sources to disrupt IIoT networks' operation by flooding them with lots of malicious requests or traffic. This type of attack can target different layers of networks, such as the transport layer, network layer, or application layer. There are ways through which the attacker can be launched, such as the specific vulnerabilities involving DNS, SMURF, or ACK mechanisms, the limited resources of IoT devices (i.e., storage, bandwidth, authentication, authorization, encryption, and firmware), or algorithms to inject noise, poison data, or manipulate feedback to learning models.

There are different ways to classify DoS attacks in IIoT networks, depending on the criteria or perspective used. For the target layer of the network~\cite{Osei2022Dec}, DoS attacks can be distinguished by three types: (1) connection-oriented services of the transport layer: SYN flooding, RST flooding, ACK flooding and FIN flooding; (2) the routing or forwarding functions of the network layer: ICMP flooding, SMURF, FRAGGLE, and LAND; and (3) application-specific services or protocols of the application layer: HTTP flooding, DNS amplification, MQTT flooding, and NTP amplification. From the perspective of source or technique~\cite{Goncalves2023Jan}, DoS attacks have three types. First, a single-source DoS attack adopts UDP flooding, ICMP flooding, and HTTP flooding methods to send a large amount of traffic or requests to a target service or network. Second, a multi-source DoS attack uses TCP-SYN flooding, DNS amplification, or SMURF to form a botnet or a network of zombies that coordinate the attack. Third, a reflective DoS attack relies on a third-party service or network as a reflector to amplify the attack traffic or requests. A typical attack example is that an attacker sends spoofed packets or requests based on the amplification mechanisms of DNS, NTP, and SSDP to the reflector, which then responds to the target with a larger amount of traffic or requests than the original ones. In addition to that, DoS attacks can also be classified based on impacts in that this affects ML-based systems~\cite{Quincozes2023Jul}, consisting of:
\begin{itemize}
    \item \textit{Poisoning data}: A volume of malicious or noisy data will be injected into learning-based systems to corrupt the training or the feedback data process, thereby degrading the performance or accuracy of the learning model or algorithm or even causing it to produce wrong or harmful outputs. For example, an attacker can poison the data of learning-based systems that control a smart traffic light by sending fake vehicle counts or speeds, causing traffic jams or accidents.

    \item \textit{Model evasion}: These attacks craft adversarial input to evade detection or classification of learning-based systems, allowing them to bypass security or functionality. For example, an attacker alters the power consumption patterns or adds noise to avoid detection of the smart grid, causing power outages or damage.

    \item \textit{Model extraction}: These attacks query learning-based systems with carefully designed inputs and observe their outputs to extract or steal the parameters or the structure, leading to critical issues of intellectual property and privacy, or even enabling further attacks such as model evasion or poisoning. For example, an attacker sends fake patient records or symptoms to extract the model of a smart health service, causing financial losses or privacy breaches.
    
\end{itemize}

DoS attacks in intelligent IoT networks might jeopardize the security, privacy, and operation of the devices, services, and applications \cite{Vishwakarma2020Jan}, and some of the potential risks that this type of attack brings, therefore, include:
\begin{itemize}
    \item \textit{Service disruption}: The availability and reliability of IoT services or smart applications (e.g., health, network, and transportation) can be disrupted or deteriorated, causing users or consumers annoyance, irritation, or even danger. For example, a DoS attack on a smart metering network can disrupt the billing and payment process, or in a connected vehicle network it can interfere with navigation and safety features. 

    \item \textit{Resource exhaustion}: IoT devices or networks with limited resources in terms of battery, memory, bandwidth, and CPU can be consumed or drained, resulting in decreased performance or functionality, or even irreversible damage. For example, a DoS attack on a smart farming network can drain the batteries of the sensors and actuators, or in an industrial automation network it can overload the controllers and actuators.

    \item \textit{Data loss of theft}: Uncertainty about the integrity or confidentiality of data collected, processed, or transmitted by IoT devices or networks can result in data corruption, deletion, leakage, or manipulation, which can violate the privacy or security of users or organizations who own or use the data.

    \item \textit{Financial losses}: The consequences involved of DoS attacks on individuals or organizations could be distinguished by: (1) direct losses, mostly the costs of repairing, replacing, and upgrading IoT devices/networks, paying fines/compensations, and losing revenues/customers; and (2) indirect losses, normally the costs of losing reputation, trust, or competitiveness in the market. 
\end{itemize}
To reduce the potential risks brought about by DoS attacks, the implementation of IIoT networks should consider some of the following possible solutions. First, building out features involved network monitoring and anomaly detection by adopting NetFlow and SNMP approaches and/or learning-based solutions, for example, {IDS} or intrusion prevention system (IPS), to identify anomalous or malicious traffic patterns (e.g., volume, frequency, source, destination, and protocol), thus being able to detect DoS attacks in time and trigger appropriate countermeasures, such as filtering, blocking, or diverting the attack traffic. Second, it is considered to employ a software-defined networking approach in order to separate the control and data planes and centralize network management and configuration, increasing the visibility and flexibility of IIoT networks, as well as enabling dynamic changes and adaptive responses to DoS attacks, for example, routing paths, bandwidth allocation, and security policies. Third, the adoption of consensus mechanisms in BCT in distributed and decentralized networks helps them verify and validate any transactions or events from the network entities, which enhances the security and trustworthiness of the network and prevents spoofing or tampering with network data or devices.

\subsubsection{Poisoning Attacks}

Poisoning attacks are a type of attack that takes advantage of vulnerabilities in learning-based systems (e.g., insecure data collection and transmission, lack of encryption and authentication, heterogeneous and distributed architecture, and limited resources and capabilities) to corrupt or degrade their ML models through introducing malicious data into the dataset. This type of attack is often executed before or during the model's learning or training phase. Usually, it takes advantage of some techniques such as label flipping, data injection, data modification, or data deletion~\cite{Fan2022Jul}.

Based on the attacker's goal and method, poisoning attacks can be distinguished by twofold. First, called poisoning attacks, the attacker aims to destroy the ML model, causing it to learn wrong or biased patterns, such as overfitting, underfitting, or skewed distribution. Automated transportation, healthcare, and industrial control systems are potential targets, where the two most frequent targets are recommender systems~\cite{Ge2022Jul} and crowdsourcing systems~\cite{Fang2021Apr}.
Second, called backdoor poisoning attacks, the attacker aims to implant a hidden functionality or behavior into the ML model, making it produce incorrect or undesirable outputs for specific inputs that trigger the backdoor, such as a predefined class or label. An illustration of this attack is that adversaries, through the aid of internal staff or intrusion systems, refine the centralized and formatted iterative parameters in centralized learning networks or pretend to be legitimate clients to upload infected parameters to IIoT servers in distributed learning networks.

The involvement of multiple IoT devices/targets and the heterogeneity of system architectures coexisting in IIoT networks make them vulnerable to poisoning attacks. Consequently, there are several potential risks that poisoning attacks pose to IIoT networks, such as
\begin{itemize}

\item \textit{Data leakage or manipulation}: Data collected and processed by IoT devices (e.g. sensors, cameras, or smart meters) may have their confidentiality and integrity compromised. In smart agriculture, industrial automation, or smart metering applications, the leakage of sensitive information from these systems leads to privacy breaches, financial losses, or operational failures. For instance, if smart farming networks train their learning model based on poisoned data, the farmers' personal information trained by these models can be disclosed, which might then be used to manipulate the crop yield predictions or sabotage the irrigation system.

\item \textit{Model misbehavior or malfunction}: It compromises the functionality and reliability of the learning model deployed on edge devices or IoT devices, such as network intrusion detection systems (NIDS), traffic classification systems, or recommendation systems. For example, when the labels or features of the training data in a medical diagnosis system are modified, the learning model may misdiagnose patients or prescribe the wrong treatments, endangering their lives. On another front, the learning model might miss real attacks, raise false alarms, or favor certain attackers when network intrusion detection in IoT environments is trained with poisoned data.

\item \textit{Physical harm or damage}: The operation of the IIoT system is not safe and secure. As training data are poisoned, the predicted outcome or guidance of the learning model can be erratic or dangerous in certain situations, such as ignoring traffic signs, colliding with obstacles, or attacking humans, leading to accidents, injuries, or deaths. For example, autonomous driving in IoT environments can ignore stop signs, swerve into other lanes, or hit pedestrians.

\item \textit{Distributed services}: The availability and scalability of the IoT network, such as cloud servers, edge servers, or IoT devices, become invalid. For example, adversarial examples generated using {GANs} might be exploited to poison the FL process among multiple IoT devices or edge servers, making them overload their CPU and memory usage, thereby slowing down their response time, causing network congestion, or even shutting down the network services.
\end{itemize}
From the above risk, it is clear that to protect and maintain the operation of IIoT networks, there are three critical works. First, training data should be cleaned of any noise, outliers, or anomalies before feeding them to the learning model, reducing all possible impacts of poisoning attacks. Second, the machine learning model is required to be able to resist poisoning attacks using different methods, such as regularization, outlier detection, or adversary training. Third, data aggregation must be protected from multiple sources in a distributed machine learning system using cryptographic protocols, such as homomorphic encryption (HE) or secret sharing.

\subsubsection{Adversarial Attacks}

Adversarial attacks are a type of attack that focuses on generating slightly modified inputs from the original ones to deceive an ML model into producing incorrect or undesirable outputs, such as misclassification, misrecognition, or misbehavior. Typically, such attacks exploit the vulnerability of IIoT networks that involve biased training data, vulnerable underlying models, and fabricated features, and these attacks mainly target the inference or testing phase in which the learning model has been trained and deployed.

Adversarial attacks can be classified in different ways, depending on the criteria used or the knowledge models. From a criterion perspective, there are four main types of adversarial attacks. First, non-targeted attacks cause learning models to generate incorrect or undesirable output without specific class or label control using small perturbations or random noise. Second, targeted attacks attempt to make learning models produce incorrect or desirable outputs using methods such as optimizing perturbations or creating adversarial examples that resemble the target class. Third, universal attacks cheat learning models to produce incorrect or undesirable outputs for input data using methods like finding perturbations or generating adversarial examples. Fourth, physical attacks aim to manipulate learning models to produce incorrect outputs for real-world input data using methods like modifying objects, adding stickers, or projecting images or patterns. In comparison, non-targeted attacks are the most general and easiest type of attack, as they do not require any knowledge or control over the output as in targeted attacks. In contrast, universal attacks do not require any knowledge or access to the input data, and thus they are more challenging and powerful than targeted or non-targeted attacks. However, physical attacks are said to be more practical than the three above attacks, as they do not require any digital manipulation of the input data. Meanwhile, the classification of adversarial attacks based on the knowledge models includes:
\begin{itemize}
    \item \textit{White-box attacks}: The attacker has full or partial access and information about the learning model (i.e., architecture, parameters, gradients, or training data), which are then exploited in conjunction with techniques, such as gradient-based methods, optimization-based methods, or transformation-based methods, to craft more effective and targeted adversarial input~\cite{Huang2022May}. For example, an attacker may use white-box attacks to fool an image classifier that uses a CNN model by using a fast gradient sign method or other gradient-based methods.

    \item \textit{Grey-box attacks}: The attacker has some access and information on learning models, including the kind of model, number of layers, or activation functions, but not enough exact weights or values to perform a white-box attack. Gray-box attacks typically use this partial knowledge in combination with zeroth-order optimization methods, decision-based methods, or score-based methods to improve the efficiency and effectiveness of black-box attacks~\cite{Takiddin2022Jul}. For example, an attacker may use gray box attacks to compromise a face recognition system that builds a deep-metric learning model by using model extraction or other optimization techniques.
    
    \item \textit{Black-box attacks}: The attacker has no or limited access to information about the learning model (e.g., input-output behavior, confidence scores, or feedback). This type of attack only relies on trial-and-error methods or surrogate models to generate adversarial input using query-based methods, transfer-based methods, or generative methods~\cite{Chi2023Jan}. For example, an attacker may employ black-box attacks to bypass a network intrusion detection system that uses a DL model in an IoT environment.
\end{itemize}

As adversarial attacks have diverse attack methods, the potential risk produced by this attack also leads to numerous disadvantages to the development of  IIoT networks, including
\begin{itemize}
    \item \textit{Data leakage or manipulation}: Data processed by intelligent networks, such as patient records, financial transactions, or personal information, can be compromised in terms of confidentiality and integrity. For example, an attacker may use structured query language injection to access or modify sensitive data in a web application's database that uses IoT devices to collect and store data.

    \item \textit{Device impersonation or hijacking}: The authenticity and availability of devices connected to the intelligent network, such as medical devices, smart meters, or cameras, are no longer guaranteed. For example, an attacker may use hard-coded keys to decrypt communications or spoof the identities of legitimate devices on an IoT medical device interface.

    \item \textit{Network intrusion or disruption}: The functionality and reliability of the intelligent network, such as NIDS, traffic classification systems, or device identification systems, can be destroyed. For example, an attacker may use adversarial examples to evade detection or mislead classification by a DL-based NIDS in IoT environments.

    \item \textit{Physical harm or damage}: The safety and security of the physical world, such as autonomous vehicles, drones, or robots, have vanished. These allow an attacker, for example, to use stickers or paint to create an adversarial stop sign that an autonomous vehicle will interpret as a yield or other sign.
\end{itemize}
To address these risks, learning-based systems are required to add some features. First, adversarial training; this approach allows the target systems to be trained from both normal and adverbial examples, and hence it can learn to resist gradient-based and black-box attacks. Second, the defensive distillation, the learning model is trained as a distilled model capable of mimicking the original model but with less sensitivity to input variations, making it harder for the attacker to generate adversarial examples. Third, randomness is added to the input or output of the target system, making it unpredictable for the black-box or gray-box attacker.

\subsubsection{Membership Inference Attacks}
This type of attack refers to the case where the adversary attempts to infer whether the data of a given IoT user are trained by a training/query learning model~\cite{Zhang2023Jan}. The information conjectured by the adversary is then applicable to launch the physical-layer authentication of heterogeneous IoT devices or service providers~\cite{Shi2022Feb}. Typically, an attack workflow includes three phases: (1) creating a shadow model dataset that emulates the original training dataset based on the knowledge of black-box mode, white-box mode, and gray-box mode; (2) training shadow models to mimic the target model; and (3) developing attack classifiers to acquire differences between the target model on training data and behavior on unseen inputs. Membership inference attacks can be classified into three levels:
\begin{itemize}
    \item \textit{Model-based attacks}: The adversary uses the outputs generated by the learning model, such as predicted class labels or probabilities, to deduce if the behavior of the model is in agreement with the individual's data during training~\cite{Chen2023Mar}. 

    \item \textit{Data-training-based attacks}: The adversary uses the model's training logs or side-channel information during the training to extract information related to the model's training data (i.e. size, structure, or distribution), from which the necessary information is inferred~\cite{Yan2022Nov}.

    \item \textit{Model-free attacks}: This form of attack does not require knowledge of the learning model or training data. Instead, it relies on statistical tools to analyze and identify discrepancies between the model output and a set of known outputs~\cite{Liu2022Dec}.
\end{itemize}

Accordingly, the potential risks of this attack on IIoT networks can be faced as follows:
\begin{itemize}
    \item \textit{Privacy violations}: By successfully inferring whether a particular user's data were part of the training set, an attacker can compromise that user's privacy, which consists of the user's behavior, habits, or other personal information.

    \item \textit{Information leakage}: Sensitive data, such as health records or financial information, could be exposed by this attack if access is successfully launched.
    
    \item \textit{Undermining integrity}: This attack can manipulate the model's output by feeding it with adversarial data, leading to incorrect or harmful decisions being made based on the manipulated data.
    
    \item \textit{Loss of trust}: If users lose confidence in IoT networks' security and privacy due to successful membership-inference attacks, they may be less likely to use or even have no adoption of smart applications in the future.
\end{itemize}
From the above risks brought by membership inference attacks, IIoT systems should consider the three techniques listed below. First, differential privacy exploitation can prevent an attacker from deducing the membership status of a data sample by adding random noise to the data or model parameters. Second, applying adversarial regularization, which trains the model to minimize the difference between the outputs of member and non-member samples, prevents an attacker from distinguishing them. Third, taking advantage of proactive defense, i.e., building a shadow model and fooling the attacker by sending false alerts or misleading information, is to reduce membership inference attack accuracy and prevent information leakage from the wireless signal classifier.

\subsubsection{Attack Measurements}

From the above-mentioned examples, it can be seen that there are various types of network attacks. Therefore, it raises the question of how to evaluate the security level of existing IoT networks. In fact, there are many ways, and they depend on each application-specific. 1) Perform a threat and risk analysis to identify assets, threats, vulnerabilities, and impacts of IoT security breaches \cite{atamli2014threat}, 2) Apply a security requirements engineering approach to evaluate and prioritize security requirements for IoT devices, software, and data \cite{Gomes2023Oct}, 3) Use a security testing framework to assess the security properties and performance of IoT devices and software under different scenarios and attacks \cite{Huang2016Nov, Bagaa2020May}, 4) Using a security evaluation standard or certification scheme to measure the security level of IoT devices and software against predefined criteria and best practices \cite{Lins2020Nov}. Moreover, the method should evaluate the security of AI models and IIoT systems before they are deployed in practice. Therefore, the security evaluation framework should be developed and designed considering the large-scale nature of IoT systems, non-IID data characteristics of AI training data, and the heterogeneous nature of network scenarios. Furthermore, it is important to develop security methods towards the standardization of IoT, AI, communications, and IIoT. There have been several standardization bodies, such as the Security Evaluation Standard for IoT Platforms (SESIP) \cite{platform2021security}, Common Criteria for Information Technology Security Evaluation (CC) \cite{infrastructure2002common}, IEEE 802.11 series \cite{potter2002802}, ETSI TS 103 645 \cite{korner2023current}, and details of other standards with specific use cases \cite{Karie2021Sep,Lee2021Mar}.
It is expected that there will be more and more collaborative research work and projects between different stakeholders in this exciting IIoT area, and this may result in the development of new standards or significant updates to the existing standards of security and privacy in IIoT.
\color{black}

\subsection{Confidentiality}
Confidentiality is one of the security criteria for IIoT networks, as data and information created by IoT devices and systems should be available only to authorized users and systems. Confidentiality can help prevent unlawful exposure or leakage of sensitive or private information, such as personal information, health records, and location data. To ensure confidentiality in IIoT networks, there are three main approaches: cryptography, access control, and information flow tracking.

\subsubsection{Cryptography}Cryptography encrypts the data using mathematical techniques to make them unreadable to unauthorized parties. For example, a user may need to have a secret key or a certificate to decrypt the data. Despite cryptography, which can protect data during transmission or storage, it cannot ensure data secrecy once it has been decrypted. Furthermore, the scarcity of resources from IoT devices limits their ability to perform complex cryptographic operations or store large quantities of keys or certificates. In this context, it is necessary to develop confidentiality solutions for IoT that are lightweight, efficient, and adaptive to resource constraints.

In terms of the above requirements, a good solution for encryption of raw IoT data during data acquisition is to use the compressive sensing scheme to increase secure interactions between IoT devices and the cloud~\cite{Zhang2019Dec}. Furthermore, the application of symmetric and asymmetric cryptosystems, that is, the advanced encryption standard (AES), the message digest algorithm (MD5), and the elliptic curve cryptography (ECC), should also be combined transmission protocols to encrypt the data during transmission~\cite{Chanal2019Jul}. 

{In particular, exploiting outsourced services, such as fog and edge computing, to reduce the computational and storage burden on IoT devices might be prone to vulnerability with sensitive information. In this context, HE emerges as a powerful technique for computing encrypted data without decrypting it. However, ensuring real-time response with HE in IoT networks is a challenge. Driven by this fact, to speed up encryption and decryption operations and reduce the power consumption of HE, some specialized hardware has been realized, such as the Cheetah acceleration architecture for server-side HE DNN inference in \cite{Reagen2021}, a new hardware architecture for Brakerski, Vaikuntanathan full HE scheme using Field Programmable Gate Array (FPGA) \cite{Behera2022Dec}, and a novel confused modulo projection-based fully HE algorithm using graphics processing units (GPUs) to support floating-point operations \cite{Li2023Sep}.  On the other hand, exploiting HE is not an easy task for IoT devices due to its complex nature. To this end, the authors of \cite{Matsumoto} proposed two HE techniques. The first technique is to encrypt plaintext with somewhat HE while converting somewhat HE ciphertext to fully HE ciphertext on the cloud service side. The second technique is to encrypt plaintext with TRIVIUM and the TRIVIUM key is then encrypted with somewhat HE. In \cite{Dar2020Apr}, the authors proposed a suite of context-sensitive encryption protocols that only encrypt data according to the device specifications and the level of confidentiality of the data, while the rest are in plaintext. In \cite{Lu2017Mar}, the authors introduced lightweight privacy-preserving data aggregation for fog computing-enhanced IoT based on the combination of homomorphic Paillier encryption, Chinese Remainder Theorem, and one-way hash chain techniques. Another illustrative scheme is to employ a new energy-efficient IoT security system in \cite{Gupta2021May}. With proportional offloading, secure MQTT protocols enable IoT devices to offload data to fog nodes and/or cloud centers according to the computational power required by message packets to effectively use energy. At fog nodes and/or the cloud, the MQTT payloads will then be encrypted using a HE technique, called ECC–ElGamal.}

\begin{figure*}[t!]
    \centering
\includegraphics[width=0.8\linewidth,keepaspectratio]{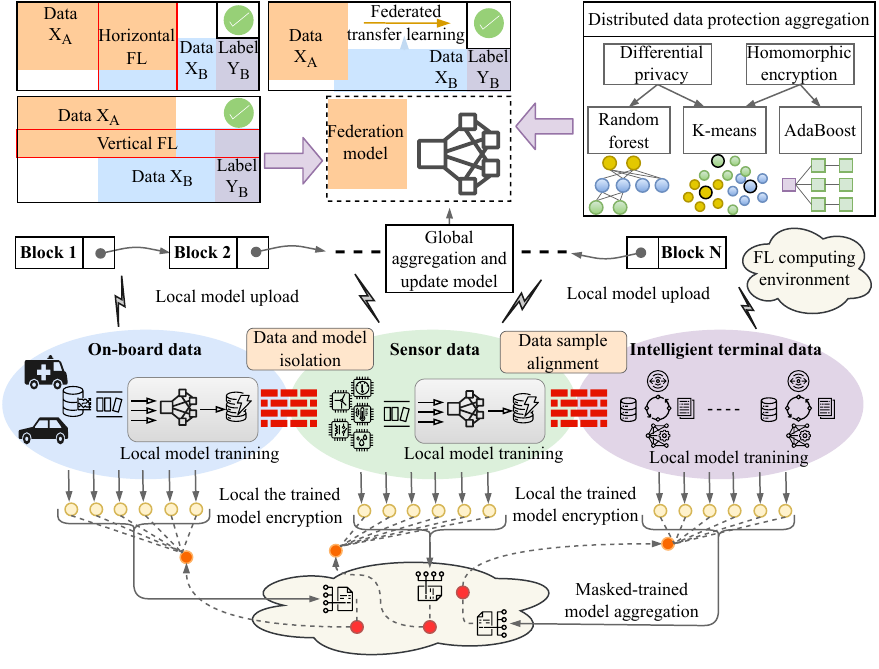}
    \caption{Data FL aggregation using blockchain-enabled EH (the top) and local encryption (the bottom).}
    \label{fig:1}
\end{figure*}

Meanwhile, data aggregation and management are becoming difficult due to the diverse types of IIoT architectures. To deal with such demands, some recent work has built several learning-based networks that combine encryption approaches during training and learning. For instance, a DL-based network has been proposed in \cite{Ding2020Jul} to encrypt and decrypt the medical image. The work in~\cite{Saba2021Jun} presented an ML-based AES algorithm against adversaries. {To prevent risks involving model extraction and reverse attacks, the work in~\cite{Song2022Feb} proposed a data aggregation framework for distributed FL IoT scenarios. As described at the bottom of Fig.~\ref{fig:1}, the user data will be locally trained and encrypted before aggregating by the server using the masked trained model to generate a global model. Unlike \cite{Song2022Feb}, the work in \cite{Jia2021Jun} designed a blockchain-enabled HE model to achieve the double goal of data privacy and model privacy. As shown in the top of Fig.~\ref{fig:1}, after locally training raw data from IoT environments, the generated local results are uploaded to the cloud. In that regard, the global aggregation and update model procedure includes two steps. First, the RAFT consensus algorithm is utilized to achieve logical separation for distributed computing, where the data executor allocates some distributed data streams on multiple nodes indexed from 1 to N with the same set of state transitions. In every node, the node manager regulates the data analysis and database as SparkContext for the Spark function, while the resource manager administers all nodes via the Spark computing platform. Second, data protection aggregation tactics are adopted, where data from multiple nodes are encrypted using differential privacy and HE based on the respective Laplace and Paillier cryptosystems. These encrypted data will then be separately trained by $K$-means, random forest, and AdaBoost. As such, all data are summarized, while private models can be aggregated. Later, these global outcomes are fed into the federation procedure to construct a federation model.}

\subsubsection{Access Control} 
Access control governs established norms and rules to ensure that only authorized entities have access to data. For example, a user may need to provide a password, a biometric feature, or a role to access the data. However, access control cannot manage how the data are used or propagated after they are accessed. Furthermore, the exchange of large volumes of data among a large number of heterogeneous IoT devices and systems is quite difficult to build and deploy scalable confidentiality solutions that can handle the diversity and complexity of IoT situations.

Following that, access control protocols must design efficient key management mechanisms capable of flexibly generating, distributing, updating, and revoking keys (e.g.,  a unique password, zero-knowledge proof, mutual authentication, public-key cryptography, and digital signatures) for IoT devices in a secure and scalable manner. Depending on each application, the realization of key management mechanisms can be built by the following criteria:
\begin{enumerate}
    \item[$a)$] \textit{Authentication factor}: IoT environments constrained by physical size, internal capacity, and other storage allocations define two authentication factors. First, identity factors refer to the use of cryptographic techniques, such as symmetric, asymmetric, and hash functions~\cite{Tanveer2022Feb}. Second, context factors are based on the basic characteristics of wireless communication to generate keys, tags, and fingerprints~\cite{Perazzone2021Jan}.
    
    \item[$b)$] \textit{Token-based authentication}: The communication session occurs for a short period of time in digital IoT contexts, which are particularly sensitive to temporal fluctuations and high-security standards. Instead of using a combination of username or password per communication, token-based authentication offers a strong link between users and smart devices, while reducing the risk of stolen authentication factors for misuse, such as autonomous vehicle systems~\cite{Manogaran2023Mar} or biometric systems~\cite{Cui2023Feb}.  

    \item[$c)$] \textit{Authentication process}: For IoT systems built with a trusted entity, one-way authentication can be applied to achieve lightweight communication, for example, using human body characteristics (e.g., voice, face, fingerprint, and gait) in biometric traits authentication~\cite{Alsellami2021Mar}. However, such a scheme is susceptible to spoofing or impersonation in complex IoT environments (e.g., smart health, smart home, smart cities). To protect IoT information between any two IoT entities and / or with third-party service providers, it is required to use two-way authentication~\cite{Leng2020Apr} or three-way authentication~\cite{Sukumaran2022Nov}.

    \item[$d)$] \textit{Authentication methodology}: To efficiently manage the identity of the accessing devices, IIoT networks tend to stipulate their architecture in terms of centralization, distribution, or decentralization. Thanks to a trusted third party (TTP), centralized authentication allows communication entity credentials to be managed and shared to ease the burden of infrastructure deployment. Meanwhile, to reduce peer-to-peer authentication failures through IoT nodes, servers, or base station in centralized fashions, distributed and decentralized authentications are administered at both nodes/devices and the network~\cite{Gautam2021Jan}.

    \item[$e)$] \textit{Authentication tier}: The authentication mechanism is distinctly specified by perception tiers, communication tiers (including data link, network, and transport layers), and application tiers~\cite{Zhao2020Aug}. The perception level interacts directly with the environment to collect raw IoT data, which requires a proper authentication mechanism for all hardware involved in the IoT network to combat tag cloning, RFID eavesdropping, fake information broadcasting, and RF jamming approaches before launching a DoS attack. At the communication level, the information collected by the perception layer will be synchronized and transferred over the mobile/private network, the wireless and wired network, and the communication protocols; however, the authentication mechanism could easily be destroyed due to network congestion with large amounts of data transmitted by DDoS attacks. Meanwhile, the application-tier interface mostly delivers custom-made services based on the individual user's requirements.   
\end{enumerate}

\subsubsection{Tracking Information Flow}This method tracks how data flows within the system and prevents it from reaching places where the confidentiality policy is violated. For example, IoT users are required to have a certain security level or permission to access the data. Information flow analysis can enforce the confidentiality policy throughout the system; however, it is difficult to implement or verify in complex or dynamic scenarios when sharing and storing data.

In this context, blockchain-based technology has been recognized as a particularly useful solution to handle sharing issues by exploiting smart contracts, distributed ledgers, or consensus mechanisms. For example, in~\cite{Ullah2022Apr}, a blockchain-based decentralized distributed storage and sharing scheme, called the \texttt{IoTChain} model, has been developed to enable data encryption and fine-grained access control. In such a model, {attribute-based encryption (ABE)} plays the role of granular permission, while the Ethereum blockchain is responsible for controlling access. Instead of relying on TTPs, the work in~\cite{Oh2022Sep} introduced secure data-sharing systems that relied on key aggregate searchable encryption and blockchain technologies, allowing owners to have all their ability to use, share and manage their data. On the other hand, {ABE} has also been attractive as it allows data to be safely stored in untrusted storage, for example, physically accessible sensors, hackable publish-subscribe brokers, and third-party cloud servers \cite{Rasori2022Feb}. In addition to that, a new deoxyribonucleic acid (DNA) cryptosystem for cloud-based IoT infrastructure has been proposed in~\cite{Namasudra2022Dec} to hide and encrypt confidential data in an image before storage using a long secret key.

\subsection{Integrity}
Integrity is a security property that means maintaining and ensuring the accuracy and completeness of data and information in IoT environments. Maintaining integrity is essential to ensure that the data and information collected and transmitted by IoT devices are trustworthy and reliable. However, it is still an open challenge because wireless transmission in IoT applications can cause errors or erasures in the data due to many reasons, such as attenuation, distortion, noise, and storage. Following that, in order to realize the integrity property for data and information, the most important task is to determine three states of information along the path between network nodes: (1) In-state motion, the data information journey is required unchanged from the IoT device to the cloud application, (2) In an idle state, stored information programs must authenticate data information by startup, (3) In state processes, checking integrity is required to perform periodically during operation and even with startup and shutdown. Accordingly, several possible methods to guarantee integrity have been developed for IIoT systems, such as a symmetric cryptographic algorithm, a secure hash algorithm, a message integrity check (MIC), and a checksum and cyclic redundancy check (CRC). However, depending on each application, these will be exploited in different manners.

Wearable sensors have attracted considerable attention in the medical field of IoT due to their ability to allow simple monitoring of human health through biochemical identification~\cite{aouedi2022handling}. However, their widespread adoption limits the ability of intelligent learning networks to perform multiple tasks and ensure data integrity. To this end, some recent efforts have focused on building automated security systems based on data impact anomaly detection and classification, checking the integrity of electrocardiogram (ECG) data~\cite{John2020Aug}, tracing the root cause of integrity compromise~\cite{Wang2022Oct}, and leveraging blockchain to improve the data integrity of food and drug administration (FDA)-approved medical wearables~\cite{Sneha2021Aug}.

A smart home can make life more convenient than ever, but in return, it also involves a lot of data integrity concerns due to the complex coexistence of devices. Furthermore, the deployment of a multistep algorithm also results in a high power consumption of the devices~\cite{Douha2022Nov}. To solve these, the work in~\cite{Tham2022Jan} proposed a protocol for smart homes based on symmetric cryptography by incorporating the BAN logic and the Scyther tool. On the other hand, research in~\cite{Kai2021Nov} argues that the use of cryptographic signatures and smart contracts still has vulnerabilities, and the use of smart home blockchain networks can preserve data integrity and robust security in different ways. In this context, \cite{Kai2021Nov} proposed four different consensus algorithms, namely concatenated hash transactions (CHT), Merkle hash tree (MHT), even modified MHT, and modified MHT. At the same time, the work in~\cite{Ammi2021May} utilized a hyper ledger fabric platform in conjunction with a hyper ledger composer to track the flow data within the smart home system. Meanwhile, the work in~\cite{Chen2022Sep} relied on the home gateway to aggregate all data information and design a data structure so that the data tag information can not only be efficiently and dynamically stored but also reduce local storage pressure. Recently, the authors of \cite{Jadav2022Apr} introduced a new secure and trusted blockchain-based onion routing framework, where the anonymity of targets is maintained by storing and tracking the onion nodes' threshold values through the blockchain network while the sensor's data requests are classified as malicious or nonmalicious through LSTM model.

In the smart grid, real-time energy management systems are in charge of control, scheduling, and decision-making tasks and rely primarily on the state estimation process. Thus, it is prone to a well-coordinated data integrity attack (DIA), causing power outages and disordered generation dispatch. In this context, the work in~\cite{Zhang2021Jan} designed a unified framework of zero-parameter information DIA and countermeasure by carefully looking at the system topology and branch parameters. However, such a system-based model approach may be limited to complex scenarios and presents new challenges in the development of appropriate algorithms. To address this limitation, some recent work is focusing on observation-driven combined data driven system models~\cite{An2022Feb,Reda2022Jun,Goyel2022Aug}. For example, the work in~\cite{An2022Feb} designed a general solution with three main steps: (1) determining all possible attack strategies related to dynamic state estimation; (2) formulating the DIA detection problem into a partially observable Markov decision process; and (3) building a DL detection framework, where the LSTM layer is used to extract anomalous state change while the DRL algorithm aims to detect DIA. In~\cite{Reda2022Jun}, a two-stage power state prediction and attack detection framework has been proposed, where the first aim is to focus on power system state prediction, while the second uses a Kullback-Leibler-based binary classifier to detect false data injection attacks. Unlike~\cite{Reda2022Jun}, the work in~\cite{Goyel2022Aug} proposed the voting-based ensemble learning technique with an imbalanced training dataset in the case of full alternating current but limited direct current state estimation information. 

However, in the IoT environment, issues such as hardware failures, software corruption, or malicious intrusions can disrupt communication. This often makes it difficult for IoT devices to detect such events in a timely manner. Consequently, ensuring data integrity verification emerges as a paramount challenge in IoT storage security. By introducing a TTP, the work in~\cite{Zhu2019Jun} demonstrated that their approach based on a short signature algorithm can effectively reduce the overhead of the hash function in the signature process. However, the authors of ~\cite{Wang2019Nov} argued that the use of TTPs might increase computational and communication overhead in large-scale IoT data, and therefore introduce a new scheme, named blockchain and bilinear mapping-based data integrity (BB-DIS). In BB-DIS, collected IoT data is divided into shards and homomorphic verifiable tags, and therefore its integrity can be easily overseen by blockchain transactions. To reduce the cost of certificate management, key escrow, or the requirement of a secret channel per user, the work in~\cite{Li2021Nov} developed an efficient certificate-based data integrity auditing protocol for cloud-assisted wireless body area networks by fixing the computational cost in tag generation for a data block so that it is independent of the size of the data block. On another front, the work in~\cite{Sang2022Aug} claimed that only considering storing a single copy in the cloud is inadequate; it would be better to store multiple replicas in the cloud. In order to simplify the authentication with only two hash operations, a provable multicopy integrity auditing scheme was suggested by leveraging the indistinguishable obfuscation technique. In summary, securing integrity checks for cloud services is critical to ensure that reliable IoT data are saved and maintained, where employing blockchain techniques is a great effort and was recently reviewed in~\cite{Han2022Oct}.

\subsection{Intrusion} 
Intrusion into intelligent IoT networks is a security threat related to unauthorized access or attack on IoT devices, data, or networks. Intrusion can jeopardize the confidentiality, integrity, and availability of the IoT system, resulting in a variety of losses such as data theft, denial of service, botnet infection, or malicious. Therefore, in IIoT networks, it is important to have an IDS that can monitor network traffic and detect any anomalous or malicious activities. Besides, the security of the IoT system can be also enhanced with IPS mechanisms in order to achieve real-time protection, reduce false positives, and adapt to dynamic environments. In what follows, we explore one by one the recent IIoT intrusion detection and prevention developments.

\subsubsection{IIoT Intrusion Detection System} 

The presence of IDS could be a device/software application that monitors any suspicious activities and gives warnings about harmful activity or policy violations. An IDS can be deployed in different locations, such as host-based, network-based, protocol-based, application protocol-based, or hybrid. There are three common detection variants that IDS employs to monitor intrusions:
\begin{itemize}
    \item \textit{Signature-based detection}: This type of attack can easily detect known cyberattacks by looking for specific patterns (e.g., byte sequences in network traffic or signatures); however, it has difficulty detecting new attacks when no pattern is available.
    \item  \textit{Anomaly-based detection}: This type of security system takes advantage of ML to create a reliable operating model and compares new behavior with the model, thereby enabling the detection of unknown attacks, in part due to the rapid development of anomalies. Compared to traditional signature-based IDSs, this approach shows better generalized properties as it is trained according to specific application and hardware configurations. However, they also yield more false positives. 

    \item  \textit{Reputation-based detection}: Identify potential cyber threats by reputation score.
\end{itemize}
Although the use of IDS does not stop the attack by itself, it can beconfiguredp quickly and easily to detect known and unknown threats and does not affect the performance or availability of the network. Thus, the functionality of IDS has been exploited and developed for distinct IIoT architectures.

In decentralized IDSs, the work in~\cite{Ioannou2021Jul} merged SVM-supervised ML into dedicated IDS agents to monitor and alert the vicinity whenever a malicious node occurs. To fine-tune the learning network's ability to spatiotemporal representations, the authors of~\cite{Abdel2021Feb} produced semi-supervised DL methods through a multi-scale residual temporal convolutional module. In addition, the work in~\cite{Roy2022Mar} proposed a novel cluster-driven ensemble learning algorithm to identify anomaly data in the sensor measurement and control signal. Based on the available data on edge devices, the FL framework has been used efficiently to detect IoT network incursions, especially when incorporating RL~\cite{Otoum2021Jun}, AE model~\cite{Regan2022Jun}, and DNN~\cite{Wang2023May}. Remarkably, FL-based RL showed superiority over the SVM-steered IDS with accuracy and detection rates of 0.985 and 96.5\%, respectively. Furthermore, FLs are particularly beneficial for softwarization of networks to achieve an accuracy of 84.32\% and detection rates up to 83.10\%~\cite{Aouedi2022May}. To improve learning networks with decentralized heterogeneous environments, some recent efforts have suggested the use of adaptive learning IDS, for example, RL in~\cite{Bikos2021Sep} to classify the monitored resource consumption parameters or segmented-FL in~\cite{Shingi2021Jul} to aggregate a global model from shared parameter among the workers and automatic segmentation of workers. However, collaborative development of IDSs in these environments can cause conflicts of interest. To this end, the work in~\cite{Liu2020May} designed decentralized Markov interaction-distribution processes to initiate the information interaction among agents. At the same time, the work in~\cite{Putra2020Apr} looked at the use of BCT to establish distributed trust between participants. On the contrary, the work in~\cite{Friha2023Apr} formulated the problem of security, decentralization, and differential privacy into three building blocks. First, a key exchange protocol is responsible for securing the communicated weights among all peers. Second, a differentially private gradient exchange scheme aims to counteract privacy leaks. Last, a decentralized FL is applied to mitigate the failure/attack risk involving the aggregation server. Consequently, this approach can achieve performance comparable to centralized learning (approximately 94.37\%) and outperforms the only FL-based approach (93.91\%) in accuracy. Compared to its FL-based IDS counterparts in terms of F1 score, recall, and precision, it, respectively, shows overall performance improvements of 12\%, 13\%, and 9\%.

In distributed IDSs, the work in~\cite{Reddy2021Aug} presented a security mechanism and a guarantee of true operation based on an exact greedy boosting ensemble method for fog computing nodes. Through the IoTID20 dataset for smart cities, the results achieved by the modeling knowledge of the proposed XGBoost algorithm provide superior attack classification ability compared to modern ML techniques in distinguishing normal and anomalies; specifically, the adoption of the XGBoost algorithm for binary and multiclassifications has the accuracy of 98\% and 99\%, respectively. Similarly, the work in~\cite{Sahi2021Oct} deployed a Raspberry Pi cluster on a local area network and demonstrated that the XGBoost can achieve a recall of 89\% in ADFA-LD datasets, a predicted inference time of 130 ms compared to the cloud with 735 ms, and an estimated running cost of 201 Indian rupees/month against the cloud cost of 2051 Indian rupees/month. To reduce the new attack detection time and complexity, the work in~\cite{Li2022May} suggested using only adaptive online ML algorithms for cooperative network IDSs. Meanwhile, the work in~\cite{Abdel2020Sep} proposed a deep-IFS framework to address current scalability and performance limitations in handling big IIoT traffic data. Experimenting with BoT-IoT datasets, this framework not only gains significant robust performance for binary classification (accuracy: 99.75; F1-score: 98.14; AUC: 99.98, training time: 135.6s) but also the multiclass scenarios (accuracy: 99.77; precision: 99.99; recall: 99.77; F1-measure: 99.88, training time: 184.8s). In~\cite{Jung2022May}, a novel DeepAuditor approach has been designed for multiple IoT devices through power auditing to detect malicious behavior with up to 98.9\% accuracy. To do this, in DeepAuditor, a distributed CNN classifier focuses on online inference in a laboratory setting, while encapsulated HE and sliding-window protocols are used against data leakage and reduce network redundancy.

\subsubsection{IIoT Intrusion Prevention System}

An IPS is a security mechanism that not only detects intrusions but also prevents or blocks them from affecting IoT systems. There are different modes in which IPS can work, such as inline, tap, or span, depending on how it interacts with the network traffic. Based on the approach that IPS systems exploit to scan network traffic in one or more detecting times, IPS can be distinguished as follows:
\begin{itemize}
    \item \textit{Signature-based detection}: This type mainly analyzes network packets and then compares them with pre-configured and predetermined attack patterns known as signatures.
    
    \item  \textit{Statistical anomaly-based detection}: This method uses an established baseline to determine whether a pattern observed in network traffic is abnormal or normal. For example, the baseline describes how much bandwidth is utilized and what protocols are used; if sample detection deviates from this baseline, it is flagged as an anomaly. This kind of protection is useful for detecting new threats, but it also produces false positives when the legitimate user uses resources that overwhelm the baseline or when the baselines are not properly established.

    \item  \textit{Stateful protocol analysis detection}: This approach detects protocol state violations by comparing observed events to predefined profiles of commonly recognized definitions of benign activity.
\end{itemize}
Although the realization of an IPS has significant hurdles (i.e., resource constraints, scalability concerns, and compatibility issues with various IoT protocols and devices), they may improve the security of an IoT system by offering real-time protection, decreasing false positives, and responding to a changing environment. Compared to IDS, an IPS can be envisaged as an extension of an IDS as it stops the potential attack from happening rather than just reporting it. For this reason, IPSs can be considered the next step after IDS in the progression of security technology.

In an effort to achieve decentralized IPSs, the authors of \cite{Chiba2022Jan} suggested the use of the multi-feature extraction (MFE) process in cloud computing and the MFE-extreme learning machine (ELM) algorithm to detect and discover network intrusions to cloud nodes. Which, ELM directly learns to use the least squares method instead of iterating to adjust the neural network weights and biases of the nodes in the hidden layer, thus gaining a faster learning speed and reducing the risk of overfitting. In \cite{Illy2022Aug}, a novel DNN-based multistage hybrid intrusion detection and prevention system framework was proposed to sequentially address the limitations of each learning stage. Initially, conventional DNN detection models learn, train, and validate the learning dataset. Misclassified or classified samples with low confidence in decision are then used to create a new dataset for next-stage DNN training. Finally, a collaborative IPS capable of emergency response will automatically mitigate attacks and detect anomalies. The numerical results confirm that such a framework achieved excellent performance in the WUSTL-IIOT-2018 dataset, with a detection accuracy of 99.99\%, a false alarm rate $\leq$0.01\%, and an undetected rate of 0. 18\% (that is, false negative rate).

In distributed IPSs, to prevent bot intrusion and control the activities of the system, the work in~\cite{Jayalaxmi2022Sep} applied a co-relation-based subset evaluation method combined with a novel cascade forward back propagation neural network model. Specifically, this approach evaluates each feature's ability to predict the class variable and measures the degree of redundancy between the features. Here, the subset is generated by selecting the variables having a high correlation with the attack type and a low correlation among other features. Later, extraneous and redundant features were removed to retain the most prominent features of the training data set. Through experiments on five popular bot datasets: NF-UNSW-NB15, NF-ToN-IoT, NF-BoT-IoT, NF-CSE-CIC-IDS2018, and ToN-IoT-Windows, this approach shows 100\% accuracy on testing of the subset and 97.3\% accuracy for the full dataset. Similarly, the work in~\cite{Onah2021Dec} also proposed a feature selection based on a genetic algorithm wrapper combined with the Nave Bayes model to detect anomalies to reduce time complexity by removing unnecessary attributes while improving accurate predictions thanks to the security laboratory knowledge discovery dataset. In~\cite{Ravi2020May}, a new attack mitigation module was integrated into the fog nodes to perform IDSs based on supervised learning. When this module is activated, a drop rule is applied to the MAC-IP address pair of the IoT device that sent the malicious packets. To avoid blocking legitimate traffic, a random timer is set before removing the drop rule. The captured traffic is analyzed by the admin and used to update the model if needed. If the mitigation module receives uncertain input, denoted by a question mark, a drop rule is set for a shorter random period. The administrator reviews the traffic to determine whether it is an attack and updates the model accordingly when necessary. 

\subsection{Countermeasures}

\subsubsection{Unreliable Communication}
In the IoT network, countless systems coexist and interact with each other to support human activities: home automation, grids, agriculture, transportation, and healthcare services. However, such complex environments generate new security threats related to data breaches, identity theft, and unauthorized access to confidential information. To address such challenges, researchers have developed corresponding countermeasures using combinations of cryptosystems with new technologies and strategies. 
Researchers in~\cite{Ullah2022Apr} introduce an attribute-based access control and AES-128 encryption schemes to better encrypt IoT streams before uploading to the interplanetary file system. The elliptic curve Diffie-Hellman key exchange protocol, a secure method of distributing private keys, enables data users to access transaction details from the Ethereum blockchain and retrieve their private keys if they forget them.

For the Internet of medical things aspect, the work in~\cite{Ding2020Jul} uses a DL-based approach to transfer the medical image into two image domains, which include the original medical image domain and the target domain (regarded as \say{hidden factors} to guide the learning model). DL-based encryption design consists of two components: a generator and a discriminator. The generator components create an image that resembles the target domain, while the discriminator component encourages the former to create images identical to those in the target domain by recognizing the generated images. In this way, the original medical image can be transformed into ciphertext in the target domain. Accordingly, these ciphertext images can be secure on the storage server from any unauthorized access.

For the aspects of the wireless sensor network (WSN), the work in~\cite{Saba2021Jun} overcomes the unpredictable nature of sensors and wireless channels using a two-phase process. In the first phase, DRL is used to identify energy-efficient and fault-tolerant routes through interactions with the environment. In the second phase, an AES cryptography-based deterministic procedure is considered to produce cipher blocks. The encryption process includes two steps: \textit{i}) a pseudorandom code, called a \texttt{keystream}, is generated based on the function $F$ given as
\begin{align}\nonumber
    F= k_s\in\{0,1\}^n+\beta,
\end{align}
where $n$ represents the size of the individual keystream while $\beta$ implies the block of bits to initiate the starting variable. \textit{ii}) the cluster head $j$ aggregates the data items $D$ and uses AES to perform an arithmetic modular operation on $D$ and $k_s$ to produce a cipher block $C_j$, which cannot be repeated twice for data encryption. 

Nevertheless, the symmetric solutions used above are still resource-intensive, leaving a significant need for lightweight cryptography for resource-constrained devices. To this end, four algorithms defined as the latest ISO/IEC 29192 standards (SPECK, SIMON, PRESENT, and CLEFIA)~\cite{Hasan2022Nov} should be considered when designing learning methods. However, the selection of these algorithms is contingent on the hardware specifications of the IoT device in use, as well as the desired degree of security. To be specific, SIMON is a Feistel network block cipher optimized for hardware-based implementations, supporting key sizes ranging from 64 to 256 bits and block sizes from 32 to 128 bits. Conversely, SPECK is tailored for software-based implementations with a low memory footprint due to its code's operations, enabling key sizes ranging from 128 to 256 bits and a block size of 128 bits. Meanwhile, CLEFIA is used for both hardware and software implementations, supporting multiple key sizes ranging from 128 to 256 bits and accepting a 128-bit plaintext block. However, it is prone to differential attacks. PRESENT, on the other hand, is a hardware-oriented block cipher belonging to the substitution-permutation network family of algorithms for various applications that do not require high levels of security, with a block size of 64 bits and key sizes of 80 and 128 bits. Although it is susceptible to various attacks such as birthday attacks, such risks can be well solved by encrypting large datasets. 

\subsubsection{Unreliable Data Aggregation}
Data generated by the multitude of local IoT devices need to be collected by cloud or edge servers to be used in training intelligent models. Despite being encrypted through various security methods before communication, aggregating data without additional protection can still result in vulnerabilities (e.g., lost data, modified data, and leaked private identity/individual information) when processing the sharing and querying tasks that impede the development of trustworthy IIoT networks. Driven by these issues, recent studies specify countermeasures by directly encrypting collected IoT data during the period of aggregation processes. Researchers in~\cite{Song2022Feb} proposed a novel encryption-based FL solution, where Shamir Secret Sharing (SSS) and the SSS homomorphism are used to back up the secret data of users whenever a user breaks down and to support two sets of shares, respectively. In particular,  the communication between the user and the server is executed in 6 rounds as follows.  \textit{Setup}: the data are prepared. \textit{Keys Distribution}: each user generates a pair of public keys and shares a secret with the other users. \textit{Keys Sharing}: the input data is encrypted by the users using secret sharing homomorphisms and then transferred to the server as ciphertext. \textit{Model Collection}: both the server and users confirm the set of active users. \textit{Message Check}: the user encrypts the input data and sends them to the server as ciphertext. \textit{Unmasking}: the server calculates an aggregated plaintext based on the collected shares and ciphertexts and stops the aggregation process if the number of receiving messages is less than the predefined number before timeout. Then, a global model with the masked trained model on the server will be returned to users for usage and retraining. 

On the other hand, the authors of~\cite{Gope2022Apr} suggested ultralightweight data aggregation frameworks with DL protection that do not require the maintenance of a secret key to communicate with the aggregator by defining a two-phase operation: enrollment and data aggregation. In the first phase, the endpoint gateways send an enrollment request to the area aggregator through a secure channel. The area aggregator first generates a set of challenge requests to the sensor via end-point gateways to collect a sufficient number of challenge-response pairs. The internal characteristics of the original {physical unclonable function (PUF)} are then estimated using an ML-based training {PUF} model built in naive Bayes/linear regression algorithms. In addition, a preselection and filtering mechanism is also considered to eliminate the impact of noisy data caused by the response of a PUF. In the end, the area aggregator generates a unique \texttt{id} for the sensor and stores it along with the trained PUF model in its secure non-volatile memory while sending a copy version to the end-point gateway. The latter will then ask the sensor to disable the PUF to prevent illegal access. In the second phase, for a particular session, a sensor randomly generates a pair of the challenge and timestamps used as input of the PUF response before encrypting them and the data collected by this session with the PUF key. Upon receiving the encrypted message from the sensor, the end-point gateway attaches this message along with \texttt{id} and then sends it to the area aggregator. For each received message, the aggregator loads the trained PUF model and generates the PUF key to decrypt this message, as well as check the validity of the timestamp and challenge. When doing the same for the other message, the aggregator can get and calculate the sum of all data reads.

However, the practical designs of the DL approaches in~\cite{Song2022Feb,Gope2022Apr} have not been focused on and are still limited by interpretability problems, requirements for large amounts of data, hard-to-design and tune parameters, and a high complexity of computation. As such, to increase the diversity of the model to improve the ability of representation and feature extraction while maintaining the original merits of the traditional model, it should be further considered with the feature of a lightweight deep model~\cite{Kong2022Mar} by constructing the hierarchical and cascaded model structure. To achieve this, the partial least squares approach is considered in order to model the association between the observed and predicted variables using latent variables when dealing with the problems of small sample capacity, high variable dimension, or bad multiple correlations.

\subsubsection{Unreliable Data Storage}
Secure data storage is always a competitive area to prevent critical problems related to secret key leakage or the replacement of sensitive data with files tampered with by attackers. Consequently, many research efforts have focused on integrating keys into data during the aggregation phase by leveraging conventional encryption mechanisms and/or combining them with BCT. However, using such cryptosystems typically requires enormous computing power in the key generation algorithms, and the dissemination of such public keys still presents risks, especially with collecting and aggregating big IoT data. These certainly affect the ability to store data securely. In this context, it is preferred to use DNA-based cryptosystems along with advanced AI technologies that allow encryption and decryption algorithms executed with simple operations, thus improving significant computational resources~\cite{Kang2023Jan}. Moreover, when dealing with untrusted third parties, masking encrypted confidential data in an image before storage with a DNA cryptosystem designed in~\cite{Namasudra2022Dec} and distributing them with multiple replicas on untrusted third parties/clouds as suggested in~\cite{Sang2022Aug} could be promising solutions not only to secure data storage but also to improve IoT network resources.

\subsubsection{Unreliable Learning}
Decentralized and distributed IoT architectures can ease the burden on central servers/clouds by allocating resources and minimizing overheads more efficiently than centralized ones. However, these also pose some challenges in creating reliable learning frameworks. This problem stems from the huge and diverse IoT data generated by various intelligent IoT applications and devices along with the presence of a multitier IoT network that can coexist with other networks. Therefore, it is difficult to build standard rules and a unified collaborative learning framework for data sharing and training among parties. Moreover, this becomes more difficult when some parties/clients are untrustworthy, as they can manipulate the outputs sent to the server and influence the training process in negative ways. In this context, exploiting FL frameworks, aware of unreliable agents~\cite{Ma2021May}, as well as combined IDP strategies~\cite{Abdel2021Feb,Shingi2021Jul,Friha2023Apr} to reduce the risk of failure/attack on the aggregation server and/or combine with IPS~\cite{Illy2022Aug,Jayalaxmi2022Sep,Ravi2020May} to solve the computational overhead in performing the training, detection, and recognition process; the reliability of learning phases can be improved. 

\subsubsection{Over-the-Air Firmware Updates}
{For IIoT networks, over-the-air firmware updates by wireless capabilities are regarded as a common, flexible, and efficient approach that enables updating new features, fixing bugs, supporting protocols, and resolving known security vulnerabilities. However, updating a large number of IoT devices like this appears to open the door for potential attackers to compromise the availability and integrity of firmware updates without access. However, securing IoT devices becomes challenging due to limited interaction abilities, battery power, sleep mode, number of devices, and different properties such as system-on-chip, manufacturers, providers, and communication protocols. These make traditional security mechanisms difficult to adopt.  To protect the integrity of firmware update processes, it is preferred to use a blockchain framework with smart contracts in \cite{He2019May}. The framework involves six steps: 1) The vendor service generates a new blockchain transaction $\texttt{ID}$ that includes the information of the target IoT device and the SHA1 hash of the new firmware update; 2) This firmware update will be sent to the target IoT device with $\texttt{ID}$; 3) After receiving the firmware update binary, the IoT device calculates the SHA1 hash; 4) The IoT device validates the update in the distributed ledger with the calculated SHA1 hash and transaction $\texttt{ID}$; 5) if validation succeeds, the IoT device applies the firmware update and sends a status update to the blockchain; otherwise, the firmware update process is aborted and a failure notice is recorded; and 6) The vendor service queries the blockchain to collect updated statistics to determine further actions.}

\subsubsection{On-Device AI}
{The growth of IIoT is expected to be widespread in both the vertical and horizontal domains. In that sense, IoT devices will be equipped with microlearning models capable of performing various tasks such as OCR, face recognition, liveness detection, identification card and bank card recognition, and translation. This approach helps IIoT networks avoid sending users' private data to the cloud, eliminates the need for a network connection, and saves the latency of back-and-forth communication. However, deploying a large number of AIs on a device without proper control raises significant security concerns regarding model privacy. Hence, it is imperative to safeguard model privacy while accessing untrusted AI accelerators for essential tasks. In this context, there are two highly recommended methods for secure on-device learning models: ShadowNet \cite{SunZhichuang2023}  and MirrorNet \cite{LiuZiyu2023}. ShadowNet is an exceptional framework developed based on CNNs that transforms the weights of linear layers before outsourcing them to the untrusted world and then restores the results inside the trusted execution environment (TEE). This allows the heavy linear layers of the model to be securely outsourced for acceleration without leaking the model weights. On the other hand, MirrorNet is an excellent alternative that uses the input DNN model as a backbone network and connects a lightweight mirror network to it, which is stored in a secure world. This approach effectively eliminates the vulnerability to model extraction and meets the computational and storage limitations of the TEE. The lightweight network is a crucial component of the entire model and allows authorized users to perform high-performance inference combined with the backbone network.}

\section{Privacy Issues in IIoT}
\label{sec:privacy}

The IIoT infrastructure for computation and communication must not only be efficient and reliable but also be trustworthy. While the efficiency of IIoT has been widely studied, critical privacy concerns have received less attention, and the phase of information collected, transmitted and analyzed, as well as the ML training, is the most vulnerable. Private information of interest might encompass \textit{payload data} acquired by sensors and {IIoT} devices, which are then transmitted through the network to a centralized processing server or retransmitted to other IIoT devices. For instance, when a medical IoT transmits monitoring data to a remote hospital or doctor's office, sensitive information such as a patient's blood pressure, sugar level, and other vital signs becomes a significant privacy concern.
At the same time, privacy concerns may also arise for \textit{context information}, e.g., the location of a sensor, and for the parameters of \textit{ML models} fed with the data from the sensors.
Given the importance of information, it requires appropriate protection from adversaries, and users should be fully aware of the processing of their private data.
While in the previous section (Section~\ref{sec:security}) we focused on measures taken to protect IIoT systems, data, and resources from unauthorized access, in this section we focus on how to keep personal data confidential and limit its collection, use, and disclosure without consent, where confidentiality is only one aspect.
As such, we will describe the definition of privacy (Section~\ref{subsec:privacy-principles}), attacks on this set of information, that is, data (Section~\ref{subsec:privacy-data}), location (Section~\ref{subsec:privacy-location}), and models (Section~\ref{subsec:privacy-model}), and possible countermeasures (Section~\ref{subsec:privacy_counter}). Finally, we summarize the main trend for preservation of privacy in IIoT in Section~\ref{subsec:lessons_privacy}.

\subsection{Privacy in IoT: Concepts and Principles}
\label{subsec:privacy-principles}

Although privacy is a very broad and multifaceted concept, privacy in IIoT refers mainly to the exposure of information from smart objects to the outside world~\cite{yang2017survey}. Therefore, it becomes crucial to acknowledge and regulate the utilization and distribution of personal information by entities beyond individual control. Privacy threats are commonly among the primary concerns of users and can significantly impact the adoption of new technologies, particularly in the context of IIoT, where devices make use of personal information.

The work in~\cite{schaar2010privacy} highlighted that privacy must be approached from a design thinking perspective. In particular, privacy must be incorporated by default into networked data systems and technologies and become integral to organizational priorities, project objectives, design processes, and planning operations. Protection of privacy in the modern era should follow the principles summarized in Table~\ref{Table:privacy_principles}. From these principles emerges how \textit{ privacy by design} requires architects and operators to keep the interests of the individual by offering measures such as strong privacy defaults, appropriate notice, and user-friendly options.

\begin{table*}[t]
    \renewcommand{\arraystretch}{1.15}
	\caption{Privacy Principles in IIoT}
	\label{Table:privacy_principles}
	\centering
	{%
	\begin{tabular}{l|p{9.5cm}}
		\hline 
		\textbf{Principle} & \textbf{Explanation} \\
		\hline
		Privacy included by design & Privacy is not an add-on feature but an intrinsic element deeply ingrained within the fundamental functionality of a computer system. It involves seamless integration with information technology, operations, and architecture, encompassing a holistic, integrated, and innovative approach. \\ \hline
		Privacy is a default setting & Users do not need to proactively take steps to protect their privacy. Instead, fair information practices will notify them about the desired specifications, limitations on data collection, restrictions on use, data retention policies, and disclosure practices.\\ \hline
		Proactive not reactive, preventative not remedial & The objective of this setup is to deter privacy intrusion by acknowledging the significance and advantages of maintaining routine. This involves having well-defined commitments at the highest level, clear communication of privacy commitments between the user community and stakeholders, and established mechanisms for identifying and addressing inadequate privacy practices. \\ \hline
		Full functionality — positive-sum & It employs a win-win approach, avoiding compromises between different goals, and strives to achieve all desired objectives (e.g., confidentiality and functionality) in a user-centric manner. \\ \hline
		End-to-end security — full lifecycle protection & Privacy are seamlessly integrated into IIoT systems right from the moment the first record is captured and throughout the data's lifecycle. This ensures that information is securely stored, processed, and appropriately disposed of at the end of the process.\\ \hline
		Visibility and transparency — keep it open & It assures all stakeholders that information is managed by stated commitments and goals, and its components are transparently visible to ensure accountability and trustworthiness.\\ \hline
	\end{tabular}
	}
\end{table*}

In addition to these principles, \cite{roman2013features} also includes six more principles:
\begin{enumerate}
    \item \textit{Data minimization:} To proactively prevent privacy risks, it is essential to systematically minimize the amount of data collected and processed. As a result, the development of software, information, and communication technologies and systems should start with non-identifiable interactions and processes. Wherever feasible, consistently identifiable and observable personal information should be limited to the minimum necessary.
    \item \textit{Informed consent:} The terms are presented in a clear, relevant, and transparent manner, enabling users to make informed decisions about sharing certain information, unless legally required. The level of data confidentiality determines the quality of consent needed from users.
    \item \textit{Transparency:} Users should be provided with an overview of how their data will be processed and used during its utilization.
    \item \textit{Verifiable preventive protection:} Implement security measures that may include verifiable measures.
    \item \textit{Accuracy:} Maintain accurate, complete, and up-to-date personal information necessary to achieve the stated objectives.
    \item \textit{Possibility to withdraw consent:} Allow users to revoke their consent at any time and provide them with the option to delete any shared information.
\end{enumerate}

Although these principles may not guarantee the absolute inviolability of users' ``personal lives'', failing to implement privacy-by-design principles leaves the IIoT system vulnerable to risks such as eavesdropping, spoofing, RF jamming, and others reviewed in the following.

In the IIoT environment, the collection, use, and exchange of user data are widespread practices. \cite{ziegeldorf2014privacy} reviews and identifies the most common threats to IoT privacy. 
\begin{enumerate}[(a)]
    \item \textit{Identification:} An attacker can link identifiers, e.g., names and addresses, to specific individuals. In traditional centralized services, a large amount of information is concentrated in a central location beyond the control of the individuals. However, in the context of IoT, especially in IIoT, the stages of interaction and data collection become crucial, as the increased frequency of interactions amplifies the risk of identification.
    \item \textit{Location and tracking:} The location of users can be determined using various tools, e.g. GPS, Internet traffic, and the location of smartphones. Instances of privacy breaches have been detected, such as GPS tracking, disclosure of personal details such as health conditions, and unease related to surveillance or oversight.
    \item \textit{Profiling:} Organizations gather relevant information through interactions with various profiles and data sources to customize electronic commerce. As IIoT evolves, not only does the volume of data grow exponentially every day, but the nature of the data also undergoes qualitative changes, as it now includes previously inaccessible aspects of an individual's personal life.
    \item \textit{Interactions and presentations:} The possibly large number of smart items and new ways of communicating between the system and the customer threatens confidentiality and personal information, as the majority of such mechanisms utilized for user interaction and feedback are inherently public in nature, which can potentially compromise individual privacy. 
    \item \textit{Life cycle transitions:} After IIoT products are sold and eventually discarded, it has been confirmed that smart devices retain a significant amount of historical data accumulated throughout their lifetime that is not deleted during transfer of ownership.
    \item \textit{Inventory attacks:} An attacker can exploit unauthorized access to personal items and features to gather information and then use inventory data to identify and destroy properties.
    \item \textit{Linking connections:} The interconnections between various systems in the recent IIoT ecosystem make unauthorized access and intrusion of personal data a significant and serious threat.
\end{enumerate}

IIoT users are confounded by the above threats, particularly when they become aware that these devices collect and transmit personal data without their full knowledge. In addition, the new scenario of IIoT poses new challenges that are unique and new compared to the traditional IoT environment.
Now we review the threats that can occur at the data level (Section~\ref{subsec:privacy-data}), location level (Section~\ref{subsec:privacy-location}), model level (Section~\ref{subsec:privacy-model}), and then some possible countermeasures (Section~\ref{subsec:privacy_counter}).

\subsection{Data Privacy Leakage}
\label{subsec:privacy-data}

Given the vast range of IIoT applications, e.g. remote patient monitoring, energy consumption control, traffic control, smart parking system, and inventory management; for all of them, users require the protection of their personal information. The general defensive approach uses encryption and authentication to protect privacy. However, because of such a wide range, the literature presents diversified attempts to address such an issue and ensure user data privacy.
In what follows, we differentiate these techniques, starting with general approaches to protect user privacy, continuing with the challenges in more specific use cases such as smart grids and UAVs, and concluding with the data streams in IIoT scenarios.

\subsubsection{User Privacy}

A significant data privacy concern involves the exposure of personal user information during its transmission over the Internet~\cite{roman2013features, porambage2016quest}. For instance, let's consider a scenario where a consumer named Bob makes a purchase using his credit card for an item with an RFID tag. An attacker could link Bob's personal information (known to cloud service providers) to the specific purchased object. Such leakage of user information poses privacy threats, including tracking, localization, and personalization risks. Additionally, if Bob owns a set of interconnected objects, adversaries who can differentiate ownership of certain items might also be able to estimate ownership of the remaining objects. 
These examples are issues that allow user profiling and tracking. Moreover, smartphones and other mobile devices connected to the Internet can reveal a user's geographical location, potentially compromising privacy. Therefore, it appears that users have varying levels of privacy awareness and concern, as information disclosure can occur to different degrees.

IIoT networks can include tens of thousands to millions of devices, each with a unique set of characteristics such as resource limitations, mobility, scalability, autonomy, and interoperability. As a result, depending on the IIoT applications used, different privacy concerns arise.
In the domain of \textit{smart healthcare}, the growing accessibility and prevalence of personal health records on the Internet (e.g., via wearable health monitors for blood pressure and heart rate, smart apparel, and fitness trackers) can also give rise to significant privacy concerns. In June 2015, a major privacy-violation attack took place, in which malware infiltrated blood gas analyzers to exploit hospital networks and illicitly obtain confidential data~\cite{storm2015medjack}. Several international regulatory laws designed to limit data access and safeguard the privacy of medical data make it even more important to protect electronic health records. For example, in the USA, the Health Insurance Portability and Accountability Act (HIPAA) \footnote{https://www.cdc.gov/phlp/publications/topic/hipaa.html} and in the European Union, the General Data Protection Regulation (GDPR)\footnote{https://gdpr-info.eu/issues/data-protection-officer/} completely rewrite the data management policy. As a result, valuable data are frequently kept on site, for example, specific hospitals, and the analysis through ML and DL models takes advantage of FL as a privacy-preserving method~\cite{aouedi2022handling} (see Section~\ref{subsec:fl_iiot}).

In \textit{smart homes},  where users can remotely control, monitor, and measure the power consumption of household appliances through the Internet, there is a potential threat to user privacy. With the use of RFID and sensing technologies, residents can monitor the conditions of their smart home and track objects~\cite{weber2010internet}. Intelligent adversaries can easily infer information about residential behavioral patterns (such as when residents are at home, away from home, or sleeping) if they eavesdrop on and collect timestamps of data transmissions over wireless channels. Attackers can also examine the transmission patterns and infer approximations of the internal layout of the home by analyzing the radio wave patterns and identifiers that are specific to sensors and RFID tags.

In particular, the communication channel in IIoT can lead to problems such as eavesdropping on adversaries. Private data transmission can be ensured by private secure communication, particularly in wireless environments. Sensitive data may be exposed to adversaries if communication channels are made available. Privacy in communication is a challenging topic because there are so many different communication standards, such as Wi-Fi, Bluetooth, ZigBee, Z-Wave, and LoRaWAN.

Finally, given the global nature of the IIoT, we can mention how national-level regulations are unacceptable for privacy. Cross-border and compliant with international law, the private sector should also enhance an adequate legal framework. Although state laws are a more complex and expensive option for protecting privacy than self-regulation, IIoT applications' extensive deployments of heterogeneous networks make self-regulation insufficient. The global distribution and marketing of IoT products, their durability, integration with ubiquitous environments, and the complexity of technological advances present the most difficult regulatory challenges.

\subsubsection{Smart Grid}

The Smart Grid is an electronic-controlled electrical system that links consumers, power producers, and distributors using information and communication technologies, where customers can be generators or producers and add their energy to the grid (prosumers)~\cite{zeadally2013towards}. The architecture provides customers with tools for energy management, improves reliability, resiliency, and power quality, and allows them to use cutting-edge technologies such as renewable energy, energy storage, and electric vehicles. Smart meters also facilitate information flow, enabling two-way communication between customers and utilities, stakeholders, and all other operators.

In managing energy consumption within smart grids, consumers share detailed information about their daily power consumption~\cite{miorandi2012internet}. This information can be leveraged to reveal their habits and behaviors, potentially leading to invasion of privacy. Unfortunately, user data regarding electricity consumption can be intercepted from any point on the Internet. As customers have less control over the data they provide to utilities, users can have concerns about unauthorized use and disclosure of personal data, data leaks or spoofing via hacking, and inferences made from new data types aggregated with other personal data. \cite{noll2014measurable} presents a new risk assessment methodology for smart grids, starting with component evaluation, for example meters, then subsystem evaluation, and finally with a complete system evaluation.
To this end, in~\cite{rahman2015secure}, it is proposed an innovative protocol is proposed that protects users' privacy, confidentiality, and integrity while sharing necessary information. They also suggested a new distributed multiparty computation (MPC) protocol that is clustering-based. The authors derive the conditions under which the smart grid can enforce cooperation among users and prevent dishonest consumption declarations by taking into account repeated interaction between honest and dishonest users. In~\cite{saputro2015privacy}, it has been shown that in a hybrid architecture comprising an IEEE 802.11s mesh-based smart meter network and an LTE-based wide area network for smart meter data collection, it is feasible to employ unique pseudonyms to translate network addresses for smart meters. With this approach, the privacy of the consumer can be ensured because the IP addresses of smart meters remain undisclosed to utility companies. Finally, the authors of~\cite{borges2015efficient} present a protocol that protects privacy between smart meters and energy suppliers. They adopted a homomorphic ECC-based encryption method to effectively protect the data collected by smart devices.

\subsubsection{Unmanned Aerial Vehicles}

Vehicular ad hoc networks (VANETs) play an important role in IoT and IIoT through intelligent transportation.
VANETs integrate an onboard unit (OBU) as an IIoT sensing layer node into the vehicle system~\cite{sutheerakul2017application}.
This node interacts with other peer vehicles and the infrastructure along the roadside. Therefore, to enable security and privacy in VANETs, two essential requirements are to establish secure communication links and provide authentication. To ensure user privacy protection to the same extent as identity and location protection, the OBU requires additional modules to support information security.

However, in addition to traditional data privacy issues, in this case, another possibility of privacy disclosure arises when aerial photos captured by UAVs are sent to the Ground Control Station (GCS). These photos often contain private information such as location and shooting time. For example, photos taken with DJI\textsuperscript{\texttrademark} equipment, such as Mavic and Phantom4 pro, are saved in JPEG format. Invisible information such as image information (manufacturer, size, etc.), camera recording information (ISO, white balance, saturation, sharpness, etc.) and GPS (shooting longitude, latitude, altitude) are also included in these images~\cite{zhi2020security}. Camera manufacturers add invisible information to the header file of the JPEG photo to better describe and illustrate the photo. Even after image modifications, such as blur, mosaic, drawing, and watermark, the header file can still convey the information of GPS and shooting time via the reader. The study in~\cite{zhi2020security} found that after image compression, most of the header file information is missing, suggesting an interesting way to solve privacy issues.

In~\cite{ch2020security}, a BCT-based solution is proposed to improve the security and privacy of UAVs, or IoT in general. Technical information about UAV instructions (or devices), authentication, integrity, and UAV reactions is stored on a cloud platform, where the Pentatope-based ECC encryption method and SHA are used to ensure privacy in data storage. The data is later stored on an Ethereum-based public blockchain to enable seamless BCT transactions. The system also uses the Ganache platform for data protection and privacy. Other simpler blockchain-based approaches, as in~\cite{hasan2018proof}, use identity-based encryption (IBE) and lightweight cryptography techniques to achieve data privacy. This specific solution is general enough that it can be used to orchestrate and govern the delivery of any digital asset or content, including streamed video and audio, which is a good fit for UAV-generated content.

\subsubsection{Data Collection and Processing in IIoT}

Due to their intrinsic data-centric perspective, IIoT applications present three critical issues: scalability, distributed processing, and real-time analytics. \textit{Scalability} matters for IIoT applications that contain numerous smart objects or manage biometric data that must be collected, processed, stored and then published. The \textit{distributed} nature of IIoT processing introduces unprecedented challenges related to privacy throughout the \textit{real-time} learning process, along with liability for data breaches (i.e., the release of secure information to distrustful entities) and different levels of data quality.
While collecting large sets of raw data, it is challenging to balance the privacy preservation in data cleaning and the intentional reduction of data quality and original purpose without losing information needed for data processing and analysis (e.g., via ML algorithms).
The most important privacy concerns in the context of IIoT applications are the gathering, sharing, and transmission of sensitive data related to people. Privacy preservation over high-dimensional datasets is also associated with computational and theoretical constraints. Records in a given dataset should be handled differently for anonymization purposes because cooperative users and individuals have different privacy restrictions. The data collected could be used and published for purposes unrelated to the original goal without the user's consent.

To solve the problem of centralized (private) data storage, which broadens the surface of attack of data leakage, FL proposes to store user-related information locally and share only model-related information \cite{pham2022energy}. On the one hand, in FL, sharing locally trained model information, e.g., parameters and gradients, allows sufficient knowledge for model training while preserving data privacy to a certain degree, since sensitive information remains undisclosed and traffic can be encrypted~\cite{rieke2020future}. On the other hand, ensuring the algorithm's optimal performance without compromising user privacy requires extra effort. Even if data are anonymized, collecting a few data attributes might still enable, for instance, the reidentification of patients~\cite{rocher2019estimating}. We discuss in Section~\ref{subsec:privacy_counter} the principal countermeasures to protect user privacy in FL-empowered IIoT and the simple ML-based learning process in IIoT.

\subsection{Location Privacy Leakage}
\label{subsec:privacy-location}

Privacy concerns may also focus on context information, such as the \textit{location} of a sensor that initiates data communication. For one, alert communication originating from a patient's heart monitor in the medical IoT is enough for an adversary to infer that the patient suffers from a heart problem. Therefore, real-world applications must consider effective countermeasures against the disclosure of private information related to data and context, such as location.

The work in~\cite{elkhodr2012review} provides an overview of the privacy issues of location found in mobile devices. Particular attention is paid to the current access-permission mechanisms used on the Android, iPhone, and Windows Mobile platforms. Actual privacy issues on mobile platforms should be inherited by IIoT and integrated with other static platforms. Integrating RFID objects into an IIoT setting enables context-aware digital representations of physical objects that possess autonomous sensing, communication, and interaction capabilities. Powerful adversaries can potentially monitor all communications, trace tags within a limited period, corrupt tags, and access side-channel information on the reader output. These privacy risks associated with RFID technology are centered on user tracking and localization, which can lead to the creation and exploitation of detailed user profiles~\cite{porambage2016quest}. Consequently, RFID systems must ensure anonymity even when the state of a tag has been exposed.

WSNs are another key underlying technology of the IoT network architecture, which exacerbates the privacy issue of location. WSN can sense the state of an object or monitor an event in the network, and then post a message in the system to notify connected people (or automatic reactive programs) about the event. Due to their self-organizing characteristics (and the uncontrollability of the environment), constraints (e.g. limited resources) and the wireless transmission medium, WSNs have inherent challenges in protecting the privacy that prevents existing techniques (such as public-key ciphers) from being directly transplanted into resource-constrained devices. This problem of location privacy has not been addressed effectively because it cannot be easily solved by encryption or authentication, and thus specific approaches are required, listed below.

Even without the ability to decrypt the transmitted data, an adversary can still compromise location information. This is particularly relevant in WSN, where transmission typically occurs hop-by-hop due to the limited transmission range of sensor nodes. By observing and analyzing traffic patterns between different hops, the adversary may determine the locations of the BS and data source, with the possibility of tracking them down. To address this challenge, context-oriented privacy protection aims to conceal the real traffic pattern~\cite{li2009privacy, sicari2015security}. Various traffic pattern obfuscation methods have been suggested in the literature to protect the location of both the data source and the BS.

\subsubsection{Location Privacy of Data Source}

\begin{figure}[t!]
    \centering
    \includegraphics[width=0.85\linewidth,keepaspectratio]{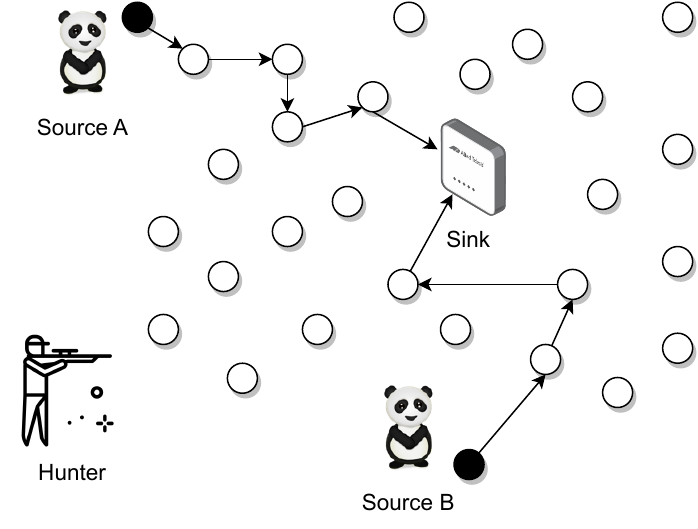}
    \caption{The general \textit{panda-hunter} model as a source location privacy-sensitive scenario in IIoT.}
    \label{fig:locationpanda}
\end{figure}

The problem of protecting the location information of the data source typically refers to the more general (and well-known in the literature) problem of the ``Panda Hunter Game''~\cite{kamat2005enhancing}, illustrated in Fig.~\ref{fig:locationpanda}. In this game, a large number of \textit{sensors} (the circles in the figure) are used to detect pandas, and once the panda is detected, they transmit event messages to the sink (or BS) by multi-hop wireless communication. Simultaneously, a \textit{panda-hunter} endeavors to locate the data source to track the panda. The defender's objective is twofold: first, to accurately gather information about the panda's movements for biological research, and second, to protect the location data from the panda hunters, who can intercept wireless communications between different sensor nodes but lack the decryption key for the payload data. The objective of the panda hunter is to compromise the location of the data source (and the panda) by analyzing transmitted traffic within the WSN. A local attacker, limited to monitoring traffic in a small local area, typically has two approaches to launch the attack: \textit{arbitrary location, }the attacker selects a location within the network and stays there to monitor the traffic; 
\textit{proximity to BS, } alternatively, the attacker may choose to position themselves near the BS with prior knowledge of its location. 

The ``Panda Hunter game'' is a problem that can happen in other IIoT applications, such as monitoring patients and doctors in a hospital~\cite{pai2008transactional}, and tracking friendly soldiers on the battlefield~\cite{chow2010privacy}, to name a few.
Similarly, VANETs integrate an OBU into the vehicle system, serving as a sensing layer node in the IIoT. As this node communicates with the roadside infrastructure and other vehicles, ensuring secure communication links and authentication become crucial requirements to guarantee security and privacy in VANETs. Consequently, the OBU requires additional modules to support information security, ensuring that user privacy is maintained at the same level as identity and location privacy.

In the past, several performance metrics for source security have been suggested~\cite{wang2008source}. One of these is the source's security time, i.e., the total number of packets the source can send out successfully before being intercepted by the adversary. Another metric is the likelihood that the adversary can find the source node in a certain amount of time. For a quantitative evaluation of the privacy of communication, these metrics are of crucial importance.

In what follows, we present five existing techniques to protect against possible disclosure of the \textit{location of the data source} in WSNs: flooding, random walk, dummy injection, cross-layer routing, and limited node detectability.

\paragraph{Data Flooding Mechanisms}

The basic concept of \textit{baseline flooding} is that each sensor node broadcasts the data it receives from one neighbor to all other neighbors~\cite{kamat2005enhancing}. The main assumption behind this approach is that, when all sensors participate in the data transmission, it becomes challenging for an attacker to trace the transmission path back to the original data source. However, the effectiveness of baseline flooding in preserving privacy depends heavily on the number of nodes on the path between the data source and BS. If the path is short and an attacker detects the arrival of the first packet at the BS, they can infer that this packet follows the shortest path between the data source and the BS. Subsequently, the adversary can trace the data source back from the last forwarding sensor along the routing path to the data source. Despite its potential privacy benefits, baseline flooding has a significant drawback, the substantial amount of energy consumed throughout the network. This high energy consumption drastically reduces the overall lifetime of the WSN.

To this end, a modified instance, \textit{probabilistic flooding}, has been proposed in which not all sensors are involved in forwarding data; instead, each node forwards a received packet with a predetermined probability~\cite{kamat2005enhancing}. This scheme can not only save significant energy but also effectively limit the adversary's ability to deterministically trace the data source. The disadvantage of probabilistic flooding is that it cannot guarantee the reception of data by the BS due to the random nature of the approach. To address these concerns about privacy and energy consumption, a routing method called Sink Toroidal Region (STaR) routing was introduced in~\cite{lightfoot2016star}. 
Provides source location privacy through a two-phase routing protocol. In the first phase, the source node forwards the message to a random intermediate node, located in a pre-defined region around the sink node. We call this region the STaR, designed to be large enough to make it impractical for an adversary to monitor the entire region. The random intermediate node serves as a false source when forwarding the message to the sink node. In the second phase, the intermediate node forwards the message to the sink node using a single-path routing. 


\paragraph{Random Walk Mechanisms}

The goal of the \textit{``random walk''} approach is to have packets following a random route through the network so that the path of a packet looks completely random to an attacker. An efficient version of the general random walk tailored to this problem is called \textit{ Phantom Routing}. The data first go through some steps of random walk from the data source and then is transmitted towards BS using a probabilistic flooding scheme. The use of random walks makes it difficult for the attacker to trace the routing path from the original data source. However, studies such as~\cite{xi2006preserving} and~\cite{ross1996stochastic} have shown that the pure random walk approach is not statistically secure when it comes to protecting the location of the data source, as it tends to stay close to the actual data source.

In~\cite{xi2006preserving}, an alternative method was introduced, known as the \textit{two-way greedy random walk (GROW)} scheme. This approach begins by initiating a random path with a specified number of hops from the {BS} and sensors located along this path act as receptors. Subsequently, each source data packet is randomly forwarded until it reaches one of the receptors. Once the packet reaches a receptor, it is then forwarded back to the BS through the pre-established path. A Bloom filter is then used in GROW to avoid repeating cycles.
Based on this concept, EDROW defines the parent nodes located closer to the BS~\cite{tan2014enhancing}. Establishing a ring as the forwarding path presents a greater challenge for adversaries to trace back to the origin of sensor communication.

Furthermore, in~\cite{tang2014cost}, a secure and efficient Cost-Aware SEcure Routing (CASER) protocol is proposed, which is based on geographical considerations. The network is partitioned into smaller meshes as if they are different clusters. In each mesh, the node possessing the highest amount of residual energy is selected as the mesh head responsible for packet transmission. These mesh heads are aware of their position and residual energy, as well as the positions and residual energies of adjacent meshes.
During packet forwarding, the CASER protocol employs a default strategy where the next-hop mesh is chosen to be the adjacent mesh closest to the BS with the highest energy level. However, to protect the privacy of the source location, instead of following the shortest path, a random walking mechanism is employed to create a routing path.

\paragraph{Dummy Injection Mechanisms}

Fake data packets can be injected into the WSN to perturb the traffic patterns observed by potential adversaries, and thus protect the location of the data source. One such approach is the \textit{short-lived fake source routing}, where each sensor has a predetermined probability of sending a fake packet. When a sensor node receives a fake packet, it is simply discarded. Although this strategy effectively perturbs the local traffic pattern visible to adversaries, it has certain limitations in relation to privacy protection. In many scenarios, to maintain energy efficiency within the WSN, the length of each path along which the fake data are forwarded is restricted to one hop only. This constraint allows adversaries to quickly identify these fake paths and exclude them from their considerations.

This version of dummy injection may prove ineffective against global adversaries capable of monitoring the transmission rate of individual sensor nodes, as they can distinguish nodes solely sending dummy data. To address this issue, a potential solution involves the global injection of dummy data while maintaining the same transmission rate for both real and dummy data. However, this approach can introduce substantial delays in actual data transmission.
A possible countermeasure is presented in~\cite{yang2013towards}, proposing to follow a certain distribution of data transmission. As long as every sensor node adheres to this distribution, the delay in real data can be minimized, while preventing the adversary from identifying the actual traffic. Alternatively, two other schemes, Proxy-based Filtering Schemes (PFS) and Tree-based Filtering Schemes (TFS), can filter partial dummy data without compromising source privacy~\cite{yang2008towards}. The entire network is divided into cells, and the proxies take responsibility for transmitting real data from nearby cells while filtering out dummy data. The proxies filter all the dummy packets from the cells, buffer the real data, and then transmit the data packets at the same transmission rate, which includes the buffered real data and newly generated dummy data. Through this filtering process, a significant amount of dummy data is removed, leading to a substantial reduction in energy consumption for overhead communication.
In~\cite{han2018source}, the authors proposed a source location protection protocol based on dynamic routing (SLPDR). In this solution, an initial node is randomly chosen from the network boundary, and then a dummy packet is transmitted to the sink along the shortest path. Every real packet takes a greedy route and then a directed route before reaching the sink. The network is divided into different rings, and dummy traffic is generated in the outer ring without significantly deteriorating the lifetime of the network. The real event packets are transmitted forward in a one-hop flooding pattern, and only the node on the greedy path will swap out the fake packet for the real one.

A similar approach is to choose one or more sensor nodes to mimic the behavior of real data sources~\cite{mehta2007location}. A solution should balance the number of fake sources since more sources lead to more security but also to more power consumption in the WSN. Moreover, fake sources must convincingly emulate the behavior of actual data sources to avoid detection. These challenges are still viewed as open problems and dependent on the IoT scenario. Based on the same idea, the Resilient Timing Analysis Protocol (TARP)~\cite{majeed2009tarp}, proposes an approach called traffic mixing, where each node sends the same number of packets at the same rate, making it difficult for attackers to distinguish between real data and fake messages. One drawback of TARP is the need to transmit a considerable number of dummy packets even when there are relatively few real packets in the network. 

To conceal the real packet with minimum energy consumption, fake packets with a low rate (FPLR) are the approach proposed in~\cite{bushnag2016source}. To obfuscate adversaries while transmitting fake packets at a low rate (and thus minimize the communication overhead), three schemes are proposed, namely Dummy Uniform Distribution (DUD), Dummy Adaptive Distribution (DAD), and Controlled Dummy Adaptive Distribution (CAD). In DUD, both real and fake packets are transmitted at the same constant rate. While this approach provides a certain level of security, it can lead to high energy consumption due to the continuous transmission of fake packets. In DAD, all nodes are initially set to a low constant transmission rate. Real packets on the network are assigned a higher transmission rate, while fake packets are assigned a lower transmission rate. This scheme improves performance and makes it difficult for eavesdroppers to distinguish between real and fake packets. Lastly, CAD incorporates packet loss considerations to further enhance the packet delivery ratio.

\paragraph{Cross-Layer Routing}

In the cross-layer routing in WSNs, nodes use different layers of the communication protocol stack in different ways. Instead of relying only on the network layer to exchange messages about events, cross-layer routing also involves control messages from the Medium Access Control (MAC) layer. This deviation from the standard network protocol stack is not uncommon in WSNs, and cross-layer optimizations are used to extend the battery life of the node and increase the network throughput~\cite{miao2009cross}. In~\cite{shao2009cross}, nodes following the IEEE 802.15.4 standard (e.g., ZigBee, WirelessHART~\cite{molisch2004ieee}) utilize beacon frames, typically employed for network maintenance, to exchange information about captured events. In this way, local adversaries, which usually monitor only network-level packets, can miss certain parts of the information exchange and fail to identify the true source of the data.

\paragraph{Limit Node Detectability}

This class of solutions limits the transmit power of the nodes, making them harder to detect. For example, localisation by silencing~\cite{dutta2010defending} and lowering radio transmit power~\cite{tavli2010mitigation}. The first solution seeks to identify a local adversary present within the WSN by assuming that the nodes possess the capability to detect adversaries using vibration or infrared sensors. 
The detecting nodes inform the others and are the only nodes transmitting, while other nodes maintain silence when the adversary is inside their grid as much as possible.



More recent approaches incorporate a combination of directional antennas (DA), transmit power control, and information compression~\cite{rana2012new}. The use of directional antennas can have two benefits: first, it can decrease the likelihood that local attackers will find the packets and second, it can make it more difficult for them to monitor the network. Additionally, as the beam width of the directional antenna decreases, the difficulty of monitoring the network increases. The network has adopted transmit power control and information compression in an effort to further reduce its energy usage. The suggested scheme can effectively reduce an adversary's ability to overhear conversations while saving energy.
Building on this foundation, an improvement called the Bypassing method based on Directional Transceivers (BDT) is introduced in~\cite{das2016bypassing}. To defend against backtracking attacks by eavesdropping, each node in the network is equipped with $4$ directional transceivers. When a sensor node detects an attacker, it sends an acknowledgment message to alert its neighbors along the opposite direction of packet transmissions. Consequently, the nodes located outside the attacker's radio range initiate a new directional transmission to avoid the adversary.

\subsubsection{Location Privacy of Base Station}

Another critical issue for context-oriented privacy is preserving the location of the BS. The BS serves as a central hub for data collection and analysis and acts as a gateway between the WSN and external wireless or wired networks. Its importance lies in its role in coordinating the network's operations, and disrupting or isolating it could result in the malfunction of the entire network.
In the following, we summarize the main privacy-preserving techniques for the location of the BS against local and global adversaries.

\paragraph{Local Adversaries and Defensive Mechanisms}

Protecting the location of the BS from local adversaries involves addressing two significant challenges: (i) Typically, the location information of the BS is contained in the data payload being transmitted, and (ii) Local adversaries can deduce the parent-child relationship (i.e., which node within two communicating nodes is closer to the BS) based on the time interval between receiving and sending data at each sensor. Using this information, the attacker can efficiently trace the path of data transmission back to the BS. To overcome the first challenge, payload confidentiality techniques can be used to hide location information from attackers. However, for the second problem, the literature proposes four main approaches to mitigate the issue and enhance the privacy of the BS's location.
\begin{itemize}
    \item \textit{Changing data appearance by re-encryption}~\cite{deng2006decorrelating}. Similarly to anonymous routing systems used in traditional wired networks~\cite{syverson2004tor}, packets in WSNs are re-encrypted hop by hop as they traverse the routing path. Changing the appearance of the data protects the disclosure of the location of the BS.
    \item \textit{Routing with multiple parents}~\cite{deng2005countermeasures}. It is based on the introduction of a multi-parent system to balance the traffic load between parents and children. In this way, an attacker cannot easily determine which sensor is closer to BS. Since this scheme allows each sensor to randomly select one of its multiple parents for data transmission, it is also harder for the attacker to determine the location of BS by tracking the data transmission.
    \item \textit{Routing with random walk}~\cite{luo2010location}. This scheme divides the neighbors of a sensor into two lists according to the number of hops from the BS -- nearer and farther lists. When the sensor sends data, it randomly selects a next-hop neighbor from the nearest list. While in the traditional random walk, the node sends to the parent with probability $p_r$ and randomly with probability $(1-p_r)$, this approach can be considered as a special case with $p_r=1/2$. On the one hand, this randomness reduces the probability of a successful analysis by the attacker. On the other hand, it delays the delivery of the data packet to the BS since it is possible to choose a node as the next hop that is further away from the sink. This increases energy consumption and there is no guarantee that the data packet will arrive at BS.
    \item \textit{De-correlating parent-child relationship by randomly selecting the sending time}~\cite{deng2006decorrelating}. Since an attacker can figure out the parent-child relationship by the short time interval between sending data to a sensor and receiving data at its neighboring node, this relationship can be decorrelated. Specifically, the parent node and the child node send packets with a randomly drawn time interval from different time windows.
\end{itemize}

An alternative approach aims to improve the resilience of a WSN after BS has been detected. For instance, in~\cite{deng2004intrusion}, the authors propose a multiple BS scheme: If one BS is detected and compromised, the other BSs can continue to collect data and support the normal operations of the WSN. However, this approach does not provide additional protection for the BSs themselves and simply delays the threat from a local adversary in identifying the location of the BSs.

\paragraph{Global Adversaries and Defensive Mechanisms}

Instead, a global adversary can monitor all traffic transmitted in the WSN, and the techniques mentioned above are ineffective. 
Due to the multi-hop scheme of WSNs (forced by the limited communication range of sensor nodes), nodes closer to BS must forward data from more distant nodes in addition to transmitting their data, resulting in a higher transmission rate. An attacker can observe the traffic patterns of different nodes throughout the network and easily identify the BS and compromise the privacy of the context, or even manipulate the sink node to prevent the proper functioning of the WSN.
Therefore, to protect against global attacks, it is necessary to inject dummy traffic and/or dummy sensor nodes, for which the following techniques have been proposed in the literature:
\begin{itemize}
    \item \textit{Hiding traffic pattern by controlling transmission rate}~\cite{deng2006decorrelating}. Since WSNs exhibit asymmetric traffic flow, where sensors close to the BS feature a high transmission rate, this privacy-preserving technique proposes to maintain the same transmission rate among all sensors by controlling the delay of the actual data.
    \item \textit{Propagating dummy data}~\cite{deng2005countermeasures}.
    The proposed solution involves injecting fake packets to prevent an adversary from discerning the real data transmission pattern, assuming that an adversary cannot distinguish between real and fake data. When a sensor detects that its neighbor is transmitting a real packet, it generates a fake packet with probability $p_c$ and forwards it to another neighboring node. The solution incorporates two fractal propagation schemes.
    In the first scheme, $p_c$ decreases as the data forwarding rate increases. The second scheme aims to create the illusion of a high transmission rate area to mislead the adversary into believing that a particular sensor node serves as the BS. However, both approaches come with the drawback of increased energy consumption due to the introduction of fake data across the network. As a result, researchers need to strike a proper trade-off between the energy consumed and the level of privacy protection.
\end{itemize}

\subsection{Model Privacy Leakage}
\label{subsec:privacy-model}

A well-performing ML model usually relies on a large volume of training data, which can be obtained thanks to IIoTs. Such huge volumes of data raise serious privacy concerns due to the risk of highly sensitive information leaking out. In addition, evolving regulatory environments that increasingly restrict access to and use of privacy-sensitive data pose a major challenge in fully benefiting from ML for data-driven applications. A trained ML model may also be vulnerable to adversarial attacks such as membership, attribute, or property inference attacks and model inversion attacks~\cite{aouedi2023f}. Therefore, it seems that well-designed privacy-preserving ML solutions are urgently needed for many new applications. There are notable research efforts in both academia and industry to integrate privacy-preserving techniques into ML pipelines or specific algorithms or to develop architectures that guarantee privacy.

FL comes with the potential for privacy preservation, in which each IoT device collaboratively trains a global model without the need to exchange private data. However, although the fact that in FL systems, there is no share of data among the participants can bring some privacy advantages, some other considerations regard the remaining privacy risks (regarded as vulnerabilities or weaknesses from multiple aspects). Thus, it appears that alternative strategies are required to protect user data and user privacy. For example, even if the data are anonymized, the collection of a few data attributes may allow re-identification of the patient in an e-health application~\cite{rocher2019estimating}. Furthermore, Carlini et al.~\cite{carlini2019secret} have shown that it is possible to extract sensitive text patterns (e.g., a specific credit card number) from a recurrent neural network trained with user speech data. The model can also be manipulated by generating additional memories through gradient-ascent style attacks. In what follows, we provide a structural overview of current methods facing the privacy protection domain, that can be combined with FL or not.


\subsection{Countermeasures}
\label{subsec:privacy_counter}

Based on an analysis of concerns and existing protection models, in what follows, we identify the key challenges that underpin the IIoT ecosystem.
The current data privacy-preserving IoT methods fall into five main categories: \textit{clustering-based methods}, \textit{differential privacy and its extensions}, \textit{lightweight cryptography}, \textit{secure multi-party computation}, and \textit{machine learning-based methods}.

\subsubsection{Clustering-based Methods}

These methods consider anonymity, dummy, generalization, and suppression via \textit{k-anonymity}. A data collection is said to have the k-anonymity property if the information for each person contained in the collection cannot be distinguished from at least $k-1$ individuals whose information also appears in the dataset. Each attribute (column) of the dataset is categorized into an identifier, non-identifier, or quasi-identifier. The basic idea is to suppress the identifiers, such as names, allow the non-identifying values, and exploit the quasi-identifier attributes to preserve the data privacy by clustering these values and ensuring that every distinct combination of quasi-identifiers designates at least $k \in R$ records in any cluster. In this way, the probability of re-identification is $1/k$. A new model can be used to improve performance by guaranteeing a $l \in R$ diversity of records within this cluster, which can resist attacks with background knowledge to some extent.
It should be noted that clustering-based methods require the distribution of the cluster to match that of the entire data set. 
Moreover, since $k$-anonymization does not include any randomization, attackers can still make inferences about data sets that may harm personal data, e.g., via down coding attack~\cite{cohen2022attacks}, and it is not considered a good method to anonymize high-dimensional datasets~\cite{aggarwal2005k}. 

In~\cite{cao2010castle}, an attempt to further improve the resistance against attacks and to anonymize data streams on the fly. These types of method can be used to generalize and normalize quasi-identifiers and thereby defend sensitive information in the IIoT.

\subsubsection{Differential Privacy}

This class of methods has been a de-facto privacy model and attempts to defeat differential attacks~\cite{dwork2006differential}. It was originally formalized to offer statistical guarantees that aggregated released data would not disclose whether specific individuals are part of or absent from the dataset. In other words, differential privacy provides solid theoretical foundations for privacy protection by guaranteeing that the difference between two adjacent datasets is not detected by statistical queries. Two datasets are considered adjacent or neighboring if their aggregated real-valued data representations contain an equal number of elements and differ in at most $m$ elements. The difference is measured using the Hamming distance, where in many cases, the value of $m$ is set to 1.

Given $\epsilon(\cdot)$ as a positive privacy budget function, $D_1$ and $D_2$ two datasets with an adjacent relationship, $M$ is a randomized algorithm that sanitizes the dataset, and such an algorithm $M$ is $\epsilon(\cdot)$-differential private on $D_1$ and $D_2$ if:
\begin{equation}\nonumber
    P[\mathcal{M}(D) \in \Omega] \leq e^{\epsilon(\cdot)} \times P[\mathcal{M}(D') \in \Omega],
\end{equation}
where the probability space $\Omega$ is taken over the randomness used by $\mathcal{M}$.

Differential privacy (DP) is widely used in IIoT applications to provide privacy protection while preserving statistical features, especially in smart communities and smart cities. Fog-based recommendation systems also use DP to hide the unique characteristics of individuals while simultaneously providing accurate recommendations. These mechanisms are probabilistic mapping functions that apply DP to the sensitive data or dataset $x$ to obtain a replacement for it. The replacement approach is based on the elaborately designed probabilistic mapping functions to ensure that the replacement of $x$ reveals the coarse-grained information about $x$ to some extent.

An improvement to this has been proposed in~\cite{mironov2017renyi}. The author proposed a new definition of differential privacy based on the R{\'e}nyi divergence. R{\'e}nyi differential privacy is a strictly stronger privacy definition that offers an operationally convenient and quantitatively accurate way of tracking the cumulative loss of privacy throughout the execution of a separate differentially private mechanism and throughout many such mechanisms. In particular, R{\'e}nyi differential privacy allows combining the intuitive and appealing concept of a privacy budget with the application of advanced composition theorems. 

A more specific problem regards the sharing of unstructured data such as images, audio, videos, and texts between a data owner and untrusted third parties on social media, smart devices, and surveillance devices. Since these unstructured data may contain privacy-sensitive information or so-called personally identifiable information (PII), for example, faces, identities, license plates, voiceprints, and authorships, the data owner may obfuscate them and publish their coarse-grained versions to protect this PII from unintended disclosure and illegal use. Although several obfuscation methods have been proposed to protect this type of data, see, e.g.,~\cite{poddar2020visor, shokri2017membership}, most of them suffer from many problems such as high computational overhead, inner attack, and nonprovable privacy. DP, when tailored to the specific input, can solve these problems by transforming unstructured data into vector representations through methods such as pixelization, singular value decomposition (SVD), bit vectorization, and word embedding, which aim to preserve the human perception of the unstructured data content as much as possible. After this precomputation process, the vectors are obfuscated in appropriate privacy models, followed by the projection of the obfuscated vectors back into the unstructured data space.
For \textit{images}, for example, some simple box-blurring algorithms introduced in the privacy-preserving Google Street View can protect human faces and license plates~\cite{frome2009large}, or a GAN -based method can pinpoint head regions while protecting their naturalness~\cite{sun2018natural}.
To preserve the perceptual quality of noisy images, the model proposed in~\cite{li2021differentially} provides DP guarantees for the latent vector representation of images. In particular, privacy budgets are assigned to the different elements in the latent space according to their weights, while the sensitivity is set to be within the maximum observed bounds. Laplace noise is then derived and added to the images in the semantic space before using GAN to synthesize realistic-looking faces.

In \textit{video} content, IBM took a pioneering role in the systematic study of privacy-preserving video surveillance with their PeopleVision system~\cite{senior2005enabling}. To protect privacy in video content, sanitization schemes, such as blurring and pixelization, were applied to regions of interest (ROIs) in individual images. With increasing sophistication in privacy requirements, sensitive information, including activities and places in videos, also came under the purview of {DP} methods~\cite{saini2014w}. The model proposed in~\cite{wang2020videodp}, applies DP to the video privacy domain by representing video data as pixels with the content of 3-dimensional vectors in the RGB color model, where adjacent inputs are defined as pairs of videos that differ in all visual elements (e.g., people) incomplete videos. Pixels are randomly sampled using DP before being classified into different RGB categories. Subsequently, pixels in the most frequent RGB categories are separately obfuscated using an appropriate Laplacian noise scale. The remaining unselected pixels are interpolated as a DP post-processing operation to reconstruct the visual elements in the video, thereby preserving privacy.

In the case of \textit{text} content, privacy is of utmost importance, especially when it involves sensitive information such as health conditions, locations, and personal lifestyle preferences. {DP} offers a valuable solution in this context. Textual data can be transformed into high-dimensional vectors using word embedding models such as GloVe, FastText, and word2vec. These vectors are then perturbed with noise vectors sampled from probabilistic distributions. The resulting noisy real-numbered vectors are projected back to their closest words in the embedding model through a discretization operation, considering the discrete nature of textual data. The pioneering solution presented in~\cite{weggenmann2018syntf} applies Hamming distance-based privacy (a traditional DP method) in the text privacy domain. It primarily focuses on concealing the term-frequency-related distributions in the bag-of-words embeddings. This is achieved by employing the exponential mechanism, which randomizes discrete terms into sets of candidate terms with associated probabilities. These probabilities are determined by a similarity metric between input and potential output terms. The similarity function is constructed as a composition of cosine similarity and Bigram overlap variables.
Other methods, such as Earth Mover’s Privacy~\cite{fernandes2019generalised}, Euclidean Privacy~\cite{feyisetan2020privacy}, and Hyperbolic Privacy~\cite{feyisetan2019leveraging}, can further enhance utility in differentially private vector embedding, leading to better privacy preservation mechanisms for text content.

\major{Currently, a rich set of work has introduced utility-enhanced mechanisms for structured data, exploring various privacy requirements for data with varying sensitivity levels~\cite{zhao2022survey}. However, future work is still needed for new rich data generated by IIoT, such as VR, gaze maps on eye tracking heatmaps, and temporal-related eye movement features. For example, while DP guarantees that attackers cannot obtain additional information by including or excluding an element, it cannot prevent attackers from gaining knowledge with publicly released data itself. Although other privacy protection methods, such as blurring of image data, have been shown to be vulnerable to DL attacks~\cite{hill2016effectiveness},
it is necessary to investigate the privacy guarantees of DP in mitigating DL-based re-identification attacks theoretically or practically. Inference attacks pose significant privacy risks to DP-protected data~\cite{hamm2017minimax}, and it may be of importance to integrate some other methods to enhance privacy in the design of DP privacy methods.}

To this end, DP finds profitable applications even in the context of \textit{FL}, a type of distributed machine learning that aims to protect clients' private data from exposure to adversaries. Since recent articles have pointed out how private information can still be divulged by analyzing client-uploaded parameters in FL, for example, weights trained in deep neural networks~\cite{mcmahan2017learning}, DP can be applied, for example, by adding artificial noise to the parameters on the client side before aggregation~\cite{wei2020federated}. In their proposed solution \texttt{NbAFL}, they quantified the trade-off between convergence performance and privacy level. Improved convergence performance results in a lower protection level, while increasing the number of clients participating in FL enhances convergence performance at a fixed privacy level. The authors also investigate the optimal number of aggregation rounds for a given protection level, which they demonstrate to exist.
\major{DP-based noise can be incorporated into the model input, gradients, and loss functions~\cite{yang2023survey, ye2022feature}, where in~\cite{ye2022feature} the authors suggest a protocol to introduce Gaussian-based noise to the output of each base model. However, their defensive tactics are limited to safeguarding categorical features.}
The authors of~\cite{li2019differentially} applied locally differential private algorithms for meta-learning, providing provable learning guarantees in convex settings. Meta-learning techniques leverage the shared knowledge gained from individual learning tasks to facilitate learning similar unseen tasks and can be applied to FL with personalization. Furthermore, DP combined with model compression algorithms can reduce communication overhead in FL while maintaining privacy benefits~\cite{agarwal2018cpsgd}.

In the context of blockchain systems, achieving client-level differential privacy can be accomplished by having each client locally add a certain amount of Gaussian noise after performing local gradient descent steps. Once the model is updated, it is submitted to the blockchain. To also protect the model from the public, the aggregated global model stored on the blockchain can be encrypted using a decryption key that is exclusively possessed by participating clients.
To enable FL with this approach, the work in~\cite{qu2020decentralized} proposed storing only the pointer of the global updates on the blockchain, while using a distributed hash table to store the actual data. This strategy ensures efficient block generation, allowing for decentralized privacy protection while preventing single-point failures.

\subsubsection{Encryption and Homomorphic Encryption}

It is one of the most important privacy protection tools, as it provides the most secure protection for point-to-point data transmission. The advantage of encryption is that it provides the highest data benefit among all privacy mechanisms because it can output the original data.
However, this technique presents some disadvantages. First, it is not easy to implement appropriate access control. Second, the original data are fully disclosed if the ciphers become decrypted. Third, efficiency. Since huge amounts of data are generated and transmitted every second in the IIoT systems and attackers can launch eavesdropping attacks during the data transmission process, traditional cryptography methods cannot efficiently solve this problem due to the large scale of the IIoT. To this end, lightweight encryption methods are proposed to partially solve the efficiency problem and to ensure secure communication and data transmission among multiple and cross-layer (cascade) networks.
Similarly, lightweight ABE can be used to provide evidence against attacks, especially collusion attacks. Finally, lightweight searchable encryption can be used to solve problems with query-based searchable content.

Current privacy-enhancing technologies for RFID technologies include limiting the distance between the reader and deactivating tags, minimalist cryptography, tag renaming and deactivation, access control, and re-encryption. It is suggested that the readers perform cryptographic calculations and store the results in tags using minimal cryptography. To prevent eavesdroppers from receiving different encrypted tag signals at different times, the reader can re-encrypt the tag with a different key and write it into its memory~\cite{gubbi2013internet}.

However, most IIoT devices are designed to be lightweight and small in size, and encryption methods must also be lightweight. In~\cite{kotamsetty2016adaptive}, the authors proposed a lightweight IBE scheme for smart homes, where public keys are just identity strings and no digital certificate is required. This approach is called the stateful IBE scheme. This solution separates the encryption process into key encryption and data encryption, emphasizing the latter because key encryption produces ciphertexts that are larger than those produced by data encryption.

In recent years, significant research efforts have been dedicated to {HE algorithms}, a type of encryption that enables computations to be performed on encrypted data without the need for prior decryption~\cite{beunardeau2016fully, natarajan2021seal, li2021lightweight, gilad2016cryptonets}.  HE allows computation over encrypted data, including multiplication (mul), addition (add), and constant-multiplication (cmul), and the decrypted result corresponds to the outcome of the operations as if they were performed on plaintext. However, although this is extremely important in IIoT, the computationally expensive traditional data encryption way (in terms of time and resources) cannot meet the privacy needs in IIoT environments.

The authors of~\cite{lu2018new} introduced a communication-efficient secure query scheme in a fog environment to ensure privacy for both the data user (e.g., application) and the data owner (e.g., an IoT device) using HE. Similarly,~\cite{li2021lightweight} presented a lightweight privacy protection protocol for data owners, third-party cloud servers, and data users. The solution is based on labeled HE, where each piece of encrypted data is associated with a unique label that is shared. This approach reduces the computational cost, eliminating the need for data users (applications) to perform expensive evaluations of the resulting ciphertexts.

An efficient privacy-preserving data aggregation method for FL is EPPDA~\cite{song2022eppda}, which can covertly aggregate user-trained models without revealing the user model and relies on secret sharing to resist reverse attacks. Utilizing homomorphisms for secret sharing protects user privacy, consumes less computational and communication resources than practical alternatives, and provides tolerable fault tolerance in case the user disconnects.
Specifically designed for healthcare data, PRCL~\cite{hao2020privacy} is a resource-efficient and privacy-conscious protocol for collaborative learning. The model, a neural network, was divided into three parts in PRLC. The first and last parts are trained on the client side, while the middle (and heavier) part is outsourced to be trained on cloud servers. To effectively perform gradient aggregation in the ciphertext context, the training data is perturbed by adding Gaussian noise before the model is trained on the client side. Additionally, packets are secured using HE. The simulation results demonstrate that PRCL (the proposed method) achieves precision comparable to other state-of-the-art techniques while reducing local training overhead by offloading the intermediate computation to a cloud server. The confidentiality of the data is preserved, as an attacker can only access the perturbed data due to the added noise, while the plaintext gradient remains inaccessible to the cloud.

\subsubsection{Secure Multi-party Computation}

Secure Multi-Party Computation (SMPC) is a generic cryptographic primitive that allows distributed parties to jointly compute arbitrary functionality without disclosing their confidential inputs and outputs~\cite{zhao2019secure}. SMPC is known to be thousands of times faster than fully implemented {HE} in typical applications. Moreover, different from differential privacy, which focuses on privacy guarantees for the entire constructed model, SMPC focuses on the privacy of intermediate steps in the computation.
SMPC addresses the problem of cooperatively computing the private data of multiple participants securely in a distributed IIoT computing environment. In the SMPC scenario, two or more parties that have private inputs want to use those inputs to jointly compute some functionality. Each participant in this task must achieve its own unique goal and nothing else to ensure security.
This is achieved by dividing the computation into multiple subtasks and distributing them among the parties who perform their computations locally on their data and share only the results of these computations, in a way that preserves privacy.
``Functionality' is a broad term that can refer to almost any cryptographic task, including encryption, authentication, zero-knowledge proofs, commitment schemes, oblivious transfer, and other protocols that are not cryptographic (i.e., application-oriented tasks include contract signing, electronic voting, machine learning, genomic data processing, and so on). In the field of cryptography, it can be said that SMPC is the most fundamental and all-encompassing theoretical research topic.
Any cryptographic task that involves multiple parties can be considered an SMPC task.

MiniONN~\cite{liu2017oblivious} and DeepSecure~\cite{rouhani2018deepsecure} for example, employ SMPC and homographic encryption and work with the existing pre-trained DNN models without changing the DNN training or the DNN structure. Although there are several techniques for creating generic SMPC protocols, they are currently very computationally intensive, making SMPC still impractical. To achieve a more computationally efficient SMPC, recent work has focused on developing secure techniques to offload the most expensive parts of computational tasks to the cloud. Rather than simply treating the cloud as a trusted party, these outsourced protocols seek to take advantage of the cloud for computation without exposing any input or output values. Cloud-assisted SMPC provides a unique but effective way to make general SMPC protocols usable and scalable by allowing parties to securely outsource their computations to a cloud provider, improving the runtime efficiency of SMPC protocols~\cite{kamara2012salus}.

The combination of cloud computing and HE leads to the development of cloud-assisted {SMPC} protocols~\cite{mukherjee2016two, peter2013efficiently}. In these protocols, participants engage in a computational task, encrypting their data using a full HE scheme and uploading the resulting ciphertexts to the cloud. The cloud then performs computations on these ciphertexts and returns the resulting ciphertexts. The level of privacy achieved depends on the security of the underlying HE process, and no participant should directly possess the secret key to the HE procedure. The challenge therefore lies in decrypting the result of the computation. The study presented in~\cite{asharov2012multiparty} addresses this problem by distributing the secret key among all participants. In this cloud-assisted SMPC protocol, all participants must generate the system parameters, including the secret keys, public keys, and evaluation keys, for each computation.

It can be observed that if applied to the context of ML applications, is very similar to FL, as both can be used to protect privacy. However, they have different methods to achieve this. FL is a more specific collaborative learning technique where each party trains the model on their local data and only sends updates to a central server. SMPC, on the other hand, divides the computation into multiple sub-tasks and distributes them among the parties, who perform their computations locally on their data and share only the results of these computations with each other in a way that preserves privacy. In~\cite{carter2016secure}, the focus was on the evaluation of secure functions for power-limited devices, such as mobile phones. They introduced a concept called ``outsourced oblivious transfer'' that allows mobile devices to delegate the task of garbled key transfer to a cloud server.

SMPC can be used to protect both data and trained models in a two-party environment. In this scenario, a cloud server (referred to as Bob) owns a dataset and trains a model using a specific ML algorithm. For proprietary reasons, Bob does not want to share this model with other companies. On the other hand, a user (Alice) is looking for personalized services based on her own data while ensuring that her private data remain confidential. This situation lends itself well to the application of a classic SMPC protocol: Alice enters her private data, and Bob enters the model. After the protocol is executed, Bob learns nothing about Alice's data, and Alice learns as little as possible about Bob's model~\cite{bost2014machine}.

However, in this field, the adoption of secure SMPC is hampered by the absence of flexible software frameworks \say{speak the language} from machine learning researchers and engineers. \texttt{CrypTen}~\cite{knott2021crypten} is a recent solution to foster the adoption of secure SMPC in ML. In particular, \texttt{CrypTen} exposes popular secure SMPC primitives via abstractions that are common in modern IIoT frameworks, such as tensor computations, automatic differentiation, and modular neural networks.

In an attempt to secure FL, it can be combined with SMPC, along with HE and DP, and can work as follows: After locally training a model on their data, users send the weights (parameters) of their model to the server using an encryption scheme that allows the server to perform computations on the encrypted data; in such a way, the server can compute a weighted average of all the encrypted weights received from users, but cannot discover the original weights for any user. In solutions that used \say{only} DP, the server would know the noisy private weights of each user. As proposed in~\cite{mugunthan2019smpai}, which combines SMPC and DP, the noisy weights sent to the server are also encrypted so that the server can only calculate the result and cannot infer anything about even the noisy weights of any particular user. By doing so, the system becomes fully private. Other approaches combining SMPC and DP, e.g.,~\cite{jayaraman2018distributed, bindschaedler2017achieving}, propose that each client chooses his/her local differential private noise. Unlike other sources, in~\cite{mugunthan2019smpai} the differentially private noise for each party is unknown to the party and is generated in a distributed manner from other parties. In particular, the combination of SMPC and DP has also found applicability in finance, with guarantees of privacy and acceptable precision~\cite{byrd2020differentially}.

\texttt{Chameleon}~\cite{riazi2018chameleon} and \texttt{Gazelle}~\cite{juvekar2018gazelle} are two works that attempt to choose between the above-mentioned cryptographic techniques (homomorphic encryption and SMPC) based on their computation and communication tradeoffs. Specifically, \texttt{Gazelle} carefully examines the tradeoffs between HE (characterized by high computation and low communication) and SMPC (involving low computation and high communication) and selects the suitable technique for each scenario accordingly. Additionally, \texttt{Gazelle} accelerates the training process by efficiently implementing cryptographic primitives, resulting in low runtime latency and communication costs, as evident from their research outcomes.

To improve scalability and reduce the communication overhead introduced by SMPC, i.e.,  all parties generate and exchange secret shares of private data with all other parties, the work in~\cite{kanagavelu2020two} proposed a two-phase SMPC-enabled FL framework. Instead of having every member generate and exchange secret shares of data across all members' lists, it proceeds to elect a subset of FL members as the model aggregation committee members out of the whole membership list. The elected committee uses SMPC to aggregate the local models of all FL parties. The 2-phase SMPC introduces a hierarchical structure, i.e., there is a need for fewer secret shares to be exchanged for privacy-preserving model aggregation.
In~\cite{sotthiwat2021partially}, instead of applying SMPC over the entire local models for secure aggregation, it is presented a partially encrypted SMPC solution by encrypting critical parts of model parameters (gradients) that are vulnerable to privacy-preserving attacks. In particular, only the first layer of local models is encrypted with the SMPC strategy, while the rest are sent directly to the centralized server. This approach can significantly reduce the extra computation and communication overhead of SMPC while inheriting SMPC benefits in both privacy-preserving and model accuracy perspectives.

\subsubsection{Machine Learning-based Methods}

In IIoT scenarios, both the complexity and volume of data are growing. The traditional security measures described previously can not cover all cases, and it becomes harder for common users and even data curators to comprehend the risk, choose the right schema, and manage their privacy.
ML and data mining methods can be utilized in both attack and defense of privacy preservation~\cite{mohassel2017secureml}. 
In terms of defense, such methods are widely employed in Cyber-physical social systems (CPSSs) to extract features of specific objects or people in social media and provide long-term privacy protection.
Instead, in terms of attacks, ML seems more applicable due to its ability to learn, e.g., to extract features of individuals and make predictions, which is not only harmful at present but also causes damage over time.

\begin{figure*}[t!]
    \centering
    \includegraphics[width=.98\linewidth,keepaspectratio]{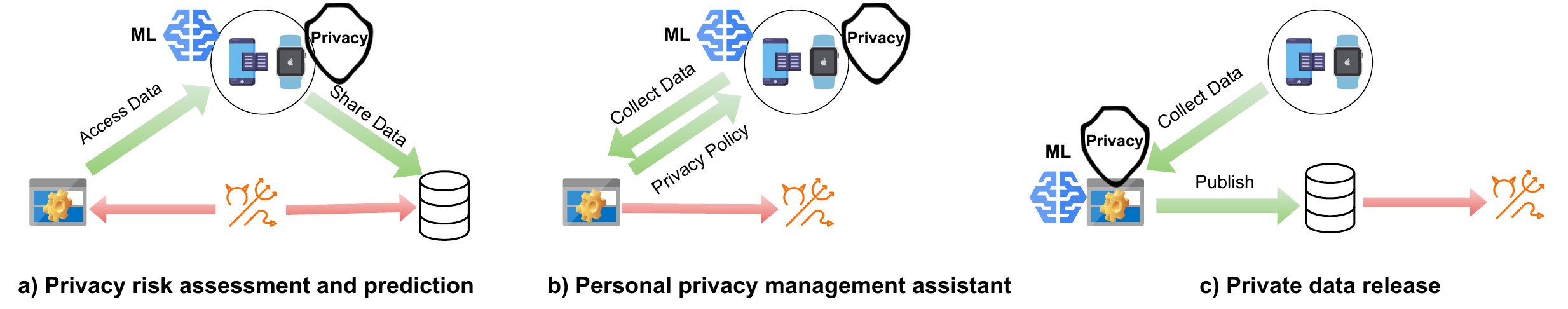}
    \caption{ML-based privacy protection schemes.}
    \label{fig:mlForPrivacy}
\end{figure*}

Regarding defense, ML has been introduced to enhance privacy protection during the past few years, where the proposed efforts include several aspects (as shown in Fig.~\ref{fig:mlForPrivacy}):
\begin{itemize}
    \item \textit{Privacy Risk Assessment and Prediction}. The objective is to detect and anticipate the user's privacy risk during the ``access'' and ``sharing'' processes. As shown in  Fig.~\ref{fig:mlForPrivacy}a, ML is leveraged to evaluate both the input and output data streams, enabling the identification of potential risks that should be considered to implement a proper privacy protection scheme.
    \item \textit{Personal Privacy Management Assistant}. This includes managing and evaluating user preferences and privacy policies, as shown in Fig.~\ref{fig:mlForPrivacy}b.
    \item \textit{Private Data Release}. Publish datasets with a guarantee of privacy, a practice commonly performed by data curators rather than individual users (Fig.~\ref{fig:mlForPrivacy}c).
\end{itemize}

In the \textit{ privacy risk assessment and prediction} methods, ML can help prevent the loss of sensitive information when the user only accesses the application (passively collected information by malicious attackers) or shares it on social networks (actively shared information). For example, in predicting privacy risks on websites and applications, a browser extension has been proposed in~\cite{sebastiani2002machine}. This extension collects information on the websites visited by users and provides feedback to users about the privacy quality of the website based on ML.
To detect and identify websites that may be malicious and risk users' privacy, a Bayesian classifier-based method has been presented in~\cite{manek2016detection}.
The suggested method analyzes online reviews to determine whether they are trustworthy or not.
The solution in~\cite{fu2019keeping} rates the privacy risks associated with applications using an SVM classifier. The results show that over 90\% accuracy can be achieved in identifying privacy risks. The use of ML to analyze the privacy risks posed by mobile applications has been taken into consideration in~\cite{avdiienko2015mining}.
Modern ML techniques can assist in identifying sensitive information when sharing data, enabling users to protect their personal information effectively. In~\cite{squicciarini2017toward}, an ML model is proposed, capable of classifying photos based on the visual content features and the metadata of the images, evaluating the sensitivity level, and making decisions considering the past choices of users.
\texttt{iPrivacy}~\cite{yu2016iprivacy} uses ML to automate this process. It identifies privacy-sensitive objects in images and categorizes them accordingly. According to the classification, \texttt{iPrivacy} alerts users to objects that should be suppressed or masked before sharing due to privacy concerns. In addition, the tool provides privacy setting recommendations based on user preferences and shared images.
In~\cite{hasan2020automatically}, an ML method is proposed to automatically identify bystanders using only the visual information available in an image.

Methods involving a \textit{Personal Privacy Management Assistant} aim to assist users in effectively managing and customizing their privacy preferences in an automated way. ML methods for privacy management can be broadly categorized into two groups: (i) privacy policy assessment and (ii) user preference prediction and management.
In the case of software and web applications, vendor privacy policies are often written using complex jargon, leading many readers to accept these policies without fully comprehending their implications. Methods falling under the first category aid users in making informed decisions about their privacy choices. 
The authors of~\cite{lebanoff2018automatic} explored the automatic detection of vague content within privacy policies. by exploiting GANs to assess and characterize the vagueness of sentences.
Methods in the second class take into account that each user has a different sensitivity and preference for privacy. 
Early studies have found that users' privacy preferences are related to some statistical and environmental parameters, e.g., the context of applications~\cite{wijesekera2015android, olejnik2017smarper}. ML can predict user privacy preferences and make privacy management decisions, as in~\cite{das2018personalized} or in the Visual Privacy Advisor~\cite{orekondy2017towards} method.

\textit{Private data release} is an important process in IIoT applications, e.g., health data from medical centers. A commonly used mechanism for the release of private data is obfuscation through the addition of noise to the original dataset, such as via ML and GANs~\cite{acs2018differentially, denton2015deep}. 
The main idea behind these approaches is that the data curator trains a deep generative model with the original data in a differentially private manner and then publishes the synthetic dataset generated by the model. The data curator may also decide to publish the deep-generative model, which may lead to an unlimited amount of synthetic data. Thus, data analysis can be performed on synthetic data instead of the real data of individuals to maintain privacy.
An alternative approach to generate synthetic patient data based on GANs and autoencoders is proposed in~\cite{choi2017generating}. The performance of the proposed generative model is also evaluated by comparing the synthetic patient records it generates with real data. In~\cite{triastcyn2018generating}, GANs are used to generate artificial data that retain the statistical properties of the real data while reducing the risk of information disclosure.
In~\cite{park2018data}, statistical similarity was demonstrated between the synthetic table data generated and the original data.
Finally, we can observe that this approach is not associated with the drawbacks associated with other traditional anonymization methods, such as the risk of background knowledge or the potential linkage of data to other external sources.

At the same time, ML can be used as an attack tool in addition to being a privacy protection tool, where the main ML attack models include \textit{re-identification attacks} and \textit{inference attacks}. In \textit{re-identification attacks}, the adversary re-identifies a certain individual using images, videos, or geolocation information. Recent advances in DL-based models have made face recognition techniques accurate~\cite{joon2015person, sun2018face}. In addition, simple schemes based on obfuscation no longer work effectively~\cite{mcpherson2016defeating, oh2016faceless}.

In \textit{inference attacks}, the adversary utilizes ML classifiers to analyze publicly available data and deduce user private attributes, such as place of residence, occupation, hobbies, and political views~\cite{cheng2010you}. Other solutions have detected the age of people~\cite{rodriguez2017age} and car license plates~\cite{zhou2012principal} from ordinary or even obfuscated images.
To defend against these attacks, the idea is to exploit the weaknesses and limitations of ML methods through adversarial machine learning. These methods share the approach of providing the ML attacking models with some well-designed inputs called \say{adversarial examples}~\cite{sharif2016accessorize}. Preliminary work in~\cite{szegedy2014intriguing} discovered that overlaying the original image with imperceptible noise would lead DNNs to misclassify it. The main reason why neural networks are prone to negative examples is related to the linear nature of neural networks. In~\cite{papernot2016limitations}, the space between adversaries and DNNs itself comes from ML techniques.
Adversarial examples demonstrate transferability, which means that if they can fool one model, they often possess the ability to deceive another model with different parameters and architectures~\cite{szegedy2014intriguing}.
Based on this idea of adversarial examples, the work in~\cite{liu2017protecting} has proposed an algorithm based on a faster RCNN framework that fools automatic detection in images. For medical text, the solution in~\cite{friedrich2019adversarial} generates a privacy-preserving shareable representation for a de-identification classifier. The work in~\cite{liu2019adversaries} employs adversarial examples in ML systems to prevent the identification of sensitive information from images, while~\cite{li2019anonymousnet} enables de-identification of faces by using adversarial perturbations. Recent work showed that by changing only one pixel, it was possible to fool the DL algorithms~\cite{su2019one}. The perturbation exploits differential evolution (DE) and, because of the inherent features of DE, can fool more types of DL models. Moreover, the authors of~\cite{aouedi2023f} proposed a Federated Blending model, called \texttt{F-BIDS}, in order to further protect the privacy of existing ML-based IDS. In contrast to the classical FL approaches, the federated meta-classifier is trained on the meta-data (composite data) instead of user-sensitive data to further enhance privacy.

\subsection{Lessons Learned}
\label{subsec:lessons_privacy}

From the state-of-the-art methods presented to protect user privacy in the IIoT, we can see that many attacks are possible in IIoT applications. Given the specificity of the scenario, {WSNs} pose critical challenges, the solutions of which are also specific and detailed in Section~\ref{subsec:privacy-location}. Therefore, attention must be paid when dealing with sensitive information to preserve user privacy. FL is an initial attempt to preserve data privacy, but it is unsatisfactory. To this end, this section presents a multitude of approaches that can be used to prevent personal data leakage and that can be combined with FL. Recent research has proposed lightweight encryption, for example, via differential privacy or secure multi-party computation, and ML-assisted privacy protection to protect both models and data. For instance, a generative neural network to create artificial datasets is a viable approach that opens up a new area of study for privacy protection, particularly for unstructured data such as images and videos. Of equal importance is the study of defense mechanisms against ML-based privacy attacks. Although this area is still in its early stages, it is expected to fly in the future, given the widespread use of AI techniques in next-generation networks. The current standard technique in this field is adversarial examples/perturbations. In conclusion, we believe that the study of attack techniques will facilitate the development of more sophisticated defensive mechanisms.

\section{Challenges and Future Directions}
\label{sec:challenges}

As discussed above, IIoT networks play an increasingly significant role in our lives in the presence of many applications. Despite its advantages, IIoT network development still faces some critical problems related to security and privacy. Furthermore, it also faces some research challenges in terms of resource management, fairness issues, economic issues, and IIoT in future wireless networks that need to be considered. Here, we further analyze and provide several possible solutions to these challenges.

\subsection{Resource Management in IIoT}
The convergence of future IIoT networks will face challenges to not only IoT network deployments but also ML models. From this perspective, we find that the following issues need to be carefully considered.

\textit{Heterogeneity}: There is a wide range of IoT systems (e.g., WSN, smart homes, industrial factories, remote healthcare, intelligent transport, smart city, smart grid, smart agriculture, etc.) and each consists of distinct IoT components with different processing capabilities, storage capacities, and power requirements. Therefore, it is necessary to standardize the resource management protocol in such networks to meet various requirements. In response to that, several protocols are now available for IoT devices, such as IPv6 over low-power wireless personal area networks, REST environments restricted by the IETF, restricted application protocol, and protocol mappings~\cite{Zahoor2021Oct}. However, it is a fact that these current protocols are independent of each other and add complexity to resource-constrained IoT devices. Thus, new resource management solutions should be able to ensure interoperability with different devices and protocols.

\textit{Scalability}: IIoT networks are projected to grow significantly in the coming years, which drives not only the massive proliferation of intelligent IoT devices/users/applications, but also large amounts of non-IID data. In order to handle these issues, the introduction of new types of integrated learning frameworks must have scalable features in response to different deployment scenarios. FedLab, a lightweight open-source framework based on FL algorithm effectiveness and communication efficiency, was proposed in~\cite{Zeng2021Jul} with the ability to add functional modules, such as unsupervised learning, semi-supervised learning, transfer learning, etc.

\textit{Dynamic Nature}: IoT networks are dynamic since apparatuses join and leave the network frequently. Therefore, management solutions must be able to adapt to these changes and ensure the efficient use of all resources. Possible is to use token-based authentication~\cite{Manogaran2023Mar, Cui2023Feb}.

\textit{Security}: IoT devices are frequently susceptible to attacks, and thus, current efforts are focused on deploying learning models on the device to perform multiple tasks: classification, detection, and data encryption. As IoT devices have limited resource capabilities, the use of these solutions can lead to data synchronization delays and high power consumption during learning/training tasks. To address these issues, optimizing/developing and realizing cooperative learning models and/or incorporating BCT becomes a possible solution~\cite{Shen2019Feb}. Adopting one of the four cryptographic techniques applied in ISO/IEC 29192~\cite{Hasan2022Nov} combined with a lightweight learning model during data encryption~\cite{Kong2022Mar} is beneficial to enhance user information security and privacy during the aggregation process of encryption and authentication attributes into IoT data. Furthermore, dividing IoT data into hard and homomorphic verifiable tags and verifying them through blockchain transactions is also essential to prevent the cloud event of hardware or software damage. 

\subsection{Learning Model Design in IIoT} 
Exploiting secure learning frameworks for IoT networks is a pivotal mission. However, the participation of different applications and network scales leads to different requirements in designing the complexity and memory of the learning models. Modern ML methods, such as DNNs, demand advanced hardware solutions such as high-power graphic processing units and tensor process units, particularly when dealing with high-dimensional data, large-size inputs, and low-latency constraints. Without these hardware solutions, achieving optimal performance and accuracy in ML tasks can be challenging. Thus, building lightweight learning models is a critical task for future IIoTs. 

To hit this goal, we recognize that three following factors should be taken into account when deploying learning models for future IIoT networks:
\begin{itemize}
    \item \textit{Model size}: This refers to the amount of memory or storage used in terms of bytes, kilobytes, or megabytes. The size of the model determines the load time, transmission cost, and energy consumption of the model needed to complete the learning process. Thus, reducing the size of the model leads to reduced memory or storage costs while improving the efficiency and portability of the model.

    \item \textit{Model complexity}: This factor showcases the amount of computation or arithmetic operations that the model performs, such as floating-point operations, multiply-accumulate operations, or tera operations per second. Thus, a lower model complexity can reduce the computational burden and improve the speed and performance of the model.

    \item \textit{Model accuracy}: This factor is the cornerstone of a model's success in completing a given task or dataset while determining its confidence and usability. Model accuracy can be evaluated using metrics such as accuracy, precision, recall, F1-score, or mean average precision.    
\end{itemize}
It is evident that there is a trade-off between the factors mentioned above. While increasing the model size can help the learning model execute faster, it also leads to higher computational costs. Similarly, to attain high accuracy in the learning model, one needs both a larger model size and a complex learning algorithm. Until now, there has been no universal standard for so-called lightweight learning models. Nevertheless, there are some possible criteria to reduce the model size and complexity for lightweight ML models.
\begin{enumerate}
    \item Using the quantization technique can reduce the precision or bit-width of the model parameters, such as weights and activations, from floating-point numbers to integers or binary numbers. This yields a smaller usage of the memory and computation requirements for the model, as well as the data transmission cost \cite{Wei2022Oct}. 

    \item Using the pruning technique can eliminate redundant or insignificant parameters of a model, such as weights or neurons, that are of low importance and contribute to the overall output of the model. This can significantly reduce the size and complexity of the model, and overfit \cite{Hu2021Oct}.

    \item Using a compression technique, such as coding, clustering, or hashing, can reduce the model size or complexity by exploiting the redundancy or similarity in the model parameters. This can reduce the memory and storage overhead, as well as the cost of data transmission \cite{Ko2021Mar}.

    \item The use of a distillation technique helps transfer the knowledge or functionality of a large or complex model (teacher) to a small or simple model (student), by training the student model to mimic the output or behavior of the teacher model. This reduces the size and complexity of the model while preserving the accuracy or performance of the model \cite{Zhu2023Aug}.

   \item Using the hyper-parameter tuning optimization approaches enables us to obtain the best possible model architecture for a given task or dataset, including the ideal number of layers, neurons, or connections that can achieve the optimal balance between model size, complexity, and accuracy. This approach can reduce the size and complexity of the model while simultaneously increasing its performance and efficiency \cite{Wang2024Mar}.
\end{enumerate}
\color{black}

\subsection{Fairness in IIoT}
The interplay between large-scale IoT, cloud demand, emerging learning solutions, and BCT has also raised another challenge related to fairness issues.

First, ensuring that each IoT device, application, or user receives a fair portion of communication resources (e.g., computing capacity, storage, energy, and bandwidth) is a critical task but challenging due to numerous factors (i.e., infrastructure, communication protocols, goals, network restrictions, and management policies)~\cite{Nguyen2021Apr}. To address this problem, a possible solution is to distribute decision-making resources to the edge network while considering the use of the FL method and decentralized BCT~\cite{Yang2022Aug}. 

Second, the collection of inefficient IoT data can lead to uncertain predictions, faulty decisions, or biases. Especially in e-health applications, even small gender-imbalanced patient datasets can result in misdiagnosis or improper treatment suggestions. Therefore, the development of a fairness-aware learning process is a critical task and there are some possible solutions as follows:
\begin{itemize}
    \item \textit{Data augmentation}: Generate new data from existing data by employing transformations (i.e., rotation, scaling, or flipping) to expand the size of the dataset as well as their diversity, potentially improving its balance~\cite{Pastaltzidis}.

    \item \textit{Data weighting}: Each dataset requires the assignment of weights to distinct samples in order to balance the spread of characteristics, features, and classes by using techniques such as oversampling or undersampling~\cite{Lo2022Jan}.

    \item \textit{Transfer learning}: To narrow the consequences of data imbalance on the learning process, knowledge models must be transferred from pre-trained models to new ones, such as learning from a smaller and more balanced dataset~\cite{Patel2022Apr}.

    \item \textit{Fairness metrics}: Use metrics such as equal opportunity or disparate impact to measure the fairness of the model predictions, thus guiding the selection of appropriate learning techniques~\cite{Li2022Mar}.
\end{itemize}

Thirdly, the consensus blockchain mechanism used to solve the trust and security issues in IIoT data sharing often creates a significant disparity in computing power as well as the diverse requirements of IIoT applications, posing two fairness factors:
\begin{itemize}
    \item \textit{Mining}: IoT nodes/devices may not have an equal opportunity to announce a new block compared to edge/cloud nodes due to limited resources and computing ability. Additionally, selfish mining attacks from edge/cloud nodes could result in unfairness in the mining process.

    \item \textit{Transaction processing}: IIoT applications may need to ensure fair transaction processing, which involves packing transactions in a desired order, such as first-in-first-out or prioritizing transactions with the highest fees.
\end{itemize}
To address these concerns, a potential solution is to adopt FruitChain's characteristics and perform a probability analysis of transaction processing effectiveness while using an Euclidean distance-based measure to assess fairness \cite{Li2022sept}.

Lastly, the centralized approach for data-intensive applications lacks marketing fairness for IoT users (who sell their data) and third-party buyers. To deal with such challenges, a possible solution in~\cite{Liu2022Feb} suggested three steps: 1) Establish a set of supervising nodes to form an efficient consortium blockchain that acts as a transparent and trusted \say{\textit{controller}} for cloud-based data marketing; 2) Manage anonymous logins for data owners using distributed and open-threshold release of credentials to ensure confidentiality and privacy; and 3) Provide financial incentives and succinct \say{\textit{commitments}} of reliable behavior based on the on/off-chain communication among the data owner, third party, and cloud server.

\begin{table*}[!th]
\centering
{\renewcommand\arraystretch{1}
\begin{tabular}{|l|p{6.75cm}|p{6.75cm}|} \hline
\textbf{Challenge in IIoT}&\multicolumn{1}{c|}{\textbf{Description}}&\multicolumn{1}{c|}{\textbf{Possible solution}}
\\ \hline\hline
Resource management& 
\begin{itemize}
    \item \textit{Heterogeneity}: The diversity of IIoT systems and their corresponding devices, users, and applications.

    \item \textit{Scalability}: The new types of integrated learning framework must meet the diverse deployment scenarios.

    \item \textit{Dynamic Nature}: IoT apparatuses frequently join and leave the network frequently.

    \item \textit{Security}: IoT devices with limited resource capacity.
\end{itemize}
&
\begin{itemize}
    \item Adoption of lightweight protocols~\cite{Zahoor2021Oct}.

    \item Development of a lightweight open-source framework: FedLab~\cite{Zeng2021Jul}.

    \item Applying token-based authentication~\cite{Manogaran2023Mar,Cui2023Feb}.

    \item Optimizing/developing/realizing cooperative learning models and/or incorporating BCT~\cite{Shen2019Feb}; Adoption of cryptographic techniques in ISO/IEC 29192~\cite{Hasan2022Nov} combined with a lightweight learning model during data encryption~\cite{Kong2022Mar}.
\end{itemize}
\\ \hline
Learning Model Design & When designing learning models, three following factors should be taken into account:
\begin{itemize}
    \item \textit{Model size}: The amount of memory or storage usage.

    \item \textit{Model complexity}: The amount of usage of computation or arithmetic operations.

    \item \textit{Model complexity}: The confidence and usability of the output generated by the learning representatives.
\end{itemize}
 &
 
 \begin{itemize}
     \item Using the quantization technique to reduce the precision or bit width of the model parameters \cite{Wei2022Oct}.
      \item Using the pruning technique to eliminate redundant or insignificant parameters of a model \cite{Hu2021Oct}.

      \item Using a compression technique to reduce model size or complexity \cite{Ko2021Mar}.
        \item Using the distillation technique to transfer the knowledge or functionality of a large or complex model (teacher) to a small or simple model (student) \cite{Zhu2023Aug}.
      \item Using hyperparameter tuning optimization approaches to obtain the best possible model architecture for a given task or dataset \cite{Wang2024Mar}.
     
 \end{itemize}

\\ \hline
Fairness issues&
\begin{itemize}
    \item \textit{Resource allocation}: Ensure that each IoT device, application, or user receives equitable communication resources.

    \item \textit{Learning process}: Develop a fairness-sensitive learning process.

    \item \textit{Blockchain consensus}: Equal opportunity between IoT nodes/devices and edge/cloud nodes when releasing a new block and guaranteeing fair transaction processing per IIoT application.

    \item \textit{Data marketing}: Guarantee fair marketing of data-intensive applications between IoT users (who sell their data) and third-party buyers.
\end{itemize}

&
\begin{itemize}
    \item Distribute decision-making resources to the edge network while considering the use of the FL method and decentralized BCT~\cite{Yang2022Aug}.

    \item Adoption techniques for data augmentation~\cite{Pastaltzidis}, data weighting~\cite{Lo2022Jan}, transfer learning~\cite{Patel2022Apr}, and fairness metrics~\cite{Li2022Mar}.

    \item Adopt FruitChain's characteristics and perform a probability analysis of transaction processing effectiveness using a Euclidean distance-based measure to assess fairness~\cite{Li2022sept}.

    \item Apply three steps in~\cite{Liu2022Feb}: 1) establish a transparent and trusted \say{\textit{controller}} for cloud-based data marketing; 2) distribute and open-threshold release of credentials to ensure confidentiality and privacy; and 3) provide financial incentives and succinct \say{\textit{commitments}} of reliable behavior.
\end{itemize}
\\ \hline
Economic issues& \multicolumn{2}{p{13.5cm}|}{
\begin{itemize}
    \item \textit{Deploying and maintaining the infrastructure}: Costs related to hardware installation (e.g. sensors, gateways, and other devices), software, and connectivity, as well as network maintenance over a long period of time~\cite{BibEntry2022Dec}.
    \item \textit{Cost of security}: Additional infrastructure investment to reduce the risks of cyberattacks and data breaches, as well as the additional costs of learning, training, testing and operating new learning frameworks~\cite{xu2022influence}.

    \item \textit{Data ownership and privacy}: Complex IoT data generated by multiple entities require efficient management mechanisms~\cite{roman2013features}.

    \item \textit{Standards and interoperability}: The existence of many different devices and technologies requires a unified standard to ensure that these devices can communicate with each other effectively~\cite{Dahmen2023Apr}.

\end{itemize}
} \\ \hline
Future mobile networks& 
The interplay between the mobile network and Metaverse environments causes:
\begin{itemize}
    \item New privacy concerns related to the use of collected user data characteristics for personalized advertising.

    \item New hardware designs for accessing Metaverse environments.
\end{itemize}
& 
Adoption of Generative AI with
\begin{itemize}
    \item Ethereum blockchain-based deep recurrent neural networks~\cite{Rabieinejad2021Dec}.

    \item Transformer-based model~\cite{Ferrag2023Mar} or DL-based model~\cite{Ferrag2023Apr}.
\end{itemize}
\\ \hline
{Model drift} & \begin{itemize}
    \item {Data Variability: The drift of the model lies in the changes in the patterns and statistical properties of the underlying data over time.}
    \item {Adaptability: Predictive models must be able to adapt to new data patterns.}
\end{itemize}
& \begin{itemize}
    \item {Online learning algorithms, can continuously assimilate and adapt to new data and offer a robust solution to the mutable patterns observed within IIoT environments~\cite{huang2021accurate}.}
    \item {Ensemble learning helps to evolving data landscapes~\cite{krawczyk2017ensemble}.}
    \item {Machine unlearning enables the selective removal or "forgetting" of data that may no longer be relevant or that could introduce biases or vulnerabilities~\cite{qu2023learn}.}
\end{itemize}
\\ \hline
\end{tabular}}
\end{table*}

\subsection{Economic Issues in IIoT}
Another possible challenge is the economic issues in IIoT networks. The first issue is the cost associated with \say{\textit{deploying and maintaining the infrastructure}}. This includes the costs related to hardware installation (e.g. sensors, gateways and other devices), software, and connectivity~\cite{BibEntry2022Dec}. In addition, when the network evolves into a larger scale or complex environment, it also requires considerable expenditures to keep the network operational for an extended period of time. Additionally, automating many routine tasks by learning solutions may result in job displacement or replacement, which can have a negative impact on the workforce, especially in industries~\cite{Pradhan2023Feb}. The second issue is the \textit{cost of security}. The risk of cyber attacks and data breaches increases in tandem with an increasing number of connected devices~\cite{xu2022influence}. To provide a robust security mechanism, additional investment in infrastructure can increase the cost of learning, training, testing, and operating new learning frameworks. The third issue is \textit{data ownership and privacy}. The massive sharing of IoT data generated by the IIoT network among different entities increases the need for efficient management mechanisms to control data ownership and protect sensitive user information~\cite{roman2013features}. The fourth issue is the \textit{standards and interoperability}. As intelligent IoT networks involve many different devices and technologies, a unified standard~\cite{Dahmen2023Apr} to ensure that these devices can communicate with each other effectively; otherwise, the development and adoption of intelligent IoT networks could be slowed or hindered~\cite{Albouq2022Mar}. 

\subsection{IIoT in Future Networks}
IIoT is critical in enabling numerous intelligent services and applications in several vertical domains. However, continuous advances in AI techniques and IoT technologies require that IIoT be further investigated in future 6G networks, particularly IIoT integration with new technologies and computing paradigms in 6G systems. \say{Metaverse}, a livened picture of potential AI visions in wireless communication networks~\cite{Chengoden2023Feb}, has recently received considerable attention in standardizing the future and advanced mobile network of 5G~\cite{Huang2023Mar}. It drives the sustainable evolution of edge intelligence and infrastructure layers~\cite{Lim2022Jul} and presents environmentally friendly networking solutions within the Metaverse~\cite{Siyue2023Jan}. Nevertheless, all these works share a common aspect, which is the rising privacy concerns related to the utilization of collected user data characteristics for personalized advertising. These characteristics include facial features, eye movements, voice patterns, and other biometric information, along with real-time location and environmental data. In particular, new hardware designs for accessing Metaverse environments generate new security challenges, for example, extracting/tracing users' fingerprints to forge passwords. To cope with such problems, \say{Generative AI}, also known as AI-generated content, has recently been recognized as a particularly efficient approach to detect cyber threats in 6G-allowed IoT networks with the combination of transformer-based model~\cite{Ferrag2023Mar} and DL model~\cite{Ferrag2023Apr} as well as the Ethereum blockchain with RNN model~\cite{Rabieinejad2021Dec}. However, these current methods still consume a significant amount of time and energy to achieve an overall detection accuracy of around 95\%, indicating the effectiveness of Generative AI in real-world applications.

\subsection{{Model Drift in IIoT Networks}}
{The evolving and dynamic nature of the data streams within the IIoT ecosystem inherently produces unpredictable behaviors at various levels of urban equipment and devices. This unpredictability is mainly due to concept drift, a phenomenon that was first identified by Schlimmer et al. in 1986~\cite{schlimmer1986incremental}. Concept drift reflects the inevitable changes in the statistical distributions of IoT data streams, which can be triggered by natural processes like the gradual wear and tear of urban infrastructure, or by unnatural events such as equipment failures or security breaches. Such changes challenge the static nature of traditional predictive models, as the ground truth on which these models were initially trained gradually becomes misaligned with new data. This misalignment leads to a reduced statistical similarity between historical data streams and current observations, necessitating adjustments in the models' classification and clustering boundaries to accommodate the dynamic, non-static nature of IoT environments.}

{Addressing these challenges necessitates the adoption of advanced methodologies, such as online learning, ensemble learning, and adaptive ML techniques~\cite{lu2018learning}. Online learning algorithms, characterized by their ability to continuously assimilate and adapt to new data, offer a robust solution to the mutable patterns observed within IIoT environments~\cite{huang2021accurate}. These algorithms proactively adjust to emerging data trends, thus efficiently counteracting the effects of concept drift. Furthermore, ensemble learning strategies have been identified as particularly effective in managing concept drift~\cite{aouedi2022ensemble}. By dynamically modifying the composition of the ensemble in response to changing data landscapes, these methods improve the resilience of the model~\cite{krawczyk2017ensemble}. Moreover, adaptive ML techniques, also known as machine unlearning, enable the selective removal or "forgetting" of data that may no longer be relevant or that could introduce biases or vulnerabilities~\cite{qu2023learn}. In addition, if data-sharing agreements change, it enables network operators to selectively remove the influence of their data. Therefore, the machine unlearning approach not only helps maintain the model's accuracy and relevance in the face of non-stationary data but also enhances privacy and security by allowing for the dynamic adjustment of the model to exclude potentially compromised data. Together, these strategies underscore a forward-thinking approach to maintaining the accuracy and reliability of predictive models in the face of the inherent variability of the IIoT.}


\section{Conclusion}
\label{sec:conclusion}

In this article, we conducted a comprehensive survey on the convergence of IoT and AI through an extensive investigation of the applications, security, and privacy issues of IIoT. A number of important application domains have been presented and analyzed, including IIoT-based smart healthcare, IIoT-based smart grid, IIoT-based smart transportation, and IIoT-based smart industry. We have then discussed several crucial security issues existing in IIoT systems and applications, from network attacks and confidentiality to integrity and intrusion. In addition, privacy concerns have been considered and investigated in three main domains, namely, data privacy leakage, location privacy leakage, and model privacy leakage. The lessons learned from our holistic discussion have been also summarized and provided. Several interesting research challenges have been identified in IIoT, along with potential solutions to stimulate further research in the interesting area of IIoT. 

\section*{Acknowledgement}
This work was partially supported by the European Union Horizon-CL4-2021 Research and Innovation Program under Grant Agreement 101070181 (TALON). The paper reflects the authors’ views, and the Commission is not responsible for any use that may be made of the information it contains.



\begin{thebibliography}{100}
\providecommand{\url}[1]{#1}
\csname url@samestyle\endcsname
\providecommand{\newblock}{\relax}
\providecommand{\bibinfo}[2]{#2}
\providecommand{\BIBentrySTDinterwordspacing}{\spaceskip=0pt\relax}
\providecommand{\BIBentryALTinterwordstretchfactor}{4}
\providecommand{\BIBentryALTinterwordspacing}{\spaceskip=\fontdimen2\font plus
\BIBentryALTinterwordstretchfactor\fontdimen3\font minus \fontdimen4\font\relax}
\providecommand{\BIBforeignlanguage}[2]{{%
\expandafter\ifx\csname l@#1\endcsname\relax
\typeout{** WARNING: IEEEtran.bst: No hyphenation pattern has been}%
\typeout{** loaded for the language `#1'. Using the pattern for}%
\typeout{** the default language instead.}%
\else
\language=\csname l@#1\endcsname
\fi
#2}}
\providecommand{\BIBdecl}{\relax}
\BIBdecl

\bibitem{broring2022intelliot}
A.~Br{\"o}ring, V.~Kulkarni, A.~Zirkler, P.~Buschmann, K.~Fysarakis, S.~Mayer, B.~Soret, L.~D. Nguyen, P.~Popovski, S.~Samarakoon \emph{et~al.}, ``{IntellIoT: Intelligent IoT Environments},'' in \emph{Global IoT Summit}.\hskip 1em plus 0.5em minus 0.4em\relax Springer, 2022, pp. 55--68.

\bibitem{zhang2021intelligent}
C.~Zhang, ``Intelligent internet of things service based on artificial intelligence technology,'' in \emph{2021 IEEE 2nd international conference on big data, artificial intelligence and internet of things engineering (ICBAIE)}.\hskip 1em plus 0.5em minus 0.4em\relax IEEE, 2021, pp. 731--734.

\bibitem{zhou2019edge}
Z.~Zhou, X.~Chen, E.~Li, L.~Zeng, K.~Luo, and J.~Zhang, ``Edge intelligence: Paving the last mile of artificial intelligence with edge computing,'' \emph{Proceedings of the IEEE}, vol. 107, no.~8, pp. 1738--1762, 2019.

\bibitem{elbir2022federated}
A.~M. Elbir, B.~Soner, S.~{\c{C}}{\"o}leri, D.~G{\"u}nd{\"u}z, and M.~Bennis, ``Federated learning in vehicular networks,'' in \emph{2022 IEEE International Mediterranean Conference on Communications and Networking (MeditCom)}.\hskip 1em plus 0.5em minus 0.4em\relax IEEE, 2022, pp. 72--77.

\bibitem{nagaty2023iot}
K.~A. Nagaty, ``IoT commercial and industrial applications and AI-powered IoT,'' in \emph{Frontiers of Quality Electronic Design (QED) AI, IoT and Hardware Security}.\hskip 1em plus 0.5em minus 0.4em\relax Springer, 2023, pp. 465--500.

\bibitem{lin2017survey}
J.~Lin, W.~Yu, N.~Zhang, X.~Yang, H.~Zhang, and W.~Zhao, ``{A Survey on Internet Of Things: Architecture, Enabling Technologies, Security and Privacy, And Applications},'' \emph{IEEE Internet of Things Journal}, vol.~4, no.~5, pp. 1125--1142, 2017.

\bibitem{al2020survey}
M.~A. Al-Garadi, A.~Mohamed, A.~K. Al-Ali, X.~Du, I.~Ali, and M.~Guizani, ``{A Survey of Machine and Deep Learning Methods For Internet Of Things (IoT) Security},'' \emph{IEEE Communications Surveys \& Tutorials}, vol.~22, no.~3, pp. 1646--1685, 2020.

\bibitem{samie2019cloud}
F.~Samie, L.~Bauer, and J.~Henkel, ``{From Cloud Down to Things: an Overview of Machine Learning in Internet of Things},'' \emph{IEEE Internet of Things Journal}, vol.~6, no.~3, pp. 4921--4934, 2019.

\bibitem{amiri2020survey}
M.~Amiri-Zarandi, R.~A. Dara, and E.~Fraser, ``{A Survey of Machine Learning-based Solutions to Protect Privacy in the Internet of Things},'' \emph{Computers \& Security}, vol.~96, p. 101921, 2020.

\bibitem{jamalipour2021taxonomy}
A.~Jamalipour and S.~Murali, ``{A Taxonomy of Machine-learning-based Intrusion Detection Systems for the Internet of Things: A Survey},'' \emph{IEEE Internet of Things Journal}, vol.~9, no.~12, pp. 9444--9466, 2021.

\bibitem{chen2021deep}
W.~Chen, X.~Qiu, T.~Cai, H.-N. Dai, Z.~Zheng, and Y.~Zhang, ``{Deep Reinforcement Learning for Internet of Things: A Comprehensive Survey},'' \emph{IEEE Communications Surveys \& Tutorials}, vol.~23, no.~3, pp. 1659--1692, 2021.

\bibitem{lei2020deep}
L.~Lei, Y.~Tan, K.~Zheng, S.~Liu, K.~Zhang, and X.~Shen, ``{Deep Reinforcement Learning for Autonomous Internet of Things: Model, Applications and Challenges},'' \emph{IEEE Communications Surveys \& Tutorials}, vol.~22, no.~3, pp. 1722--1760, 2020.

\bibitem{wu2020research}
H.~Wu, H.~Han, X.~Wang, and S.~Sun, ``{Research on Artificial Intelligence Enhancing Internet of Things Security: A Survey},'' \emph{IEEE Access}, vol.~8, pp. 153\,826--153\,848, 2020.

\bibitem{koroniotis2019forensics}
N.~Koroniotis, N.~Moustafa, and E.~Sitnikova, ``{Forensics and Deep Learning Mechanisms for Botnets in Internet of Things: A Survey of Challenges and Solutions},'' \emph{IEEE Access}, vol.~7, pp. 61\,764--61\,785, 2019.

\bibitem{khalil2021deep}
R.~A. Khalil, N.~Saeed, M.~Masood, Y.~M. Fard, M.-S. Alouini, and T.~Y. Al-Naffouri, ``{Deep Learning in the Industrial Internet of Things: Potentials, Challenges, and Emerging Applications},'' \emph{IEEE Internet of Things Journal}, vol.~8, no.~14, pp. 11\,016--11\,040, 2021.

\bibitem{amin2020edge}
S.~U. Amin and M.~S. Hossain, ``{Edge Intelligence and Internet of Things in Healthcare: A Survey},'' \emph{IEEE Access}, vol.~9, pp. 45--59, 2020.

\bibitem{ferrag2023edgelearning}
M.~A. Ferrag, O.~Friha, B.~Kantarci, N.~Tihanyi, L.~Cordeiro, M.~Debbah, D.~Hamouda, M.~Al-Hawawreh, and K.-K. Raymond~Choo, ``{Edge Learning for 6G-enabled Internet of Things: A Comprehensive Survey of Vulnerabilities, Datasets, and Defenses},'' \emph{arXiv preprint arXiv:2306.10309}, 06 2023.

\bibitem{russell2010artificial}
S.~J. Russell, \emph{Artificial intelligence a modern approach}.\hskip 1em plus 0.5em minus 0.4em\relax Pearson Education, Inc., 2010.

\bibitem{jordan2015machine}
M.~I. Jordan and T.~M. Mitchell, ``Machine learning: Trends, perspectives, and prospects,'' \emph{Science}, vol. 349, no. 6245, pp. 255--260, 2015.

\bibitem{aouedi2022intelligent}
O.~Aouedi, K.~Piamrat, and B.~Parrein, ``Intelligent traffic management in next-generation networks,'' \emph{Future internet}, vol.~14, no.~2, p.~44, 2022.

\bibitem{b137}
Y.~Bengio, A.~Courville, and P.~Vincent, ``"{R}epresentation learning: A review and new perspectives",'' \emph{IEEE transactions on pattern analysis and machine intelligence}, vol.~35, no.~8, pp. 1798--1828, 2013.

\bibitem{goodfellow2016deep}
I.~Goodfellow, Y.~Bengio, and A.~Courville, \emph{Deep learning}.\hskip 1em plus 0.5em minus 0.4em\relax MIT press, 2016.

\bibitem{b136}
R.~S. Sutton, A.~G. Barto \emph{et~al.}, \emph{"{I}ntroduction to reinforcement learning"}.\hskip 1em plus 0.5em minus 0.4em\relax MIT press Cambridge, 1998, vol. 135.

\bibitem{sacco2022partially}
A.~Sacco, M.~Flocco, F.~Esposito, and G.~Marchetto, ``Partially oblivious congestion control for the internet via reinforcement learning,'' \emph{IEEE Transactions on Network and Service Management}, 2022.

\bibitem{kokkonen2022autonomy}
H.~Kokkonen, L.~Lov{\'e}n, N.~H. Motlagh, J.~Partala, A.~Gonz{\'a}lez-Gil, E.~Sola, I.~Angulo, M.~Liyanage, T.~Lepp{\"a}nen, T.~Nguyen \emph{et~al.}, ``Autonomy and intelligence in the computing continuum: Challenges, enablers, and future directions for orchestration,'' \emph{arXiv preprint arXiv:2205.01423}, 2022.

\bibitem{nguyen2021federated}
D.~C. Nguyen, M.~Ding, P.~N. Pathirana, A.~Seneviratne, J.~Li, and H.~V. Poor, ``{Federated Learning for Internet of Things: A Comprehensive Survey},'' \emph{IEEE Communications Surveys \& Tutorials}, vol.~23, no.~3, pp. 1622--1658, 2021.

\bibitem{le2023applications}
M.~Le, T.~Huynh-The, T.~Do-Duy, T.-H. Vu, W.-J. Hwang, and Q.-V. Pham, ``Applications of distributed machine learning for the internet-of-things: A comprehensive survey,'' \emph{arXiv preprint arXiv:2310.10549}, 2023.

\bibitem{mcmahan2017communication}
B.~McMahan, E.~Moore, D.~Ramage, S.~Hampson, and B.~A. y~Arcas, ``Communication-efficient learning of deep networks from decentralized data,'' in \emph{Artificial intelligence and statistics}.\hskip 1em plus 0.5em minus 0.4em\relax PMLR, 2017, pp. 1273--1282.

\bibitem{konevcny2016federated}
J.~Kone{\v{c}}n{\`y}, H.~B. McMahan, D.~Ramage, and P.~Richt{\'a}rik, ``{Federated Optimization: Distributed Machine Learning for on-device Intelligence},'' \emph{arXiv preprint arXiv:1610.02527}, 2016.

\bibitem{lim2020federated}
W.~Y.~B. Lim, N.~C. Luong, D.~T. Hoang, Y.~Jiao, Y.-C. Liang, Q.~Yang, D.~Niyato, and C.~Miao, ``Federated learning in mobile edge networks: A comprehensive survey,'' \emph{IEEE Communications Surveys \& Tutorials}, vol.~22, no.~3, pp. 2031--2063, 2020.

\bibitem{aouedi2022federated}
O.~Aouedi, K.~Piamrat, G.~Muller, and K.~Singh, ``{Federated Semisupervised Learning for Attack Detection in Industrial Internet of Things},'' \emph{IEEE Transactions on Industrial Informatics}, vol.~19, no.~1, pp. 286--295, 2022.

\bibitem{feng2020multi}
S.~Feng and H.~Yu, ``Multi-participant multi-class vertical federated learning,'' \emph{arXiv preprint arXiv:2001.11154}, 2020.

\bibitem{liu2020secure}
Y.~Liu, Y.~Kang, C.~Xing, T.~Chen, and Q.~Yang, ``A secure federated transfer learning framework,'' \emph{IEEE Intelligent Systems}, vol.~35, no.~4, pp. 70--82, 2020.

\bibitem{mohammadi2018deep}
M.~Mohammadi, A.~Al-Fuqaha, S.~Sorour, and M.~Guizani, ``Deep learning for iot big data and streaming analytics: A survey,'' \emph{IEEE Communications Surveys \& Tutorials}, vol.~20, no.~4, pp. 2923--2960, 2018.

\bibitem{le2023wirelessly}
M.~Le, D.~T. Hoang, D.~N. Nguyen, W.-J. Hwang, and Q.-V. Pham, ``Wirelessly powered federated learning networks: Joint power transfer, data sensing, model training, and resource allocation,'' \emph{IEEE Internet of Things Journal}, 2023.

\bibitem{zhong2021multi}
R.~Zhong, X.~Liu, Y.~Liu, and Y.~Chen, ``Multi-agent reinforcement learning in noma-aided uav networks for cellular offloading,'' \emph{IEEE Transactions on Wireless Communications}, vol.~21, no.~3, pp. 1498--1512, 2021.

\bibitem{aouedi2023f}
O.~Aouedi and K.~Piamrat, ``{F-BIDS: Federated-Blending based Intrusion Detection System},'' \emph{Pervasive and Mobile Computing}, p. 101750, 2023.

\bibitem{aouedi2024towards}
O.~Aouedi, ``{Towards a Scalable and Energy-Efficient Framework for Industrial Cloud-Edge-IoT Continuum},'' \emph{IEEE Internet of Things Magazine}, 2024.

\bibitem{al2015internet}
A.~Al-Fuqaha, M.~Guizani, M.~Mohammadi, M.~Aledhari, and M.~Ayyash, ``{Internet of Things: A Survey on Enabling Technologies, Protocols, and Applications},'' \emph{IEEE Communications Surveys \& Tutorials}, vol.~17, no.~4, pp. 2347--2376, 2015.

\bibitem{walinjkar2017personalized}
A.~Walinjkar and J.~Woods, ``{Personalized Wearable Systems for Real-time ECG Classification and Healthcare Interoperability: Real-time ECG Classification and FHIR Interoperability},'' in \emph{2017 Internet Technologies and Applications (ITA)}.\hskip 1em plus 0.5em minus 0.4em\relax IEEE, 2017, pp. 9--14.

\bibitem{hong2017multivariate}
J.~Hong and J.~Yoon, ``{Multivariate Time-series Classification of Sleep Patterns Using a Hybrid Deep Learning Architecture},'' in \emph{2017 IEEE 19th international conference on e-Health networking, applications and services (Healthcom)}.\hskip 1em plus 0.5em minus 0.4em\relax IEEE, 2017, pp. 1--6.

\bibitem{matar2016internet}
G.~Matar, J.-M. Lina, J.~Carrier, A.~Riley, and G.~Kaddoum, ``{Internet of Things in Sleep Monitoring: An Application for Posture Recognition Using Supervised Learning},'' in \emph{2016 IEEE 18th International conference on e-Health networking, applications and services (Healthcom)}.\hskip 1em plus 0.5em minus 0.4em\relax IEEE, 2016, pp. 1--6.

\bibitem{zeng2014convolutional}
M.~Zeng, L.~T. Nguyen, B.~Yu, O.~J. Mengshoel, J.~Zhu, P.~Wu, and J.~Zhang, ``{Convolutional Neural Networks for Human Activity Recognition Using Mobile Sensors},'' in \emph{6th international conference on mobile computing, applications and services}.\hskip 1em plus 0.5em minus 0.4em\relax IEEE, 2014, pp. 197--205.

\bibitem{yang2015deep}
J.~Yang, M.~N. Nguyen, P.~P. San, X.~L. Li, and S.~Krishnaswamy, ``{Deep Convolutional Neural Networks on Multichannel Time Series for Human Activity Recognition},'' in \emph{Twenty-fourth International Joint Conference on Artificial Intelligence (ICAI)}, vol.~15, 2015, pp. 3995--4001.

\bibitem{panwar2017cnn}
M.~Panwar, S.~R. Dyuthi, K.~C. Prakash, D.~Biswas, A.~Acharyya, K.~Maharatna, A.~Gautam, and G.~R. Naik, ``{CNN Based Approach for Activity Recognition Using a Wrist-worn Accelerometer},'' in \emph{2017 39th Annual International Conference of the IEEE Engineering in Medicine and Biology Society (EMBC)}.\hskip 1em plus 0.5em minus 0.4em\relax IEEE, 2017, pp. 2438--2441.

\bibitem{cheng2022real}
X.~Cheng, L.~Zhang, Y.~Tang, Y.~Liu, H.~Wu, and J.~He, ``{Real-time Human Activity Recognition Using Conditionally Parametrized Convolutions on Mobile and Wearable Devices},'' \emph{IEEE Sensors Journal}, vol.~22, no.~6, pp. 5889--5901, 2022.

\bibitem{ordonez2016deep}
F.~J. Ord{\'o}{\~n}ez and D.~Roggen, ``{Deep Convolutional and LSTM Recurrent Neural Networks for Multimodal Wearable Activity Recognition},'' \emph{Sensors}, vol.~16, no.~1, p. 115, 2016.

\bibitem{drolet2017electronic}
B.~C. Drolet, J.~S. Marwaha, B.~Hyatt, P.~E. Blazar, and S.~D. Lifchez, ``{Electronic Communication of Protected Health Information: Privacy, Security, and HIPAA Compliance},'' \emph{The Journal of Hand Surgery}, vol.~42, no.~6, pp. 411--416, 2017.

\bibitem{aouedi2022handling}
O.~Aouedi, A.~Sacco, K.~Piamrat, and G.~Marchetto, ``{Handling Privacy-Sensitive Medical Data With Federated Learning: Challenges and Future Directions},'' \emph{IEEE Journal of Biomedical and Health Informatics}, vol.~27, no.~2, pp. 790--803, 2022.

\bibitem{zehtabian2021privacy}
S.~Zehtabian, S.~Khodadadeh, L.~B{\"o}l{\"o}ni, and D.~Turgut, ``Privacy-preserving learning of human activity predictors in smart environments,'' in \emph{IEEE INFOCOM 2021-IEEE Conference on Computer Communications}.\hskip 1em plus 0.5em minus 0.4em\relax IEEE, 2021, pp. 1--10.

\bibitem{presotto2021semi}
R.~Presotto, ``{Semi-supervised Methodologies to Tackle the Annotated Data Scarcity Problem in the Field of HAR},'' in \emph{2021 22nd IEEE International Conference on Mobile Data Management (MDM)}.\hskip 1em plus 0.5em minus 0.4em\relax IEEE, 2021, pp. 269--271.

\bibitem{zhao2020semi}
Y.~Zhao, H.~Liu, H.~Li, P.~Barnaghi, and H.~Haddadi, ``{Semi-supervised Federated Learning for Activity Recognition},'' \emph{arXiv preprint arXiv:2011.00851}, 2020.

\bibitem{bettini2021personalized}
C.~Bettini, G.~Civitarese, and R.~Presotto, ``Personalized semi-supervised federated learning for human activity recognition,'' \emph{arXiv preprint arXiv:2104.08094}, 2021.

\bibitem{wu2020personalized}
Q.~Wu, K.~He, and X.~Chen, ``{Personalized Federated Learning for Intelligent IoT Applications: A Cloud-edge Based Framework},'' \emph{IEEE Open Journal of the Computer Society}, vol.~1, pp. 35--44, 2020.

\bibitem{imteaj2021fedparl}
A.~Imteaj and M.~H. Amini, ``{FedPARL: Client Activity and Resource-oriented Lightweight Federated Learning Model for Resource-constrained Heterogeneous IoT Environment},'' \emph{Frontiers in Communications and Networks}, p.~10, 2021.

\bibitem{shaik2022fedstack}
T.~Shaik, X.~Tao, N.~Higgins, R.~Gururajan, Y.~Li, X.~Zhou, and U.~R. Acharya, ``{FedStack: Personalized Activity Monitoring Using Stacked Federated Learning},'' \emph{Knowledge-Based Systems}, p. 109929, 2022.

\bibitem{pham2020artificial}
Q.-V. Pham, D.~C. Nguyen, T.~Huynh-The, W.-J. Hwang, and P.~N. Pathirana, ``{Artificial Intelligence (AI) and Big Data for Coronavirus (COVID-19) Pandemic: A Survey on the State-of-the-arts},'' \emph{IEEE access}, vol.~8, p. 130820, 2020.

\bibitem{shi2020review}
F.~Shi, J.~Wang, J.~Shi, Z.~Wu, Q.~Wang, Z.~Tang, K.~He, Y.~Shi, and D.~Shen, ``{Review of Artificial Intelligence Techniques in Imaging Data Acquisition, Segmentation, and Diagnosis for COVID-19},'' \emph{IEEE reviews in biomedical engineering}, vol.~14, pp. 4--15, 2020.

\bibitem{barstugan2020coronavirus}
M.~Barstugan, U.~Ozkaya, and S.~Ozturk, ``{Coronavirus (COVID-19) Classification Using CT Images by Machine Learning Methods},'' \emph{arXiv preprint arXiv:2003.09424}, 2020.

\bibitem{zheng2020deep}
C.~Zheng, X.~Deng, Q.~Fu, Q.~Zhou, J.~Feng, H.~Ma, W.~Liu, and X.~Wang, ``{Deep Learning-based Detection for COVID-19 From Chest CT Using Weak Label},'' \emph{MedRxiv}, 2020.

\bibitem{narin2021automatic}
A.~Narin, C.~Kaya, and Z.~Pamuk, ``{Automatic Detection of Coronavirus Disease (COVID-19) Using X-ray Images and Deep Convolutional Neural Networks},'' \emph{Pattern Analysis and Applications}, vol.~24, no.~3, pp. 1207--1220, 2021.

\bibitem{cohen2020covid}
J.~P. Cohen, P.~Morrison, and L.~Dao, ``{COVID-19 Image Data Collection},'' \emph{arXiv preprint arXiv:2003.11597}, 2020.

\bibitem{apostolopoulos2020covid}
I.~D. Apostolopoulos and T.~A. Mpesiana, ``{COVID-19: Automatic Detection From X-Ray Images Utilizing Transfer Learning With Convolutional Neural Networks},'' \emph{Physical and Engineering Sciences in Medicine}, vol.~43, no.~2, pp. 635--640, 2020.

\bibitem{pathak2020deep}
Y.~Pathak, P.~K. Shukla, A.~Tiwari, S.~Stalin, and S.~Singh, ``{Deep Transfer Learning Based Classification Model for COVID-19 Disease},'' \emph{Irbm}, 2020.

\bibitem{feki2021federated}
I.~Feki, S.~Ammar, Y.~Kessentini, and K.~Muhammad, ``{Federated Learning for COVID-19 Screening From Chest X-ray Images},'' \emph{Applied Soft Computing}, vol. 106, p. 107330, 2021.

\bibitem{wang2021auxiliary}
R.~Wang, J.~Xu, Y.~Ma, M.~Talha, M.~S. Al-Rakhami, and A.~Ghoneim, ``{Auxiliary Diagnosis of COVID-19 Based on 5G-Enabled Federated Learning},'' \emph{IEEE Network}, vol.~35, no.~3, pp. 14--20, 2021.

\bibitem{zhang2021dynamic}
W.~Zhang, T.~Zhou, Q.~Lu, X.~Wang, C.~Zhu, H.~Sun, Z.~Wang, S.~K. Lo, and F.-Y. Wang, ``{Dynamic-fusion-based Federated Learning for COVID-19 Detection},'' \emph{IEEE Internet of Things Journal}, vol.~8, no.~21, pp. 15\,884--15\,891, 2021.

\bibitem{sarumi2022potential}
O.~A. Sarumi, O.~Aouedi, and L.~J. Muhammad, ``{Potential of Deep Learning Algorithms in Mitigating the Spread of COVID-19},'' \emph{Understanding COVID-19: The Role of Computational Intelligence}, pp. 225--244, 2022.

\bibitem{alazab2020covid}
M.~Alazab, A.~Awajan, A.~Mesleh, A.~Abraham, V.~Jatana, and S.~Alhyari, ``{COVID-19 Prediction and Detection Using Deep Learning},'' \emph{International Journal of Computer Information Systems and Industrial Management Applications}, vol.~12, no. June, pp. 168--181, 2020.

\bibitem{syed2020smart}
D.~Syed, A.~Zainab, A.~Ghrayeb, S.~S. Refaat, H.~Abu-Rub, and O.~Bouhali, ``{Smart Grid Big Data Analytics: Survey of Technologies, Techniques, and Applications},'' \emph{IEEE Access}, vol.~9, pp. 59\,564--59\,585, 2020.

\bibitem{baumeister2010literature}
T.~Baumeister, ``{Literature Review on Smart Grid Cyber Security},'' \emph{Collaborative Software Development Laboratory at the University of Hawaii}, vol. 650, 2010.

\bibitem{reka2018future}
S.~S. Reka and T.~Dragicevic, ``{Future Effectual Role of Energy Delivery: A Comprehensive Review of Internet of Things and Smart Grid},'' \emph{Renewable and Sustainable Energy Reviews}, vol.~91, pp. 90--108, 2018.

\bibitem{ahmed2020machine}
W.~Ahmed, H.~Ansari, B.~Khan, Z.~Ullah, S.~M. Ali, C.~A.~A. Mehmood, M.~B. Qureshi, I.~Hussain, M.~Jawad, M.~U.~S. Khan \emph{et~al.}, ``Machine learning based energy management model for smart grid and renewable energy districts,'' \emph{IEEE Access}, vol.~8, pp. 185\,059--185\,078, 2020.

\bibitem{alazab2020multidirectional}
M.~Alazab, S.~Khan, S.~S.~R. Krishnan, Q.-V. Pham, M.~P.~K. Reddy, and T.~R. Gadekallu, ``A multidirectional lstm model for predicting the stability of a smart grid,'' \emph{IEEE Access}, vol.~8, pp. 85\,454--85\,463, 2020.

\bibitem{zhai2021dynamic}
S.~Zhai, X.~Jin, L.~Wei, H.~Luo, and M.~Cao, ``Dynamic federated learning for gmec with time-varying wireless link,'' \emph{IEEE Access}, vol.~9, pp. 10\,400--10\,412, 2021.

\bibitem{cao2020ifed}
H.~Cao, S.~Liu, R.~Zhao, and X.~Xiong, ``Ifed: A novel federated learning framework for local differential privacy in power internet of things,'' \emph{International Journal of Distributed Sensor Networks}, vol.~16, no.~5, p. 1550147720919698, 2020.

\bibitem{su2021secure}
Z.~Su, Y.~Wang, T.~H. Luan, N.~Zhang, F.~Li, T.~Chen, and H.~Cao, ``Secure and efficient federated learning for smart grid with edge-cloud collaboration,'' \emph{IEEE Transactions on Industrial Informatics}, vol.~18, no.~2, pp. 1333--1344, 2021.

\bibitem{tun2021federated}
Y.~L. Tun, K.~Thar, C.~M. Thwal, and C.~S. Hong, ``Federated learning based energy demand prediction with clustered aggregation,'' in \emph{2021 IEEE International Conference on Big Data and Smart Computing (BigComp)}.\hskip 1em plus 0.5em minus 0.4em\relax IEEE, 2021, pp. 164--167.

\bibitem{zhao2020application}
L.~Zhao, T.~Dai, Z.~Qiao, P.~Sun, J.~Hao, and Y.~Yang, ``Application of artificial intelligence to wastewater treatment: A bibliometric analysis and systematic review of technology, economy, management, and wastewater reuse,'' \emph{Process Safety and Environmental Protection}, vol. 133, pp. 169--182, 2020.

\bibitem{hou2021machine}
X.~Hou, J.~Wang, Z.~Fang, X.~Zhang, S.~Song, X.~Zhang, and Y.~Ren, ``{Machine-Learning-Aided Mission-Critical Internet of Underwater Things},'' \emph{IEEE Network}, vol.~35, no.~4, pp. 160--166, 2021.

\bibitem{ktari2022lightweight}
J.~Ktari, T.~Frikha, M.~Hamdi, H.~Elmannai, and H.~Hmam, ``Lightweight ai framework for industry 4.0 case study: water meter recognition,'' \emph{Big Data and Cognitive Computing}, vol.~6, no.~3, p.~72, 2022.

\bibitem{baek2020prediction}
S.-S. Baek, J.~Pyo, and J.~A. Chun, ``Prediction of water level and water quality using a cnn-lstm combined deep learning approach,'' \emph{Water}, vol.~12, no.~12, p. 3399, 2020.

\bibitem{saraiva2020automatic}
M.~Saraiva, {\'E}.~Protas, M.~Salgado, and C.~Souza~Jr, ``Automatic mapping of center pivot irrigation systems from satellite images using deep learning,'' \emph{Remote Sensing}, vol.~12, no.~3, p. 558, 2020.

\bibitem{wang2017water}
Y.~Wang, J.~Zhou, K.~Chen, Y.~Wang, and L.~Liu, ``{Water Quality Prediction Method Based on LSTM Neural Network},'' in \emph{2017 12th international conference on intelligent systems and knowledge engineering (ISKE)}.\hskip 1em plus 0.5em minus 0.4em\relax IEEE, 2017, pp. 1--5.

\bibitem{prasad2022analysis}
D.~V.~V. Prasad, L.~Y. Venkataramana, P.~S. Kumar, G.~Prasannamedha, S.~Harshana, S.~J. Srividya, K.~Harrinei, and S.~Indraganti, ``Analysis and prediction of water quality using deep learning and auto deep learning techniques,'' \emph{Science of The Total Environment}, vol. 821, p. 153311, 2022.

\bibitem{victor2022federated}
N.~Victor, M.~Alazab, S.~Bhattacharya, S.~Magnusson, P.~K.~R. Maddikunta, K.~Ramana, T.~R. Gadekallu \emph{et~al.}, ``{Federated Learning for IoUT: Concepts, Applications, Challenges and Opportunities},'' \emph{arXiv preprint arXiv:2207.13976}, 2022.

\bibitem{zhao2021federated}
H.~Zhao, F.~Ji, Q.~Guan, Q.~Li, S.~Wang, H.~Dong, and M.~Wen, ``{Federated Meta Learning Enhanced Acoustic Radio Cooperative Framework for Ocean of Things Underwater Acoustic Communications},'' \emph{arXiv preprint arXiv:2105.13296}, 2021.

\bibitem{balali2015detection}
V.~Balali, A.~Ashouri~Rad, and M.~Golparvar-Fard, ``{Detection, Classification, and Mapping of Us Traffic Signs Using Google Street View Images for Roadway Inventory Management},'' \emph{Visualization in Engineering}, vol.~3, pp. 1--18, 2015.

\bibitem{zaklouta2012real}
F.~Zaklouta and B.~Stanciulescu, ``{Real-time Traffic-sign Recognition Using Tree Classifiers},'' \emph{IEEE Transactions on Intelligent Transportation Systems}, vol.~13, no.~4, pp. 1507--1514, 2012.

\bibitem{chen2019fire}
Y.~Chen, W.~Xu, J.~Zuo, and K.~Yang, ``{The Fire Recognition Algorithm Using Dynamic Feature Fusion and IV-SVM Classifier},'' \emph{Cluster Computing}, vol.~22, pp. 7665--7675, 2019.

\bibitem{tabernik2019deep}
D.~Tabernik and D.~Sko{\v{c}}aj, ``Deep learning for large-scale traffic-sign detection and recognition,'' \emph{IEEE transactions on intelligent transportation systems}, vol.~21, no.~4, pp. 1427--1440, 2019.

\bibitem{yang2018deep}
T.~Yang, X.~Long, A.~K. Sangaiah, Z.~Zheng, and C.~Tong, ``Deep detection network for real-life traffic sign in vehicular networks,'' \emph{Computer Networks}, vol. 136, pp. 95--104, 2018.

\bibitem{arcos2018deep}
{\'A}.~Arcos-Garc{\'\i}a, J.~A. Alvarez-Garcia, and L.~M. Soria-Morillo, ``Deep neural network for traffic sign recognition systems: An analysis of spatial transformers and stochastic optimisation methods,'' \emph{Neural Networks}, vol.~99, pp. 158--165, 2018.

\bibitem{zhang2020lightweight}
J.~Zhang, W.~Wang, C.~Lu, J.~Wang, and A.~K. Sangaiah, ``Lightweight deep network for traffic sign classification,'' \emph{Annals of Telecommunications}, vol.~75, pp. 369--379, 2020.

\bibitem{xie2022efficient}
K.~Xie, Z.~Zhang, B.~Li, J.~Kang, D.~Niyato, S.~Xie, and Y.~Wu, ``Efficient federated learning with spike neural networks for traffic sign recognition,'' \emph{IEEE Transactions on Vehicular Technology}, vol.~71, no.~9, pp. 9980--9992, 2022.

\bibitem{b35}
\BIBentryALTinterwordspacing
``{The Growth of Electric Vehicles},'' [Accessed 14. Apr. 2023]. [Online]. Available: \url{https://www.cnbc.com/2018/05/30/electric-vehicles-will-grow- from-3-million-to-125-million-by-2030-iea.html}
\BIBentrySTDinterwordspacing

\bibitem{majidpour2014fast}
M.~Majidpour, C.~Qiu, P.~Chu, R.~Gadh, and H.~R. Pota, ``{Fast Prediction for Sparse Time Series: Demand Forecast of EV Charging Stations for Cell Phone Applications},'' \emph{IEEE Transactions on Industrial Informatics}, vol.~11, no.~1, pp. 242--250, 2014.

\bibitem{saputra2019energy}
Y.~M. Saputra, D.~T. Hoang, D.~N. Nguyen, E.~Dutkiewicz, M.~D. Mueck, and S.~Srikanteswara, ``{Energy Demand Prediction With Federated Learning for Electric Vehicle Networks},'' in \emph{2019 IEEE global communications conference (GLOBECOM)}.\hskip 1em plus 0.5em minus 0.4em\relax IEEE, 2019, pp. 1--6.

\bibitem{chics2016reinforcement}
A.~Chi{\c{s}}, J.~Lund{\'e}n, and V.~Koivunen, ``{Reinforcement Learning-Based Plug-in Electric Vehicle Charging With Forecasted Price},'' \emph{IEEE Transactions on Vehicular Technology}, vol.~66, no.~5, pp. 3674--3684, 2016.

\bibitem{jahangir2020plug}
H.~Jahangir, S.~S. Gougheri, B.~Vatandoust, M.~A. Golkar, A.~Ahmadian, and A.~Hajizadeh, ``{Plug-in Electric Vehicle Behavior Modeling in Energy Market: A Novel Deep Learning-Based Approach With Clustering Technique},'' \emph{IEEE Transactions on Smart Grid}, vol.~11, no.~6, pp. 4738--4748, 2020.

\bibitem{liu2022mobile}
L.~Liu, Z.~Xi, K.~Zhu, R.~Wang, and E.~Hossain, ``{Mobile Charging Station Placements in Internet of Electric Vehicles: A Federated Learning Approach},'' \emph{IEEE Transactions on Intelligent Transportation Systems}, vol.~23, no.~12, pp. 24\,561--24\,577, 2022.

\bibitem{shi2018application}
J.~Shi, F.~Qiao, Q.~Li, L.~Yu, and Y.~Hu, ``Application and evaluation of the reinforcement learning approach to eco-driving at intersections under infrastructure-to-vehicle communications,'' \emph{Transportation Research Record}, vol. 2672, no.~25, pp. 89--98, 2018.

\bibitem{wegener2021automated}
M.~Wegener, L.~Koch, M.~Eisenbarth, and J.~Andert, ``{Automated Eco-Driving in Urban Scenarios Using Deep Reinforcement Learning},'' \emph{Transportation Research Part C: Emerging Technologies}, vol. 126, p. 102967, 2021.

\bibitem{pozzi2020ecological}
A.~Pozzi, S.~Bae, Y.~Choi, F.~Borrelli, D.~M. Raimondo, and S.~Moura, ``Ecological velocity planning through signalized intersections: A deep reinforcement learning approach,'' in \emph{2020 59th IEEE Conference on Decision and Control (CDC)}.\hskip 1em plus 0.5em minus 0.4em\relax IEEE, 2020, pp. 245--252.

\bibitem{yuan2021survey}
H.~Yuan and G.~Li, ``A survey of traffic prediction: from spatio-temporal data to intelligent transportation,'' \emph{Data Science and Engineering}, vol.~6, pp. 63--85, 2021.

\bibitem{cai2016spatiotemporal}
P.~Cai, Y.~Wang, G.~Lu, P.~Chen, C.~Ding, and J.~Sun, ``A spatiotemporal correlative k-nearest neighbor model for short-term traffic multistep forecasting,'' \emph{Transportation Research Part C: Emerging Technologies}, vol.~62, pp. 21--34, 2016.

\bibitem{johansson2014regression}
U.~Johansson, H.~Bostr{\"o}m, T.~L{\"o}fstr{\"o}m, and H.~Linusson, ``Regression conformal prediction with random forests,'' \emph{Machine learning}, vol.~97, pp. 155--176, 2014.

\bibitem{lara2021experimental}
P.~Lara-Ben{\'\i}tez, M.~Carranza-Garc{\'\i}a, and J.~C. Riquelme, ``An experimental review on deep learning architectures for time series forecasting,'' \emph{International Journal of Neural Systems}, vol.~31, no.~03, p. 2130001, 2021.

\bibitem{sun2020city}
S.~Sun, H.~Wu, and L.~Xiang, ``City-wide traffic flow forecasting using a deep convolutional neural network,'' \emph{Sensors}, vol.~20, no.~2, p. 421, 2020.

\bibitem{zhao2019deep}
W.~Zhao, Y.~Gao, T.~Ji, X.~Wan, F.~Ye, and G.~Bai, ``Deep temporal convolutional networks for short-term traffic flow forecasting,'' \emph{IEEE Access}, vol.~7, pp. 114\,496--114\,507, 2019.

\bibitem{kipf2016semi}
T.~N. Kipf and M.~Welling, ``Semi-supervised classification with graph convolutional networks,'' \emph{arXiv preprint arXiv:1609.02907}, 2016.

\bibitem{li2017diffusion}
Y.~Li, R.~Yu, C.~Shahabi, and Y.~Liu, ``Diffusion convolutional recurrent neural network: Data-driven traffic forecasting,'' \emph{arXiv preprint arXiv:1707.01926}, 2017.

\bibitem{wu2019graph}
Z.~Wu, S.~Pan, G.~Long, J.~Jiang, and C.~Zhang, ``Graph wavenet for deep spatial-temporal graph modeling,'' \emph{arXiv preprint arXiv:1906.00121}, 2019.

\bibitem{guo2020optimized}
K.~Guo, Y.~Hu, Z.~Qian, H.~Liu, K.~Zhang, Y.~Sun, J.~Gao, and B.~Yin, ``Optimized graph convolution recurrent neural network for traffic prediction,'' \emph{IEEE Transactions on Intelligent Transportation Systems}, vol.~22, no.~2, pp. 1138--1149, 2020.

\bibitem{zheng2020gman}
C.~Zheng, X.~Fan, C.~Wang, and J.~Qi, ``Gman: A graph multi-attention network for traffic prediction,'' in \emph{Proceedings of the AAAI conference on artificial intelligence}, vol.~34, no.~01, 2020, pp. 1234--1241.

\bibitem{li2019hybrid}
Z.~Li, G.~Xiong, Y.~Chen, Y.~Lv, B.~Hu, F.~Zhu, and F.-Y. Wang, ``A hybrid deep learning approach with gcn and lstm for traffic flow prediction,'' in \emph{2019 IEEE intelligent transportation systems conference (ITSC)}.\hskip 1em plus 0.5em minus 0.4em\relax IEEE, 2019, pp. 1929--1933.

\bibitem{wang2018crowd}
L.~Wang, X.~Geng, X.~Ma, F.~Liu, and Q.~Yang, ``{Crowd Flow Prediction by Deep Spatio-temporal Transfer Learning},'' \emph{arXiv preprint arXiv:1802.00386}, 2018.

\bibitem{liu2020privacy}
Y.~Liu, J.~James, J.~Kang, D.~Niyato, and S.~Zhang, ``{Privacy-Preserving Traffic Flow Prediction: A Federated Learning Approach},'' \emph{IEEE Internet of Things Journal}, vol.~7, no.~8, pp. 7751--7763, 2020.

\bibitem{yuan2022fedstn}
X.~Yuan, J.~Chen, J.~Yang, N.~Zhang, T.~Yang, T.~Han, and A.~Taherkordi, ``{FedSTN: Graph Representation Driven Federated Learning for Edge Computing Enabled Urban Traffic Flow Prediction},'' \emph{IEEE Transactions on Intelligent Transportation Systems}, 2022.

\bibitem{zhang2021fastgnn}
C.~Zhang, S.~Zhang, J.~James, and S.~Yu, ``Fastgnn: A topological information protected federated learning approach for traffic speed forecasting,'' \emph{IEEE Transactions on Industrial Informatics}, vol.~17, no.~12, pp. 8464--8474, 2021.

\bibitem{atallah2016reinforcement}
R.~F. Atallah, C.~M. Assi, and J.~Y. Yu, ``A reinforcement learning technique for optimizing downlink scheduling in an energy-limited vehicular network,'' \emph{IEEE Transactions on Vehicular Technology}, vol.~66, no.~6, pp. 4592--4601, 2016.

\bibitem{yan2018smart}
M.~Yan, G.~Feng, J.~Zhou, and S.~Qin, ``{Smart Multi-RAT Access Based on Multiagent Reinforcement Learning},'' \emph{IEEE Transactions on Vehicular Technology}, vol.~67, no.~5, pp. 4539--4551, 2018.

\bibitem{ge2017data}
Z.~Ge, Z.~Song, S.~X. Ding, and B.~Huang, ``{Data Mining and Analytics in the Process Industry: The Role of Machine Learning},'' \emph{IEEE Access}, vol.~5, pp. 20\,590--20\,616, 2017.

\bibitem{maddikunta2022industry}
P.~K.~R. Maddikunta, Q.-V. Pham, B.~Prabadevi \emph{et~al.}, ``Industry 5.0: A survey on enabling technologies and potential applications,'' \emph{Journal of Industrial Information Integration}, vol.~26, p. 100257, 2022.

\bibitem{penumuru2020identification}
D.~P. Penumuru, S.~Muthuswamy, and P.~Karumbu, ``Identification and classification of materials using machine vision and machine learning in the context of industry 4.0,'' \emph{Journal of Intelligent Manufacturing}, vol.~31, no.~5, pp. 1229--1241, 2020.

\bibitem{candanedo2018machine}
I.~S. Candanedo, E.~H. Nieves, S.~R. Gonz{\'a}lez, M.~T.~S. Mart{\'\i}n, and A.~G. Briones, ``Machine learning predictive model for industry 4.0,'' in \emph{Knowledge Management in Organizations: 13th International Conference, KMO 2018, {\v{Z}}ilina, Slovakia, August 6--10, 2018, Proceedings 13}.\hskip 1em plus 0.5em minus 0.4em\relax Springer, 2018, pp. 501--510.

\bibitem{calabrese2020sophia}
M.~Calabrese, M.~Cimmino, F.~Fiume, M.~Manfrin, L.~Romeo, S.~Ceccacci, M.~Paolanti, G.~Toscano, G.~Ciandrini, A.~Carrotta \emph{et~al.}, ``{SOPHIA: An event-based IoT and machine learning architecture for predictive maintenance in industry 4.0},'' \emph{Information}, vol.~11, no.~4, p. 202, 2020.

\bibitem{bajic2018machine}
B.~Bajic, I.~Cosic, M.~Lazarevic, N.~Sremcev, and A.~Rikalovic, ``Machine learning techniques for smart manufacturing: Applications and challenges in industry 4.0,'' \emph{Department of Industrial Engineering and Management Novi Sad, Serbia}, vol.~29, 2018.

\bibitem{mivskuf2016comparison}
M.~Mi{\v{s}}kuf and I.~Zolotov{\'a}, ``Comparison between multi-class classifiers and deep learning with focus on industry 4.0,'' in \emph{2016 Cybernetics \& Informatics (K\&I)}.\hskip 1em plus 0.5em minus 0.4em\relax IEEE, 2016, pp. 1--5.

\bibitem{subakti2018indoor}
H.~Subakti and J.-R. Jiang, ``Indoor augmented reality using deep learning for industry 4.0 smart factories,'' in \emph{2018 IEEE 42nd Annual Computer Software and Applications Conference (COMPSAC)}, vol.~2.\hskip 1em plus 0.5em minus 0.4em\relax IEEE, 2018, pp. 63--68.

\bibitem{mozaffar2018data}
M.~Mozaffar, A.~Paul, R.~Al-Bahrani, S.~Wolff, A.~Choudhary, A.~Agrawal, K.~Ehmann, and J.~Cao, ``Data-driven prediction of the high-dimensional thermal history in directed energy deposition processes via recurrent neural networks,'' \emph{Manufacturing letters}, vol.~18, pp. 35--39, 2018.

\bibitem{francis2019deep}
J.~Francis and L.~Bian, ``Deep learning for distortion prediction in laser-based additive manufacturing using big data,'' \emph{Manufacturing Letters}, vol.~20, pp. 10--14, 2019.

\bibitem{cao2020multiagent}
Z.~Cao, P.~Zhou, R.~Li, S.~Huang, and D.~Wu, ``Multiagent deep reinforcement learning for joint multichannel access and task offloading of mobile-edge computing in industry 4.0,'' \emph{IEEE Internet of Things Journal}, vol.~7, no.~7, pp. 6201--6213, 2020.

\bibitem{messaoud2020deep}
S.~Messaoud, A.~Bradai, O.~B. Ahmed, P.~T.~A. Quang, M.~Atri, and M.~S. Hossain, ``Deep federated q-learning-based network slicing for industrial iot,'' \emph{IEEE Transactions on Industrial Informatics}, vol.~17, no.~8, pp. 5572--5582, 2020.

\bibitem{qu2020blockchained}
Y.~Qu, S.~R. Pokhrel, S.~Garg, L.~Gao, and Y.~Xiang, ``A blockchained federated learning framework for cognitive computing in industry 4.0 networks,'' \emph{IEEE Transactions on Industrial Informatics}, vol.~17, no.~4, pp. 2964--2973, 2020.

\bibitem{arachchige2020trustworthy}
P.~C.~M. Arachchige, P.~Bertok, I.~Khalil, D.~Liu, S.~Camtepe, and M.~Atiquzzaman, ``A trustworthy privacy preserving framework for machine learning in industrial iot systems,'' \emph{IEEE Transactions on Industrial Informatics}, vol.~16, no.~9, pp. 6092--6102, 2020.

\bibitem{mozaffar2021geometry}
M.~Mozaffar, S.~Liao, H.~Lin, K.~Ehmann, and J.~Cao, ``Geometry-agnostic data-driven thermal modeling of additive manufacturing processes using graph neural networks,'' \emph{Additive Manufacturing}, vol.~48, p. 102449, 2021.

\bibitem{dong2023graph}
G.~Dong, M.~Tang, Z.~Wang, J.~Gao, S.~Guo, L.~Cai, R.~Gutierrez, B.~Campbel, L.~E. Barnes, and M.~Boukhechba, ``Graph neural networks in iot: A survey,'' \emph{ACM Transactions on Sensor Networks}, vol.~19, no.~2, pp. 1--50, 2023.

\bibitem{zhou2022graph}
L.~Zhou, V.~D. Sharma, Q.~Li, A.~Prorok, A.~Ribeiro, P.~Tokekar, and V.~Kumar, ``Graph neural networks for decentralized multi-robot target tracking,'' in \emph{2022 IEEE International Symposium on Safety, Security, and Rescue Robotics (SSRR)}.\hskip 1em plus 0.5em minus 0.4em\relax IEEE, 2022, pp. 195--202.

\bibitem{kapelyukh2022my}
I.~Kapelyukh and E.~Johns, ``My house, my rules: Learning tidying preferences with graph neural networks,'' in \emph{Conference on Robot Learning}.\hskip 1em plus 0.5em minus 0.4em\relax PMLR, 2022, pp. 740--749.

\bibitem{ruan2019mobile}
X.~Ruan, D.~Ren, X.~Zhu, and J.~Huang, ``Mobile robot navigation based on deep reinforcement learning,'' in \emph{2019 Chinese control and decision conference (CCDC)}.\hskip 1em plus 0.5em minus 0.4em\relax IEEE, 2019, pp. 6174--6178.

\bibitem{liu2020federated}
B.~Liu, L.~Wang, M.~Liu, and C.-Z. Xu, ``Federated imitation learning: A novel framework for cloud robotic systems with heterogeneous sensor data,'' \emph{IEEE Robotics and Automation Letters}, vol.~5, no.~2, pp. 3509--3516, 2020.

\bibitem{liu2019lifelong}
B.~Liu, L.~Wang, and M.~Liu, ``Lifelong federated reinforcement learning: a learning architecture for navigation in cloud robotic systems,'' \emph{IEEE Robotics and Automation Letters}, vol.~4, no.~4, pp. 4555--4562, 2019.

\bibitem{pham2020survey}
Q.-V. Pham, F.~Fang, V.~N. Ha, M.~J. Piran, M.~Le, L.~B. Le, W.-J. Hwang, and Z.~Ding, ``A survey of multi-access edge computing in {5G} and beyond: Fundamentals, technology integration, and state-of-the-art,'' \emph{IEEE access}, vol.~8, pp. 116\,974--117\,017, 2020.

\bibitem{Osei2022Dec}
A.~B. Osei, S.~R. Yeginati, Y.~Al~Mtawa, and T.~Halabi, ``{Optimized Moving Target Defense Against DDoS Attacks in IoT Networks: When to Adapt?}'' in \emph{{GLOBECOM 2022 - 2022 IEEE Global Communications Conference}}.\hskip 1em plus 0.5em minus 0.4em\relax IEEE, Dec. 2022, pp. 2782--2787.

\bibitem{Goncalves2023Jan}
D.~S.~M. Gon{\ifmmode\mbox{\c{c}}\else\c{c}\fi}alves, R.~S. Couto, and M.~G. Rubinstein, ``{A Protection System Against HTTP Flood Attacks Using Software Defined Networking},'' \emph{J. Netw. Syst. Manage.}, vol.~31, no.~1, pp. 1--23, Jan. 2023.

\bibitem{Quincozes2023Jul}
S.~E. Quincozes, J.~F. Kazienko, and V.~E. Quincozes, ``{An extended evaluation on machine learning techniques for Denial-of-Service detection in Wireless Sensor Networks},'' \emph{Internet of Things}, vol.~22, p. 100684, Jul. 2023.

\bibitem{Vishwakarma2020Jan}
R.~Vishwakarma and A.~K. Jain, ``{A survey of DDoS attacking techniques and defence mechanisms in the IoT network},'' \emph{Telecommunication Systems}, vol.~73, no.~1, pp. 3--25, Jan. 2020.

\bibitem{Fan2022Jul}
J.~Fan, Q.~Yan, M.~Li, G.~Qu, and Y.~Xiao, ``{A Survey on Data Poisoning Attacks and Defenses},'' in \emph{{2022 7th IEEE International Conference on Data Science in Cyberspace (DSC)}}.\hskip 1em plus 0.5em minus 0.4em\relax IEEE, Jul. 2022, pp. 48--55.

\bibitem{Ge2022Jul}
Y.~Ge, S.~Liu, Z.~Fu, J.~Tan, Z.~Li \emph{et~al.}, ``{A Survey on Trustworthy Recommender Systems},'' \emph{arXiv}, Jul. 2022.

\bibitem{Fang2021Apr}
M.~Fang, M.~Sun, Q.~Li, N.~Z. Gong, J.~Tian \emph{et~al.}, ``{Data Poisoning Attacks and Defenses to Crowdsourcing Systems},'' in \emph{{WWW '21: Proceedings of the Web Conference 2021}}.\hskip 1em plus 0.5em minus 0.4em\relax New York, NY, USA: Association for Computing Machinery, Apr. 2021, pp. 969--980.

\bibitem{Huang2022May}
Y.~Huang and C.~Chen, ``{Smart App Attack: Hacking Deep Learning Models in Android Apps},'' \emph{IEEE Transactions on Information Forensics and Security}, vol.~17, pp. 1827--1840, May 2022.

\bibitem{Takiddin2022Jul}
A.~Takiddin, M.~Ismail, and E.~Serpedin, ``{Robust Data-Driven Detection of Electricity Theft Adversarial Evasion Attacks in Smart Grids},'' \emph{IEEE Transactions on Smart Grid}, vol.~14, no.~1, pp. 663--676, Jan. 2022.

\bibitem{Chi2023Jan}
L.~Chi, M.~Msahli, G.~Memmi, and H.~Qiu, ``{Public-attention-based Adversarial Attack on Traffic Sign Recognition},'' in \emph{{2023 IEEE 20th Consumer Communications {\&} Networking Conference (CCNC)}}.\hskip 1em plus 0.5em minus 0.4em\relax IEEE, Jan 2023, pp. 740--745.

\bibitem{Zhang2023Jan}
X.~Zhang, C.~Chen, Y.~Xie, X.~Chen, J.~Zhang \emph{et~al.}, ``{A survey on privacy inference attacks and defenses in cloud-based Deep Neural Network},'' \emph{Computer Standards {\&} Interfaces}, vol.~83, p. 103672, Jan. 2023.

\bibitem{Shi2022Feb}
Y.~Shi and Y.~Sagduyu, ``{Membership Inference Attack and Defense for Wireless Signal Classifiers with Deep Learning},'' \emph{IEEE Trans. Mob. Comput.}, p.~1, Feb. 2022.

\bibitem{Chen2023Mar}
Q.~Chen, Y.~Wu, X.~Wang, Z.~L. Jiang, W.~Zhang \emph{et~al.}, ``{A Generic Cryptographic Deep-Learning Inference Platform for Remote Sensing Scenes},'' \emph{IEEE J. Sel. Top. Appl. Earth Obs. Remote Sens.}, vol.~16, pp. 3309--3321, Mar. 2023.

\bibitem{Yan2022Nov}
H.~Yan, S.~Li, Y.~Wang, Y.~Zhang, K.~Sharif \emph{et~al.}, ``{Membership Inference Attacks Against Deep Learning Models Via Logits Distribution},'' \emph{IEEE Trans. Dependable Secure Comput.}, Nov. 2022, in press.

\bibitem{Liu2022Dec}
Y.~Liu, H.~Li, G.~Huang, and W.~Hua, ``{OPUPO: Defending Against Membership Inference Attacks With Order-Preserving and Utility-Preserving Obfuscation},'' \emph{IEEE Trans. Dependable Secure Comput.}, p. in press, Dec. 2022.

\bibitem{atamli2014threat}
A.~W. Atamli and A.~Martin, ``Threat-based security analysis for the internet of things,'' in \emph{2014 International Workshop on Secure Internet of Things}.\hskip 1em plus 0.5em minus 0.4em\relax IEEE, 2014, pp. 35--43.

\bibitem{Gomes2023Oct}
D.~R. Gomes, F.~A.~A. Lins, O.~O. N{\ifmmode\acute{o}\else\'{o}\fi}brega, E.~F. Felix, B.~A. Jesus \emph{et~al.}, ``Security evaluation of authentication requirements in {IoT} gateways,'' \emph{J. Netw. Syst. Manage.}, vol.~31, no.~4, pp. 1--24, Oct. 2023.

\bibitem{Huang2016Nov}
X.~Huang, P.~Craig, H.~Lin, and Z.~Yan, ``{SecIoT: A} security framework for the internet of things,'' \emph{Secur. Commun. Netw.}, vol.~9, no.~16, pp. 3083--3094, Nov. 2016.

\bibitem{Bagaa2020May}
M.~Bagaa, T.~Taleb, J.~B. Bernabe, and A.~Skarmeta, ``{A Machine Learning Security Framework for IoT Systems},'' \emph{IEEE Access}, vol.~8, pp. 114\,066--114\,077, May 2020.

\bibitem{Lins2020Nov}
F.~A.~A. Lins and M.~Vieira, ``{Security Requirements and Solutions for IoT Gateways: A Comprehensive Study},'' \emph{IEEE IoT J.}, vol.~8, no.~11, pp. 8667--8679, Nov. 2020.

\bibitem{platform2021security}
G.~Platform, ``Security evaluation standard for {IoT} platforms v1. 1 {(SESIP)},'' \emph{Global Platform Standard}, 2021.

\bibitem{infrastructure2002common}
P.~K. Infrastructure and T.~P. Profile, ``Common criteria for information technology security evaluation,'' \emph{National Security Agency}, 2002.

\bibitem{potter2002802}
B.~Potter and B.~Fleck, \emph{{802.11 Security}}.\hskip 1em plus 0.5em minus 0.4em\relax " O'Reilly Media, Inc.", 2002.

\bibitem{korner2023current}
F.~K{\"o}rner, ``Current challenges of implementing {ETSI EN} 303 645 as a baseline security standard for consumer {IoT} security certification,'' \emph{Authorea Preprints}, 2023.

\bibitem{Karie2021Sep}
N.~M. Karie, N.~M. Sahri, W.~Yang, C.~Valli, and V.~R. Kebande, ``{A Review of Security Standards and Frameworks for IoT-Based Smart Environments},'' \emph{IEEE Access}, vol.~9, pp. 121\,975--121\,995, Sep. 2021.

\bibitem{Lee2021Mar}
E.~Lee, Y.-D. Seo, S.-R. Oh, and Y.-G. Kim, ``{A Survey on Standards for Interoperability and Security in the Internet of Things},'' \emph{IEEE Commun. Surv. Tutorials}, vol.~23, no.~2, pp. 1020--1047, Mar. 2021.

\bibitem{Zhang2019Dec}
Y.~Zhang, Q.~He, G.~Chen, X.~Zhang, and Y.~Xiang, ``{A Low-overhead, Confidentiality-assured, and Authenticated Data Acquisition Framework for IoT},'' \emph{IEEE Transactions on Industrial Informatics}, vol.~16, no.~12, pp. 7566--7578, Dec. 2019.

\bibitem{Chanal2019Jul}
P.~M. Chanal and M.~S. Kakkasageri, ``{Hybrid Algorithm for Data Confidentiality in Internet of Things},'' in \emph{{2019 10th International Conference on Computing, Communication and Networking Technologies (ICCCNT)}}.\hskip 1em plus 0.5em minus 0.4em\relax IEEE, Jul. 2019, pp. 1--5.

\bibitem{Reagen2021}
\emph{{Cheetah: Optimizing} and Accelerating Homomorphic Encryption for Private Inference}.\hskip 1em plus 0.5em minus 0.4em\relax IEEE.

\bibitem{Behera2022Dec}
S.~Behera and J.~R. Prathuri, ``Design of novel hardware architecture for fully homomorphic encryption algorithms in {FPGA} for real-time data in cloud computing,'' \emph{IEEE Access}, vol.~10, pp. 131\,406--131\,418, Dec. 2022.

\bibitem{Li2023Sep}
X.~Li, H.~Gao, J.~Zhang, S.~Yang, X.~Jin \emph{et~al.}, ``{GPU} accelerated full homomorphic encryption cryptosystem, library, and applications for {IoT} systems,'' \emph{IEEE IoT J.}, vol.~11, no.~4, pp. 6893--6903, Sep. 2023.

\bibitem{Matsumoto}
\emph{Speeding Up Encryption on {IoT} Devices Using Homomorphic Encryption}.\hskip 1em plus 0.5em minus 0.4em\relax IEEE.

\bibitem{Dar2020Apr}
Z.~Dar, A.~Ahmad, F.~A. Khan, F.~Zeshan, R.~Iqbal \emph{et~al.}, ``A context-aware encryption protocol suite for edge computing-based {IoT devices},'' \emph{J. Supercomput.}, vol.~76, no.~4, pp. 2548--2567, Apr. 2020.

\bibitem{Lu2017Mar}
R.~Lu, K.~Heung, A.~H. Lashkari, and A.~A. Ghorbani, ``A lightweight privacy-preserving data aggregation scheme for fog computing-enhanced {IoT},'' \emph{IEEE Access}, vol.~5, pp. 3302--3312, Mar. 2017.

\bibitem{Gupta2021May}
S.~Gupta, R.~Garg, N.~Gupta, W.~S. Alnumay, U.~Ghosh \emph{et~al.}, ``Energy-efficient dynamic homomorphic security scheme for fog computing in {IoT networks},'' \emph{Journal of Information Security and Applications}, vol.~58, p. 102768, May 2021.

\bibitem{Ding2020Jul}
Y.~Ding, G.~Wu, D.~Chen, N.~Zhang, L.~Gong \emph{et~al.}, ``{DeepEDN: A deep-learning-based image encryption and decryption network for internet of medical things},'' \emph{IEEE Internet of Things Journal}, vol.~8, no.~3, pp. 1504--1518, July 2020.

\bibitem{Saba2021Jun}
T.~Saba, K.~Haseeb, A.~A. Shah, A.~Rehman, U.~Tariq \emph{et~al.}, ``{A Machine-Learning-Based Approach for Autonomous IoT Security},'' \emph{IT Prof.}, vol.~23, no.~3, pp. 69--75, Jun. 2021.

\bibitem{Song2022Feb}
J.~Song, W.~Wang, T.~R. Gadekallu, J.~Cao, and Y.~Liu, ``{EPPDA: An Efficient Privacy-Preserving Data Aggregation Federated Learning Scheme},'' \emph{IEEE Trans. Network Sci. Eng.}, p.~1, Feb. 2022.

\bibitem{Jia2021Jun}
B.~Jia, X.~Zhang, J.~Liu, Y.~Zhang, K.~Huang \emph{et~al.}, ``{Blockchain-Enabled Federated Learning Data Protection Aggregation Scheme With Differential Privacy and Homomorphic Encryption in IIoT},'' \emph{IEEE Transactions on Industrial Informatics}, vol.~18, no.~6, pp. 4049--4058, Jun. 2021.

\bibitem{Tanveer2022Feb}
M.~Tanveer, A.~Alkhayyat, A.~Naushad, A.~U. Khan, N.~Kumar \emph{et~al.}, ``{RUAM-IoD: A Robust User Authentication Mechanism for the Internet of Drones},'' \emph{IEEE Access}, vol.~10, pp. 19\,836--19\,851, Feb. 2022.

\bibitem{Perazzone2021Jan}
J.~B. Perazzone, P.~L. Yu, B.~M. Sadler, and R.~S. Blum, ``{Artificial Noise-Aided MIMO Physical Layer Authentication With Imperfect CSI},'' \emph{IEEE Transactions on Information Forensics and Security}, vol.~16, pp. 2173--2185, Jan. 2021.

\bibitem{Manogaran2023Mar}
G.~Manogaran, B.~S. Rawal, V.~Saravanan, P.~M~K, Q.~Xin \emph{et~al.}, ``{Token-Based Authorization and Authentication for Secure Internet of Vehicles Communication},'' \emph{ACM Transactions on Internet Technology}, vol.~22, no.~4, pp. 1--20, Mar. 2023.

\bibitem{Cui2023Feb}
H.~Cui, X.~Yang, W.~Yang, B.~Qin, and X.~Yi, ``{Token-Based Biometric Enhanced Key Derivation for Authentication Over Wireless Networks},'' \emph{IEEE Transactions on Network Science and Engineering}, pp. 1--12, Feb. 2023.

\bibitem{Alsellami2021Mar}
B.~M. Alsellami and P.~D. Deshmukh, ``{The Recent Trends in Biometric Traits Authentication Based on Internet of Things (IoT)},'' in \emph{{2021 International Conference on Artificial Intelligence and Smart Systems (ICAIS)}}.\hskip 1em plus 0.5em minus 0.4em\relax IEEE, Mar. 2021, pp. 1359--1365.

\bibitem{Leng2020Apr}
J.~Leng, Z.~Lin, and P.~Wang, ``{Poster Abstract: An Implementation of an Internet of Things System for Smart Hospitals},'' in \emph{{2020 IEEE/ACM Fifth International Conference on Internet-of-Things Design and Implementation (IoTDI)}}.\hskip 1em plus 0.5em minus 0.4em\relax IEEE, Apr. 2020, pp. 254--255.

\bibitem{Sukumaran2022Nov}
R.~P. Sukumaran and S.~Benedict, ``{Authentication and Cryptography solutions for Industrial IoT - A Study},'' in \emph{{2022 Sixth International Conference on I-SMAC (IoT in Social, Mobile, Analytics and Cloud) (I-SMAC)}}.\hskip 1em plus 0.5em minus 0.4em\relax IEEE, Nov. 2022, pp. 76--81.

\bibitem{Gautam2021Jan}
A.~K. Gautam and R.~Kumar, ``{A comprehensive study on key management, authentication and trust management techniques in wireless sensor networks},'' \emph{SN Appl. Sci.}, vol.~3, no.~1, pp. 1--27, Jan. 2021.

\bibitem{Zhao2020Aug}
W.~Zhao, S.~Yang, and X.~Luo, ``{On Threat Analysis of IoT-Based Systems: A Survey},'' in \emph{{2020 IEEE International Conference on Smart Internet of Things (SmartIoT)}}.\hskip 1em plus 0.5em minus 0.4em\relax IEEE, Aug. 2020, pp. 205--212.

\bibitem{Ullah2022Apr}
Z.~Ullah, B.~Raza, H.~Shah, S.~Khan, and A.~Waheed, ``{Towards Blockchain-Based Secure Storage and Trusted Data Sharing Scheme for IoT Environment},'' \emph{IEEE Access}, vol.~10, pp. 36\,978--36\,994, Apr. 2022.

\bibitem{Oh2022Sep}
J.~Oh, J.~Lee, M.~Kim, Y.~Park, K.~Park \emph{et~al.}, ``{A Secure Data Sharing Based on Key Aggregate Searchable Encryption in Fog-Enabled IoT Environment},'' \emph{IEEE Trans. Network Sci. Eng.}, vol.~9, no.~6, pp. 4468--4481, Sep. 2022.

\bibitem{Rasori2022Feb}
M.~Rasori, M.~La~Manna, P.~Perazzo, and G.~Dini, ``{A Survey on Attribute-Based Encryption Schemes Suitable for the Internet of Things},'' \emph{IEEE Internet of Things Journal}, vol.~9, no.~11, pp. 8269--8290, Feb. 2022.

\bibitem{Namasudra2022Dec}
S.~Namasudra, ``{A secure cryptosystem using DNA cryptography and DNA steganography for the cloud-based IoT infrastructure},'' \emph{Computers and Electrical Engineering}, vol. 104, p. 108426, Dec. 2022.

\bibitem{John2020Aug}
A.~John, R.~C. Panicker, B.~Cardiff, Y.~Lian, and D.~John, ``{Binary Classifiers for Data Integrity Detection in Wearable IoT Edge Devices},'' \emph{IEEE Open J. Circuits Syst.}, vol.~1, pp. 88--99, Aug. 2020.

\bibitem{Wang2022Oct}
Y.~Wang and T.~Liao, ``{Data Integrity and Causation Analysis for Wearable Devices in 5G},'' in \emph{{2022 IEEE International Conference on E-health Networking, Application {\&} Services (HealthCom)}}.\hskip 1em plus 0.5em minus 0.4em\relax IEEE, Oct. 2022, pp. 142--148.

\bibitem{Sneha2021Aug}
S.~Sneha, A.~Panjwani, B.~Lade, J.~Randolph, and M.~Vickery, ``{Alleviating Challenges Related to FDA-Approved Medical Wearables Using Blockchain Technology},'' \emph{IT Prof.}, vol.~23, no.~4, pp. 21--27, Aug. 2021.

\bibitem{Douha2022Nov}
N.~Y.-R. Douha, M.~Bhuyan, S.~Kashihara, D.~Fall, Y.~Taenaka \emph{et~al.}, ``{A survey on blockchain, SDN and NFV for the smart-home security},'' \emph{Internet of Things}, vol.~20, p. 100588, Nov. 2022.

\bibitem{Tham2022Jan}
C.~Thammarat and C.~Techapanupreeda, ``{Secure Key Establishment Protocol for Smart Homes Based on Symmetric Cryptography},'' in \emph{{2022 International Conference on Information Networking (ICOIN)}}.\hskip 1em plus 0.5em minus 0.4em\relax IEEE, Jan. 2022, pp. 46--51.

\bibitem{Kai2021Nov}
A.~R. Kairaldeen, N.~F. Abdullah, A.~Abu-Samah, and R.~Nordin, ``{Data Integrity Time Optimization of a Blockchain IoT Smart Home Network Using Different Consensus and Hash Algorithms},'' \emph{Wireless Commun. Mobile Comput.}, vol. 2021, Nov. 2021.

\bibitem{Ammi2021May}
M.~Ammi, S.~Alarabi, and E.~Benkhelifa, ``{Customized blockchain-based architecture for secure smart home for lightweight IoT},'' \emph{Information Processing {\&} Management}, vol.~58, no.~3, p. 102482, May 2021.

\bibitem{Chen2022Sep}
C.~Chen, L.~Wang, Y.~Long, Y.~Luo, and K.~Chen, ``{A blockchain-based dynamic and traceable data integrity verification scheme for smart homes},'' \emph{J. Syst. Archit.}, vol. 130, p. 102677, Sep. 2022.

\bibitem{Jadav2022Apr}
N.~K. Jadav, R.~Gupta, M.~D. Alshehri, H.~Mankodiya, S.~Tanwar \emph{et~al.}, ``{Deep Learning and Onion Routing-Based Collaborative Intelligence Framework for Smart Homes Underlying 6G Networks},'' \emph{IEEE Transactions on Network and Service Management}, vol.~19, no.~3, pp. 3401--3412, Apr. 2022.

\bibitem{Zhang2021Jan}
Z.~Zhang, R.~Deng, D.~K.~Y. Yau, and P.~Chen, ``{Zero-Parameter-Information Data Integrity Attacks and Countermeasures in IoT-Based Smart Grid},'' \emph{IEEE Internet of Things Journal}, vol.~8, no.~8, pp. 6608--6623, Jan. 2021.

\bibitem{An2022Feb}
D.~An, F.~Zhang, Q.~Yang, and C.~Zhang, ``{Data Integrity Attack in Dynamic State Estimation of Smart Grid: Attack Model and Countermeasures},'' \emph{IEEE Transactions on Automation Science and Engineering}, vol.~19, no.~3, pp. 1631--1644, Feb. 2022.

\bibitem{Reda2022Jun}
H.~T. Reda, A.~Anwar, A.~Mahmood, and N.~Chilamkurti, ``{Data-driven Approach for State Prediction and Detection of False Data Injection Attacks in Smart Grid},'' \emph{J. Mod. Power Syst. Clean Energy}, vol.~11, no.~2, pp. 455--467, Jun. 2022.

\bibitem{Goyel2022Aug}
H.~Goyel and K.~S. Swarup, ``{Data Integrity Attack Detection Using Ensemble-Based Learning for Cyber{\textendash}Physical Power Systems},'' \emph{IEEE Transactions on Smart Grid}, vol.~14, no.~2, pp. 1198--1209, Aug. 2022.

\bibitem{Zhu2019Jun}
H.~Zhu, Y.~Yuan, Y.~Chen, Y.~Zha, W.~Xi \emph{et~al.}, ``{A Secure and Efficient Data Integrity Verification Scheme for Cloud-IoT Based on Short Signature},'' \emph{IEEE Access}, vol.~7, pp. 90\,036--90\,044, Jun. 2019.

\bibitem{Wang2019Nov}
H.~Wang and J.~Zhang, ``{Blockchain Based Data Integrity Verification for Large-Scale IoT Data},'' \emph{IEEE Access}, vol.~7, pp. 164\,996--165\,006, Nov. 2019.

\bibitem{Li2021Nov}
Y.~Li and F.~Zhang, ``{An Efficient Certificate-Based Data Integrity Auditing Protocol for Cloud-Assisted WBANs},'' \emph{IEEE Internet of Things Journal}, vol.~9, no.~13, pp. 11\,513--11\,523, Nov. 2021.

\bibitem{Sang2022Aug}
T.~Sang, P.~Zeng, and K.-K.~R. Choo, ``{Provable Multiple-Copy Integrity Auditing Scheme for Cloud-Based IoT},'' \emph{IEEE Syst. J.}, vol.~17, no.~1, pp. 224--233, Aug. 2022.

\bibitem{Han2022Oct}
H.~Han, S.~Fei, Z.~Yan, and X.~Zhou, ``{A survey on blockchain-based integrity auditing for cloud data},'' \emph{Digital Communications and Networks}, vol.~8, no.~5, pp. 591--603, Oct. 2022.

\bibitem{Ioannou2021Jul}
C.~Ioannou, A.~Charalambus, and V.~Vassiliou, ``{Decentralized Dedicated Intrusion Detection Security Agents for IoT Networks},'' in \emph{{2021 17th International Conference on Distributed Computing in Sensor Systems (DCOSS)}}.\hskip 1em plus 0.5em minus 0.4em\relax IEEE, Jul. 2021, pp. 414--419.

\bibitem{Abdel2021Feb}
M.~Abdel-Basset, H.~Hawash, R.~K. Chakrabortty, and M.~J. Ryan, ``{Semi-Supervised Spatiotemporal Deep Learning for Intrusions Detection in IoT Networks},'' \emph{IEEE Internet of Things Journal}, vol.~8, no.~15, pp. 12\,251--12\,265, Feb. 2021.

\bibitem{Roy2022Mar}
S.~D. Roy, S.~Debbarma, and A.~Iqbal, ``{A Decentralized Intrusion Detection System for Security of Generation Control},'' \emph{IEEE Internet of Things Journal}, vol.~9, no.~19, pp. 18\,924--18\,933, Mar. 2022.

\bibitem{Otoum2021Jun}
S.~Otoum, N.~Guizani, and H.~Mouftah, ``{Federated Reinforcement Learning-Supported IDS for IoT-steered Healthcare Systems},'' in \emph{{ICC 2021 - IEEE International Conference on Communications}}.\hskip 1em plus 0.5em minus 0.4em\relax IEEE, Jun. 2021, pp. 1--6.

\bibitem{Regan2022Jun}
C.~Regan, M.~Nasajpour, R.~M. Parizi, S.~Pouriyeh, A.~Dehghantanha \emph{et~al.}, ``{Federated IoT attack detection using decentralized edge data},'' \emph{Machine Learning with Applications}, vol.~8, p. 100263, Jun. 2022.

\bibitem{Wang2023May}
X.~Wang, Y.~Wang, Z.~Javaheri, L.~Almutairi, N.~Moghadamnejad \emph{et~al.}, ``{Federated deep learning for anomaly detection in the internet of things},'' \emph{Comput. Electr. Eng.}, vol. 108, p. 108651, May 2023.

\bibitem{Aouedi2022May}
O.~Aouedi, K.~Piamrat, G.~Muller, and K.~Singh, ``{Intrusion detection for Softwarized Networks with Semi-supervised Federated Learning},'' in \emph{{ICC 2022 - IEEE International Conference on Communications}}.\hskip 1em plus 0.5em minus 0.4em\relax IEEE, May 2022, pp. 5244--5249.

\bibitem{Bikos2021Sep}
A.~N. Bikos and S.~Kumar, ``{Reinforcement Learning-Based Anomaly Detection for Internet of Things Distributed Ledger Technology},'' in \emph{{2021 IEEE Symposium on Computers and Communications (ISCC)}}.\hskip 1em plus 0.5em minus 0.4em\relax IEEE, Sep. 2021, pp. 1--7.

\bibitem{Shingi2021Jul}
G.~Shingi, H.~Saglani, and P.~Jain, ``{Segmented Federated Learning for Adaptive Intrusion Detection System},'' \emph{arXiv}, Jul. 2021.

\bibitem{Liu2020May}
M.~Liu, L.~Ma, C.~Li, and R.~Li, ``{Fortified Network Security Perception: A Decentralized Multiagent Coordination Perspective},'' in \emph{{2020 IEEE 3rd International Conference on Electronics Technology (ICET)}}.\hskip 1em plus 0.5em minus 0.4em\relax IEEE, May 2020, pp. 746--750.

\bibitem{Putra2020Apr}
G.~D. Putra, V.~Dedeoglu, S.~S. Kanhere, and R.~Jurdak, ``{Poster Abstract: Towards Scalable and Trustworthy Decentralized Collaborative Intrusion Detection System for IoT},'' in \emph{{2020 IEEE/ACM Fifth International Conference on Internet-of-Things Design and Implementation (IoTDI)}}.\hskip 1em plus 0.5em minus 0.4em\relax IEEE, Apr. 2020, pp. 256--257.

\bibitem{Friha2023Apr}
O.~Friha, M.~A. Ferrag, M.~Benbouzid, T.~Berghout, B.~Kantarci \emph{et~al.}, ``{2DF-IDS: Decentralized and differentially private federated learning-based intrusion detection system for industrial IoT},'' \emph{Computers {\&} Security}, vol. 127, p. 103097, Apr. 2023.

\bibitem{Reddy2021Aug}
D.~K.~K. Reddy, H.~S. Behera, J.~Nayak, B.~Naik, U.~Ghosh \emph{et~al.}, ``{Exact greedy algorithm based split finding approach for intrusion detection in fog-enabled IoT environment},'' \emph{Journal of Information Security and Applications}, vol.~60, p. 102866, Aug. 2021.

\bibitem{Sahi2021Oct}
M.~Sahi, M.~Soni, and N.~Auluck, ``{An Intrusion Detection System on Fog Architecture},'' in \emph{{2021 IEEE 18th International Conference on Mobile Ad Hoc and Smart Systems (MASS)}}.\hskip 1em plus 0.5em minus 0.4em\relax IEEE, Oct. 2021, pp. 591--596.

\bibitem{Li2022May}
S.~Li, Y.~Lu, and J.~Li, ``{CAD-IDS: A Cooperative Adaptive Distributed Intrusion Detection System with Fog Computing},'' in \emph{{2022 IEEE 25th International Conference on Computer Supported Cooperative Work in Design (CSCWD)}}.\hskip 1em plus 0.5em minus 0.4em\relax IEEE, May 2022, pp. 635--640.

\bibitem{Abdel2020Sep}
M.~Abdel-Basset, V.~Chang, H.~Hawash, R.~K. Chakrabortty, and M.~Ryan, ``{Deep-IFS: Intrusion Detection Approach for Industrial Internet of Things Traffic in Fog Environment},'' \emph{IEEE Transactions on Industrial Informatics}, vol.~17, no.~11, pp. 7704--7715, Sep. 2020.

\bibitem{Jung2022May}
W.~Jung, Y.~Feng, S.~A. Khan, C.~Xin, D.~Zhao \emph{et~al.}, ``{DeepAuditor: Distributed Online Intrusion Detection System for IoT Devices via Power Side-channel Auditing},'' in \emph{{2022 21st ACM/IEEE International Conference on Information Processing in Sensor Networks (IPSN)}}.\hskip 1em plus 0.5em minus 0.4em\relax IEEE, May 2022, pp. 415--427.

\bibitem{Chiba2022Jan}
Z.~Chiba, N.~Abghour, K.~Moussaid, O.~Lifandali, and R.~Kinta, ``{A Deep Study of Novel Intrusion Detection Systems and Intrusion Prevention Systems for Internet of Things Networks},'' \emph{Procedia Comput. Sci.}, vol. 210, pp. 94--103, Jan. 2022.

\bibitem{Illy2022Aug}
P.~Illy, G.~Kaddoum, P.~F. de~Araujo-Filho, K.~Kaur, and S.~Garg, ``{A Hybrid Multistage DNN-Based Collaborative IDPS for High-Risk Smart Factory Networks},'' \emph{IEEE Transactions on Network and Service Management}, vol.~19, no.~4, pp. 4273--4283, Aug. 2022.

\bibitem{Jayalaxmi2022Sep}
P.~L.~S. Jayalaxmi, G.~Kumar, R.~Saha, M.~Conti, T.-h. Kim \emph{et~al.}, ``{DeBot: A deep learning-based model for bot detection in industrial internet-of-things},'' \emph{Computers and Electrical Engineering}, vol. 102, p. 108214, Sep. 2022.

\bibitem{Onah2021Dec}
J.~O. Onah, S.~M. Abdulhamid, M.~Abdullahi, I.~H. Hassan, and A.~Al-Ghusham, ``{Genetic Algorithm based feature selection and Na{\ifmmode\ddot{\imath}\else\"{\i}\fi}ve Bayes for anomaly detection in fog computing environment},'' \emph{Machine Learning with Applications}, vol.~6, p. 100156, Dec. 2021.

\bibitem{Ravi2020May}
N.~Ravi and S.~M. Shalinie, ``{Semisupervised-Learning-Based Security to Detect and Mitigate Intrusions in IoT Network},'' \emph{IEEE Internet of Things Journal}, vol.~7, no.~11, pp. 11\,041--11\,052, May 2020.

\bibitem{Hasan2022Nov}
H.~Hasan, G.~Ali, W.~Elmedany, and C.~Balakrishna, ``{Lightweight Encryption Algorithms for Internet of Things: A Review on Security and Performance Aspects},'' in \emph{{2022 International Conference on Innovation and Intelligence for Informatics, Computing, and Technologies (3ICT)}}.\hskip 1em plus 0.5em minus 0.4em\relax IEEE, Nov. 2022, pp. 239--244.

\bibitem{Gope2022Apr}
P.~Gope, P.~K. Sharma, and B.~Sikdar, ``{An Ultra-Lightweight Data-Aggregation Scheme with Deep Learning Security for Smart Grid},'' \emph{IEEE Wireless Commun.}, vol.~29, no.~2, pp. 30--36, Apr. 2022.

\bibitem{Kong2022Mar}
X.~Kong and Z.~Ge, ``{Deep PLS: A Lightweight Deep Learning Model for Interpretable and Efficient Data Analytics},'' \emph{IEEE Trans. Neural Networks Learn. Syst.}, pp. 1--15, Mar. 2022.

\bibitem{Kang2023Jan}
S.~Kang, Y.~Gao, J.~Jeong, S.-J. Park, J.-W. Kim \emph{et~al.}, ``{Generative adversarial networks for DNA storage channel simulator},'' \emph{IEEE Access}, vol.~11, pp. 3781--3793, Jan. 2023.

\bibitem{Ma2021May}
C.~Ma, J.~Li, M.~Ding, K.~Wei, W.~Chen \emph{et~al.}, ``{Federated Learning With Unreliable Clients: Performance Analysis and Mechanism Design},'' \emph{IEEE Internet of Things Journal}, vol.~8, no.~24, pp. 17\,308--17\,319, May 2021.

\bibitem{He2019May}
X.~He, S.~Alqahtani, R.~Gamble, and M.~Papa, ``Securing over-the-air {IoT} firmware updates using blockchain,'' in \emph{{COINS '19: Proceedings of the International Conference on Omni-Layer Intelligent Systems}}.\hskip 1em plus 0.5em minus 0.4em\relax New York, NY, USA: Association for Computing Machinery, May 2019, pp. 164--171.

\bibitem{SunZhichuang2023}
Z.~Sun, R.~Sun, C.~Liu, A.~R. Chowdhury, L.~Lu \emph{et~al.}, ``{ShadowNet: A} secure and efficient on-device model inference system for convolutional neural networks,'' in \emph{{2023 IEEE Symposium on Security and Privacy (SP)}}.\hskip 1em plus 0.5em minus 0.4em\relax IEEE, pp. 21--25.

\bibitem{LiuZiyu2023}
Z.~Liu, Y.~Luo., S.~Duan, T.~Zhou, and X.~Xu, ``{MirrorNet: A TEE}-friendly framework for secure on-device {DNN} inference,'' in \emph{{2023 IEEE/ACM International Conference on Computer Aided Design (ICCAD)}}.\hskip 1em plus 0.5em minus 0.4em\relax IEEE, pp. 2023--02.

\bibitem{yang2017survey}
Y.~Yang, L.~Wu, G.~Yin, L.~Li, and H.~Zhao, ``{A Survey on Security and Privacy Issues in Internet-of-Things},'' \emph{IEEE Internet of Things Journal}, vol.~4, no.~5, pp. 1250--1258, 2017.

\bibitem{schaar2010privacy}
P.~Schaar, ``{Privacy by Design},'' \emph{Identity in the Information Society}, vol.~3, no.~2, pp. 267--274, 2010.

\bibitem{roman2013features}
R.~Roman, J.~Zhou, and J.~Lopez, ``{On the Features and Challenges of Security and Privacy in Distributed Internet of Things},'' \emph{Computer Networks}, vol.~57, no.~10, pp. 2266--2279, 2013.

\bibitem{ziegeldorf2014privacy}
J.~H. Ziegeldorf, O.~G. Morchon, and K.~Wehrle, ``{Privacy in the Internet of Things: Threats and Challenges},'' \emph{Security and Communication Networks}, vol.~7, no.~12, pp. 2728--2742, 2014.

\bibitem{porambage2016quest}
P.~Porambage, M.~Ylianttila, C.~Schmitt, P.~Kumar, A.~Gurtov, and A.~V. Vasilakos, ``{The Quest for Privacy in the Internet of Things},'' \emph{IEEE Cloud Computing}, vol.~3, no.~2, pp. 36--45, 2016.

\bibitem{storm2015medjack}
D.~Storm, ``{MEDJACK: Hackers Hijacking Medical Devices to Create Backdoors in Hospital Networks},'' \emph{Computerworld}, vol.~8, p.~42, 2015.

\bibitem{weber2010internet}
R.~H. Weber, ``{Internet of Things--New Security and Privacy Challenges},'' \emph{Computer law \& security review}, vol.~26, no.~1, pp. 23--30, 2010.

\bibitem{zeadally2013towards}
S.~Zeadally, A.-S.~K. Pathan, C.~Alcaraz, and M.~Badra, ``{Towards Privacy Protection in Smart Grid},'' \emph{Wireless personal communications}, vol.~73, pp. 23--50, 2013.

\bibitem{miorandi2012internet}
D.~Miorandi, S.~Sicari, F.~De~Pellegrini, and I.~Chlamtac, ``{Internet of Things: Vision, Applications and Research Challenges},'' \emph{Ad hoc networks}, vol.~10, no.~7, pp. 1497--1516, 2012.

\bibitem{noll2014measurable}
J.~Noll, I.~Garitano, S.~Fayyad, {\~A}.~Erik, H.~Abie \emph{et~al.}, ``{Measurable Security, Privacy and Dependability in Smart Grids},'' \emph{Journal of Cyber Security and Mobility}, vol.~3, no.~4, pp. 371--398, 2014.

\bibitem{rahman2015secure}
M.~A. Rahman, M.~H. Manshaei, E.~Al-Shaer, and M.~Shehab, ``Secure and private data aggregation for energy consumption scheduling in smart grids,'' \emph{IEEE Transactions on Dependable and Secure Computing}, vol.~14, no.~2, pp. 221--234, 2015.

\bibitem{saputro2015privacy}
N.~Saputro, K.~Akkaya, and I.~Guvenc, ``{Privacy-aware Communication Protocol for Hybrid IEEE 802.11 s/LTE Smart Grid Architectures},'' in \emph{IEEE 40th Local Computer Networks Conference Workshops (LCN Workshops)}.\hskip 1em plus 0.5em minus 0.4em\relax IEEE, 2015, pp. 905--911.

\bibitem{borges2015efficient}
F.~Borges, F.~Volk, and M.~M{\"u}hlh{\"a}user, ``{Efficient, Verifiable, Secure, and Privacy-friendly Computations for the Smart Grid},'' in \emph{IEEE Power \& Energy Society Innovative Smart Grid Technologies Conference (ISGT)}.\hskip 1em plus 0.5em minus 0.4em\relax IEEE, 2015, pp. 1--5.

\bibitem{sutheerakul2017application}
C.~Sutheerakul, N.~Kronprasert, M.~Kaewmoracharoen, and P.~Pichayapan, ``{Application of Unmanned Aerial Vehicles to Pedestrian Traffic Monitoring and Management for Shopping Streets},'' \emph{Transportation research procedia}, vol.~25, pp. 1717--1734, 2017.

\bibitem{zhi2020security}
Y.~Zhi, Z.~Fu, X.~Sun, and J.~Yu, ``{Security and Privacy Issues of UAV: A Survey},'' \emph{Mobile Networks and Applications}, vol.~25, pp. 95--101, 2020.

\bibitem{ch2020security}
R.~Ch, G.~Srivastava, T.~R. Gadekallu, P.~K.~R. Maddikunta, and S.~Bhattacharya, ``{Security and Privacy of UAV Data Using Blockchain Technology},'' \emph{Journal of Information security and Applications}, vol.~55, p. 102670, 2020.

\bibitem{hasan2018proof}
H.~R. Hasan and K.~Salah, ``{Proof of Delivery of Digital Assets Using Blockchain and Smart Contracts},'' \emph{IEEE Access}, vol.~6, pp. 65\,439--65\,448, 2018.

\bibitem{pham2022energy}
Q.-V. Pham, M.~Le, T.~Huynh-The, Z.~Han, and W.-J. Hwang, ``Energy-efficient federated learning over {UAV}-enabled wireless powered communications,'' \emph{IEEE Transactions on Vehicular Technology}, vol.~71, no.~5, pp. 4977--4990, 2022.

\bibitem{rieke2020future}
N.~Rieke, J.~Hancox, W.~Li, F.~Milletari, H.~R. Roth, S.~Albarqouni, S.~Bakas, M.~N. Galtier, B.~A. Landman, K.~Maier-Hein \emph{et~al.}, ``{The Future of Digital Health With Federated Learning},'' \emph{NPJ digital medicine}, vol.~3, no.~1, pp. 1--7, 2020.

\bibitem{rocher2019estimating}
L.~Rocher, J.~M. Hendrickx, and Y.-A. De~Montjoye, ``{Estimating the Success of Re-identifications in Incomplete Datasets Using Generative Models},'' \emph{Nature communications}, vol.~10, no.~1, pp. 1--9, 2019.

\bibitem{elkhodr2012review}
M.~Elkhodr, S.~Shahrestani, and H.~Cheung, ``{A Review of Mobile Location Privacy in the Internet of Things},'' in \emph{Tenth International Conference on ICT and Knowledge Engineering}.\hskip 1em plus 0.5em minus 0.4em\relax IEEE, 2012, pp. 266--272.

\bibitem{li2009privacy}
N.~Li, N.~Zhang, S.~K. Das, and B.~Thuraisingham, ``{Privacy Preservation in Wireless Sensor Networks: A State-of-the-art Survey},'' \emph{Ad Hoc Networks}, vol.~7, no.~8, pp. 1501--1514, 2009.

\bibitem{sicari2015security}
S.~Sicari, A.~Rizzardi, L.~A. Grieco, and A.~Coen-Porisini, ``{Security, Privacy and Trust in Internet of Things: The Road Ahead},'' \emph{Computer Networks}, vol.~76, pp. 146--164, 2015.

\bibitem{kamat2005enhancing}
P.~Kamat, Y.~Zhang, W.~Trappe, and C.~Ozturk, ``{Enhancing Source-location Privacy in Sensor Network Routing},'' in \emph{25th IEEE international conference on distributed computing systems (ICDCS '05)}.\hskip 1em plus 0.5em minus 0.4em\relax IEEE, 2005, pp. 599--608.

\bibitem{pai2008transactional}
S.~Pai, M.~Meingast, T.~Roosta, S.~Bermudez, S.~B. Wicker, D.~K. Mulligan, and S.~Sastry, ``{Transactional Confidentiality in Sensor Networks},'' \emph{IEEE Security \& Privacy}, vol.~6, no.~4, pp. 28--35, 2008.

\bibitem{chow2010privacy}
C.-Y. Chow, M.~F. Mokbel, and T.~He, ``{A Privacy-preserving Location Monitoring System for Wireless Sensor Networks},'' \emph{IEEE Transactions on Mobile Computing}, vol.~10, no.~1, pp. 94--107, 2010.

\bibitem{wang2008source}
W.-P. Wang, L.~Chen, and J.-X. Wang, ``{A Source-location Privacy Protocol in WSN Based on Locational Angle},'' in \emph{IEEE International Conference on Communications (ICC '08)}.\hskip 1em plus 0.5em minus 0.4em\relax IEEE, 2008, pp. 1630--1634.

\bibitem{lightfoot2016star}
L.~Lightfoot, Y.~Li, and J.~Ren, ``{STaR: Design and Quantitative Measurement of Source-location Privacy for Wireless Sensor Networks},'' \emph{Security and Communication Networks}, vol.~9, no.~3, pp. 220--228, 2016.

\bibitem{xi2006preserving}
Y.~Xi, L.~Schwiebert, and W.~Shi, ``Preserving source location privacy in monitoring-based wireless sensor networks,'' in \emph{Proceedings 20th IEEE International Parallel \& Distributed Processing Symposium (IPDPS '06)}.\hskip 1em plus 0.5em minus 0.4em\relax IEEE, 2006, pp. 8--pp.

\bibitem{ross1996stochastic}
S.~M. Ross, J.~J. Kelly, R.~J. Sullivan, W.~J. Perry, D.~Mercer, R.~M. Davis, T.~D. Washburn, E.~V. Sager, J.~B. Boyce, and V.~L. Bristow, \emph{{Stochastic Processes}}.\hskip 1em plus 0.5em minus 0.4em\relax Wiley New York, 1996, vol.~2.

\bibitem{tan2014enhancing}
G.~Tan, W.~Li, and J.~Song, ``Enhancing source location privacy in energy-constrained wireless sensor networks,'' in \emph{Proceedings of International Conference on Computer Science and Information Technology (CSAIT '13)}.\hskip 1em plus 0.5em minus 0.4em\relax Springer, 2014, pp. 279--289.

\bibitem{tang2014cost}
D.~Tang, T.~Li, J.~Ren, and J.~Wu, ``{Cost-aware Secure Routing (CASER) Protocol Design for Wireless Sensor Networks},'' \emph{IEEE Transactions on Parallel and Distributed Systems}, vol.~26, no.~4, pp. 960--973, 2014.

\bibitem{yang2013towards}
Y.~Yang, M.~Shao, S.~Zhu, and G.~Cao, ``{Towards Statistically Strong Source Anonymity for Sensor Networks},'' \emph{ACM Transactions on Sensor Networks (TOSN)}, vol.~9, no.~3, pp. 1--23, 2013.

\bibitem{yang2008towards}
Y.~Yang, M.~Shao, S.~Zhu, B.~Urgaonkar, and G.~Cao, ``{Towards Event Source Unobservability With Minimum Network Traffic in Sensor Networks},'' in \emph{Proceedings of the first ACM conference on Wireless network security (WiSec)}, 2008, pp. 77--88.

\bibitem{han2018source}
G.~Han, L.~Zhou, H.~Wang, W.~Zhang, and S.~Chan, ``{A source location protection protocol based on dynamic routing in WSNs for the Social Internet of Things},'' \emph{Future Generation Computer Systems}, vol.~82, pp. 689--697, 2018.

\bibitem{mehta2007location}
K.~Mehta, D.~Liu, and M.~Wright, ``{Location Privacy in Sensor Networks Against a Global Eavesdropper},'' in \emph{IEEE International Conference on Network Protocols (ICNP '07)}.\hskip 1em plus 0.5em minus 0.4em\relax IEEE, 2007, pp. 314--323.

\bibitem{majeed2009tarp}
A.~Majeed, K.~Liu, and N.~Abu-Ghazaleh, ``{TARP: Timing Analysis Resilient Protocol for Wireless Sensor Networks},'' in \emph{IEEE International Conference on Wireless and Mobile Computing, Networking and Communications}.\hskip 1em plus 0.5em minus 0.4em\relax IEEE, 2009, pp. 85--90.

\bibitem{bushnag2016source}
A.~Bushnag, A.~Abuzneid, and A.~Mahmood, ``{Source Anonymity in WSNs Against Global Adversary Utilizing Low Transmission Rates With Delay Constraints},'' \emph{Sensors}, vol.~16, no.~7, p. 957, 2016.

\bibitem{miao2009cross}
G.~Miao, N.~Himayat, Y.~Li, and A.~Swami, ``{Cross-layer Optimization for Energy-efficient Wireless Communications: A Survey},'' \emph{Wireless Communications and Mobile Computing}, vol.~9, no.~4, pp. 529--542, 2009.

\bibitem{shao2009cross}
M.~Shao, W.~Hu, S.~Zhu, G.~Cao, S.~Krishnamurth, and T.~La~Porta, ``{Cross-Layer Enhanced Source Location Privacy in Sensor Networks},'' in \emph{6th Annual IEEE Communications Society Conference on Sensor, Mesh and Ad Hoc Communications and Networks (SECON 09)}.\hskip 1em plus 0.5em minus 0.4em\relax IEEE, 2009, pp. 1--9.

\bibitem{molisch2004ieee}
A.~F. Molisch, K.~Balakrishnan, C.-C. Chong, S.~Emami, A.~Fort, J.~Karedal, J.~Kunisch, H.~Schantz, U.~Schuster, and K.~Siwiak, ``{IEEE 802.15.4: A Channel Model-final Report},'' \emph{IEEE P802}, vol.~15, no.~04, p. 0662, 2004.

\bibitem{dutta2010defending}
N.~Dutta, A.~Saxena, and S.~Chellappan, ``{Defending Wireless Sensor Networks Against Adversarial Localization},'' in \emph{Eleventh International Conference on Mobile Data Management (MDM 10)}.\hskip 1em plus 0.5em minus 0.4em\relax IEEE, 2010, pp. 336--341.

\bibitem{tavli2010mitigation}
B.~Tavli, M.~M. Ozciloglu, and K.~Bicakci, ``{Mitigation of Compromising Privacy by Transmission Range Control in Wireless Sensor Networks},'' \emph{IEEE Communications Letters}, vol.~14, no.~12, pp. 1104--1106, 2010.

\bibitem{rana2012new}
S.~S. Rana and N.~H. Vaidya, ``A new ‘direction’for source location privacy in wireless sensor networks','' in \emph{IEEE Global Communications Conference (GLOBECOM '12)}.\hskip 1em plus 0.5em minus 0.4em\relax IEEE, 2012, pp. 342--347.

\bibitem{das2016bypassing}
A.~Das and L.~Moharana, ``{Bypassing Using Directional Transceivers: A Design for Anti-tracking Source Location Privacy Protection in WSNs},'' in \emph{Second International Conference on Research in Computational Intelligence and Communication Networks (ICRCICN '16)}.\hskip 1em plus 0.5em minus 0.4em\relax IEEE, 2016, pp. 39--44.

\bibitem{deng2006decorrelating}
J.~Deng, R.~Han, and S.~Mishra, ``{Decorrelating Wireless Sensor Network Traffic to Inhibit Traffic Analysis Attack}s,'' \emph{Pervasive and Mobile Computing}, vol.~2, no.~2, pp. 159--186, 2006.

\bibitem{syverson2004tor}
P.~Syverson, R.~Dingledine, and N.~Mathewson, ``{Tor: The Secondgeneration Onion Router},'' in \emph{Proceedings of the 13th USENIX Security Symposium}, 2004, pp. 303--320.

\bibitem{deng2005countermeasures}
J.~Deng, R.~Han, and S.~Mishra, ``{Countermeasures Against Traffic Analysis Attacks in Wireless Sensor Networks},'' in \emph{First International Conference on Security and Privacy for Emerging Areas in Communications Networks (SECURECOMM'05)}.\hskip 1em plus 0.5em minus 0.4em\relax IEEE, 2005, pp. 113--126.

\bibitem{luo2010location}
X.~Luo, X.~Ji, and M.-S. Park, ``{Location Privacy Against Traffic Analysis Attacks in Wireless Sensor Networks},'' in \emph{International Conference on Information Science and Applications (ICISA '10)}.\hskip 1em plus 0.5em minus 0.4em\relax IEEE, 2010, pp. 1--6.

\bibitem{deng2004intrusion}
J.~Deng, R.~Han, and S.~Mishra, ``{Intrusion Tolerance and Anti-traffic Analysis Strategies for Wireless Sensor Networks},'' in \emph{International Conference on Dependable Systems and Networks (DSN 04)}.\hskip 1em plus 0.5em minus 0.4em\relax IEEE, 2004, pp. 637--646.

\bibitem{carlini2019secret}
N.~Carlini, C.~Liu, {\'U}.~Erlingsson, J.~Kos, and D.~Song, ``{The Secret Sharer: Evaluating and Testing Unintended Memorization in Neural Networks},'' in \emph{28th {USENIX} Security Symposium (Security '19)}.\hskip 1em plus 0.5em minus 0.4em\relax {USENIX} Association, 2019, pp. 267--284.

\bibitem{cohen2022attacks}
A.~Cohen, ``{Attacks on Deidentification's Defenses},'' in \emph{31st {USENIX} Security Symposium (Security '22)}.\hskip 1em plus 0.5em minus 0.4em\relax {USENIX} Association, 2022, pp. 1469--1486.

\bibitem{aggarwal2005k}
C.~C. Aggarwal, ``{On k-anonymity and the Curse of Dimensionality},'' in \emph{VLDB}, vol.~5, 2005, pp. 901--909.

\bibitem{cao2010castle}
J.~Cao, B.~Carminati, E.~Ferrari, and K.-L. Tan, ``{Castle: Continuously Anonymizing Data Streams},'' \emph{IEEE Transactions on Dependable and Secure Computing}, vol.~8, no.~3, pp. 337--352, 2010.

\bibitem{dwork2006differential}
C.~Dwork, ``{Differential Privacy},'' in \emph{Automata, Languages and Programming: 33rd International Colloquium, ICALP 2006, Venice, Italy, July 10-14, Proceedings, Part II 33}.\hskip 1em plus 0.5em minus 0.4em\relax Springer, 2006, pp. 1--12.

\bibitem{mironov2017renyi}
I.~Mironov, ``{R{\'e}nyi Differential Privacy},'' in \emph{IEEE 30th computer security foundations symposium (CSF)}.\hskip 1em plus 0.5em minus 0.4em\relax IEEE, 2017, pp. 263--275.

\bibitem{poddar2020visor}
R.~Poddar, G.~Ananthanarayanan, S.~Setty, S.~Volos, and R.~A. Popa, ``{Visor: Privacy-preserving Video Analytics as a Cloud Service},'' in \emph{Proceedings of the 29th USENIX Conference on Security Symposium}, 2020, pp. 1039--1056.

\bibitem{shokri2017membership}
R.~Shokri, M.~Stronati, C.~Song, and V.~Shmatikov, ``{Membership Inference Attacks Against Machine Learning Models},'' in \emph{38th IEEE symposium on security and privacy (SP)}.\hskip 1em plus 0.5em minus 0.4em\relax IEEE, 2017, pp. 3--18.

\bibitem{frome2009large}
A.~Frome, G.~Cheung, A.~Abdulkader, M.~Zennaro \emph{et~al.}, ``{Large-scale Privacy Protection in Google Street View},'' in \emph{IEEE 12th international conference on computer vision (ICCV '09)}.\hskip 1em plus 0.5em minus 0.4em\relax IEEE, 2009, pp. 2373--2380.

\bibitem{sun2018natural}
Q.~Sun, L.~Ma, S.~J. Oh, L.~Van~Gool, B.~Schiele, and M.~Fritz, ``{Natural and Effective Obfuscation by Head Inpainting},'' in \emph{Proceedings of the IEEE Conference on Computer Vision and Pattern Recognition (CVPR '18)}.\hskip 1em plus 0.5em minus 0.4em\relax IEEE, 2018, pp. 5050--5059.

\bibitem{li2021differentially}
T.~Li and C.~Clifton, ``{Differentially Private Imaging via Latent Space Manipulation},'' \emph{arXiv preprint arXiv:2103.05472}, 2021.

\bibitem{senior2005enabling}
A.~Senior, S.~Pankanti, A.~Hampapur, L.~Brown, Y.-L. Tian, A.~Ekin, J.~Connell, C.~F. Shu, and M.~Lu, ``{Enabling Video Privacy Through Computer Vision},'' \emph{IEEE Security \& Privacy}, vol.~3, no.~3, pp. 50--57, 2005.

\bibitem{saini2014w}
M.~Saini, P.~K. Atrey, S.~Mehrotra, and M.~Kankanhalli, ``{W^3-privacy: Understanding What, When, and Where Inference Channels in Multi-camera Surveillance Video},'' \emph{Multimedia Tools and Applications}, vol.~68, pp. 135--158, 2014.

\bibitem{wang2020videodp}
H.~Wang, S.~Xie, and Y.~Hong, ``{VideoDP: A Flexible Platform for Video Analytics With Differential Privacy},'' \emph{Proceedings on Privacy Enhancing Technologies}, vol. 2020, no.~4, pp. 277--296, 2020.

\bibitem{weggenmann2018syntf}
B.~Weggenmann and F.~Kerschbaum, ``{Syntf: Synthetic and Differentially Private Term Frequency Vectors for Privacy-preserving Text Mining},'' in \emph{The 41st International ACM SIGIR Conference on Research \& Development in Information Retrieval (SIGIR '18)}, 2018, pp. 305--314.

\bibitem{fernandes2019generalised}
N.~Fernandes, M.~Dras, and A.~McIver, ``{Generalised Differential Privacy for Text Document Processing},'' in \emph{8th International Conference on Principles of Security and Trust (POST 19)}.\hskip 1em plus 0.5em minus 0.4em\relax Springer International Publishing, 2019, pp. 123--148.

\bibitem{feyisetan2020privacy}
O.~Feyisetan, B.~Balle, T.~Drake, and T.~Diethe, ``{Privacy-and Utility-preserving Textual Analysis via Calibrated Multivariate Perturbations},'' in \emph{Proceedings of the 13th ACM International Conference on Web Search and Data Mining (WSDM '20)}, 2020, pp. 178--186.

\bibitem{feyisetan2019leveraging}
O.~Feyisetan, T.~Diethe, and T.~Drake, ``{Leveraging Hierarchical Representations for Preserving Privacy and Utility in Text},'' in \emph{2019 IEEE International Conference on Data Mining (ICDM '19)}.\hskip 1em plus 0.5em minus 0.4em\relax IEEE, 2019, pp. 210--219.

\bibitem{zhao2022survey}
Y.~Zhao and J.~Chen, ``{A Survey on Differential Privacy for Unstructured Data Content},'' \emph{ACM Computing Surveys (CSUR)}, vol.~54, no. 10s, pp. 1--28, 2022.

\bibitem{hill2016effectiveness}
S.~Hill, Z.~Zhou, L.~K. Saul, and H.~Shacham, ``{On the (In) effectiveness of Mosaicing and Blurring as Tools for Document Redaction},'' \emph{Proceedings on Privacy Enhancing Technologies}, vol. 2016, no.~4, pp. 403--417, 2016.

\bibitem{hamm2017minimax}
J.~Hamm, ``{Minimax Filter: Learning to Preserve Privacy From Inference Attacks},'' \emph{The Journal of Machine Learning Research}, vol.~18, no.~1, pp. 4704--4734, 2017.

\bibitem{mcmahan2017learning}
H.~B. McMahan, D.~Ramage, K.~Talwar, and L.~Zhang, ``{Learning Differentially Private Recurrent Language Models},'' \emph{arXiv preprint arXiv:1710.06963}, 2017.

\bibitem{wei2020federated}
K.~Wei, J.~Li, M.~Ding, C.~Ma, H.~H. Yang, F.~Farokhi, S.~Jin, T.~Q. Quek, and H.~V. Poor, ``{Federated Learning With Differential Privacy: Algorithms and Performance Analysis},'' \emph{IEEE Transactions on Information Forensics and Security}, vol.~15, pp. 3454--3469, 2020.

\bibitem{yang2023survey}
L.~Yang, D.~Chai, J.~Zhang, Y.~Jin, L.~Wang, H.~Liu, H.~Tian, Q.~Xu, and K.~Chen, ``{A Survey on Vertical Federated Learning: From a Layered Perspective},'' \emph{arXiv preprint arXiv:2304.01829}, 2023.

\bibitem{ye2022feature}
P.~Ye, Z.~Jiang, W.~Wang, B.~Li, and B.~Li, ``{Feature Reconstruction Attacks and Countermeasures of DNN training in Vertical Federated Learning},'' \emph{arXiv preprint arXiv:2210.06771}, 2022.

\bibitem{li2019differentially}
J.~Li, M.~Khodak, S.~Caldas, and A.~Talwalkar, ``{Differentially Private Meta-Learning},'' in \emph{International Conference on Learning Representations (ICLR '20)}, 2020.

\bibitem{agarwal2018cpsgd}
N.~Agarwal, A.~T. Suresh, F.~X.~X. Yu, S.~Kumar, and B.~McMahan, ``{cpSGD: Communication-efficient and differentially-private distributed SGD},'' \emph{Advances in Neural Information Processing Systems (NIPS '18)}, vol.~31, pp. 7564--7575, 2018.

\bibitem{qu2020decentralized}
Y.~Qu, L.~Gao, T.~H. Luan, Y.~Xiang, S.~Yu, B.~Li, and G.~Zheng, ``{Decentralized Privacy Using Blockchain-enabled Federated Learning in Fog Computing},'' \emph{IEEE Internet of Things Journal}, vol.~7, no.~6, pp. 5171--5183, 2020.

\bibitem{gubbi2013internet}
J.~Gubbi, R.~Buyya, S.~Marusic, and M.~Palaniswami, ``{Internet of Things (IoT): A Vision, Architectural Elements, and Future Directions},'' \emph{Future generation computer systems}, vol.~29, no.~7, pp. 1645--1660, 2013.

\bibitem{kotamsetty2016adaptive}
R.~Kotamsetty and M.~Govindarasu, ``{Adaptive Latency-aware Query Processing on Encrypted Data for the Internet of Things},'' in \emph{2016 25th International Conference on Computer Communication and Networks (ICCCN)}.\hskip 1em plus 0.5em minus 0.4em\relax IEEE, 2016, pp. 1--7.

\bibitem{beunardeau2016fully}
M.~Beunardeau, A.~Connolly, R.~Geraud, and D.~Naccache, ``{Fully Homomorphic Encryption: Computations With a Blindfold},'' \emph{IEEE Security \& Privacy}, vol.~14, no.~1, pp. 63--67, 2016.

\bibitem{natarajan2021seal}
D.~Natarajan and W.~Dai, ``{Seal-embedded: A Homomorphic Encryption Library for the Internet of Things},'' \emph{IACR Transactions on Cryptographic Hardware and Embedded Systems}, pp. 756--779, 2021.

\bibitem{li2021lightweight}
S.~Li, S.~Zhao, G.~Min, L.~Qi, and G.~Liu, ``{Lightweight Privacy-preserving Scheme Using Homomorphic Encryption in Industrial Internet of Things},'' \emph{IEEE Internet of Things Journal}, vol.~9, no.~16, pp. 14\,542--14\,550, 2021.

\bibitem{gilad2016cryptonets}
R.~Gilad-Bachrach, N.~Dowlin, K.~Laine, K.~Lauter, M.~Naehrig, and J.~Wernsing, ``{CryptoNets: Applying Neural Networks to Encrypted Data with High Throughput and Accuracy},'' in \emph{International conference on machine learning (ICML '16)}.\hskip 1em plus 0.5em minus 0.4em\relax PMLR, 2016, pp. 201--210.

\bibitem{lu2018new}
R.~Lu, ``{A New Communication-efficient Privacy-preserving Range Query Scheme in Fog-enhanced Iot},'' \emph{IEEE Internet of Things Journal}, vol.~6, no.~2, pp. 2497--2505, 2018.

\bibitem{song2022eppda}
J.~Song, W.~Wang, T.~R. Gadekallu, J.~Cao, and Y.~Liu, ``{Eppda: an Efficient Privacy-preserving Data Aggregation Federated Learning Scheme},'' \emph{IEEE Transactions on Network Science and Engineering}, 2022.

\bibitem{hao2020privacy}
M.~Hao, H.~Li, G.~Xu, Z.~Liu, and Z.~Chen, ``{Privacy-aware and Resource-saving Collaborative Learning for Healthcare in Cloud Computing},'' in \emph{ICC 2020- IEEE International Conference on Communications (ICC)}.\hskip 1em plus 0.5em minus 0.4em\relax IEEE, 2020, pp. 1--6.

\bibitem{zhao2019secure}
C.~Zhao, S.~Zhao, M.~Zhao, Z.~Chen, C.-Z. Gao, H.~Li, and Y.-a. Tan, ``{Secure Multi-party Computation: Theory, Practice and Applications},'' \emph{Information Sciences}, vol. 476, pp. 357--372, 2019.

\bibitem{liu2017oblivious}
J.~Liu, M.~Juuti, Y.~Lu, and N.~Asokan, ``{Oblivious Neural Network Predictions via MiniONN Transformations},'' in \emph{Proceedings of the 2017 ACM SIGSAC conference on computer and communications security (CCS '17)}.\hskip 1em plus 0.5em minus 0.4em\relax ACM, 2017, pp. 619--631.

\bibitem{rouhani2018deepsecure}
B.~D. Rouhani, M.~S. Riazi, and F.~Koushanfar, ``{Deepsecure: scalable provably-secure deep learning},'' in \emph{Proceedings of the 55th annual Design Automation Conference (DAC '18)}.\hskip 1em plus 0.5em minus 0.4em\relax ACM, 2018, pp. 1--6.

\bibitem{kamara2012salus}
S.~Kamara, P.~Mohassel, and B.~Riva, ``{Salus: A System for Server-aided Secure Function Evaluation},'' in \emph{Proceedings of the ACM conference on Computer and Communications Security (CCS '12)}.\hskip 1em plus 0.5em minus 0.4em\relax ACM, 2012, pp. 797--808.

\bibitem{mukherjee2016two}
P.~Mukherjee and D.~Wichs, ``{Two Round Multiparty Computation via Multi-Key FHE},'' in \emph{Advances in Cryptology--EUROCRYPT 2016: 35th Annual International Conference on the Theory and Applications of Cryptographic Techniques}.\hskip 1em plus 0.5em minus 0.4em\relax Springer, 2016, pp. 735--763.

\bibitem{peter2013efficiently}
A.~Peter, E.~Tews, and S.~Katzenbeisser, ``{Efficiently Outsourcing Multiparty Computation Under Multiple Keys},'' \emph{IEEE Transactions on Information Forensics and Security}, vol.~8, no.~12, pp. 2046--2058, 2013.

\bibitem{asharov2012multiparty}
G.~Asharov, A.~Jain, A.~L{\'o}pez-Alt, E.~Tromer, V.~Vaikuntanathan, and D.~Wichs, ``{Multiparty Computation With Low Communication, Computation and Interaction via Threshold FHE},'' in \emph{Advances in Cryptology--EUROCRYPT 2012: 31st Annual International Conference on the Theory and Applications of Cryptographic Techniques}.\hskip 1em plus 0.5em minus 0.4em\relax Springer, 2012, pp. 483--501.

\bibitem{carter2016secure}
H.~Carter, B.~Mood, P.~Traynor, and K.~Butler, ``{Secure Outsourced Garbled Circuit Evaluation for Mobile Devices},'' \emph{Journal of Computer Security}, vol.~24, no.~2, pp. 137--180, 2016.

\bibitem{bost2014machine}
\BIBentryALTinterwordspacing
R.~Bost, R.~A. Popa, S.~Tu, and S.~Goldwasser, ``{Machine Learning Classification over Encrypted Data},'' Cryptology ePrint Archive, Paper 2014/331, 2014. [Online]. Available: \url{https://eprint.iacr.org/2014/331}
\BIBentrySTDinterwordspacing

\bibitem{knott2021crypten}
B.~Knott, S.~Venkataraman, A.~Hannun, S.~Sengupta, M.~Ibrahim, and L.~van~der Maaten, ``{Crypten: Secure Multi-party Computation Meets Machine Learning},'' \emph{Advances in Neural Information Processing Systems (NeurIPS '21)}, vol.~34, pp. 4961--4973, 2021.

\bibitem{mugunthan2019smpai}
V.~Mugunthan, A.~Polychroniadou, D.~Byrd, and T.~H. Balch, ``{Smpai: Secure Multi-party Computation for Federated Learning},'' in \emph{Proceedings of the NeurIPS 2019 Workshop on Robust AI in Financial Services}, 2019.

\bibitem{jayaraman2018distributed}
B.~Jayaraman, L.~Wang, D.~Evans, and Q.~Gu, ``{Distributed Learning Without Distress: Privacy-preserving Empirical Risk Minimization},'' \emph{Advances in Neural Information Processing Systems (NeurIPS '18)}, vol.~31, 2018.

\bibitem{bindschaedler2017achieving}
V.~Bindschaedler, S.~Rane, A.~E. Brito, V.~Rao, and E.~Uzun, ``{Achieving Differential Privacy in Secure Multiparty Data Aggregation Protocols on Star Networks},'' in \emph{Proceedings of the Seventh ACM on Conference on Data and Application Security and Privacy (CODASPY '17)}.\hskip 1em plus 0.5em minus 0.4em\relax ACM, 2017, pp. 115--125.

\bibitem{byrd2020differentially}
D.~Byrd and A.~Polychroniadou, ``{Differentially Private Secure Multi-Party Computation for Federated Learning in Financial Applications},'' in \emph{Proceedings of the First ACM International Conference on AI in Finance (ICAIF '20)}.\hskip 1em plus 0.5em minus 0.4em\relax ACM, 2020, pp. 1--9.

\bibitem{riazi2018chameleon}
M.~S. Riazi, C.~Weinert, O.~Tkachenko, E.~M. Songhori, T.~Schneider, and F.~Koushanfar, ``{Chameleon: A Hybrid Secure Computation Framework for Machine Learning Applications},'' in \emph{Proceedings of the 2018 on Asia Conference on Computer and Communications Security (ASIACCS '18)}.\hskip 1em plus 0.5em minus 0.4em\relax ACM, 2018, pp. 707--721.

\bibitem{juvekar2018gazelle}
C.~Juvekar, V.~Vaikuntanathan, and A.~Chandrakasan, ``{GAZELLE: A Low Latency Framework for Secure Neural Network Inference},'' in \emph{27th {USENIX} Security Symposium ({USENIX} Security '18)}, 2018, pp. 1651--1669.

\bibitem{kanagavelu2020two}
R.~Kanagavelu, Z.~Li, J.~Samsudin, Y.~Yang, F.~Yang, R.~S.~M. Goh, M.~Cheah, P.~Wiwatphonthana, K.~Akkarajitsakul, and S.~Wang, ``{Two-Phase Multi-Party Computation Enabled Privacy-Preserving Federated Learning},'' in \emph{20th IEEE/ACM International Symposium on Cluster, Cloud and Internet Computing (CCGrid '20)}.\hskip 1em plus 0.5em minus 0.4em\relax IEEE, 2020, pp. 410--419.

\bibitem{sotthiwat2021partially}
E.~Sotthiwat, L.~Zhen, Z.~Li, and C.~Zhang, ``{Partially Encrypted Multi-Party Computation for Federated Learning},'' in \emph{21st IEEE/ACM International Symposium on Cluster, Cloud and Internet Computing (CCGrid '21)}.\hskip 1em plus 0.5em minus 0.4em\relax IEEE, 2021, pp. 828--835.

\bibitem{mohassel2017secureml}
P.~Mohassel and Y.~Zhang, ``{SecureML: A System for Scalable Privacy-preserving Machine Learning},'' in \emph{2017 IEEE symposium on security and privacy (SP)}.\hskip 1em plus 0.5em minus 0.4em\relax IEEE, 2017, pp. 19--38.

\bibitem{sebastiani2002machine}
F.~Sebastiani, ``Machine learning in automated text categorization,'' \emph{ACM computing surveys (CSUR)}, vol.~34, no.~1, pp. 1--47, 2002.

\bibitem{manek2016detection}
A.~S. Manek, P.~D. Shenoy, M.~C. Mohan, and K.~Venugopal, ``Detection of fraudulent and malicious websites by analysing user reviews for online shopping websites,'' \emph{International Journal of Knowledge and Web Intelligence}, vol.~5, no.~3, pp. 171--189, 2016.

\bibitem{fu2019keeping}
H.~Fu, Z.~Zheng, S.~Zhu, and P.~Mohapatra, ``Keeping context in mind: Automating mobile app access control with user interface inspection,'' in \emph{IEEE INFOCOM 2019-IEEE Conference on Computer Communications}.\hskip 1em plus 0.5em minus 0.4em\relax IEEE, 2019, pp. 2089--2097.

\bibitem{avdiienko2015mining}
V.~Avdiienko, K.~Kuznetsov, A.~Gorla, A.~Zeller, S.~Arzt, S.~Rasthofer, and E.~Bodden, ``Mining apps for abnormal usage of sensitive data,'' in \emph{IEEE/ACM 37th IEEE international conference on software engineering (ICSE '15)}, vol.~1.\hskip 1em plus 0.5em minus 0.4em\relax IEEE, 2015, pp. 426--436.

\bibitem{squicciarini2017toward}
A.~Squicciarini, C.~Caragea, and R.~Balakavi, ``Toward automated online photo privacy,'' \emph{ACM Transactions on the Web (TWEB)}, vol.~11, no.~1, pp. 1--29, 2017.

\bibitem{yu2016iprivacy}
J.~Yu, B.~Zhang, Z.~Kuang, D.~Lin, and J.~Fan, ``iprivacy: image privacy protection by identifying sensitive objects via deep multi-task learning,'' \emph{IEEE Transactions on Information Forensics and Security}, vol.~12, no.~5, pp. 1005--1016, 2016.

\bibitem{hasan2020automatically}
R.~Hasan, D.~Crandall, M.~Fritz, and A.~Kapadia, ``Automatically detecting bystanders in photos to reduce privacy risks,'' in \emph{IEEE Symposium on Security and Privacy (SP)}.\hskip 1em plus 0.5em minus 0.4em\relax IEEE, 2020, pp. 318--335.

\bibitem{lebanoff2018automatic}
L.~Lebanoff and F.~Liu, ``Automatic detection of vague words and sentences in privacy policies,'' in \emph{Proceedings of the 2018 Conference on Empirical Methods in Natural Language Processing}, 2018, pp. 3508--3517.

\bibitem{wijesekera2015android}
P.~Wijesekera, A.~Baokar, A.~Hosseini, S.~Egelman, D.~Wagner, and K.~Beznosov, ``Android permissions remystified: A field study on contextual integrity,'' in \emph{24th {USENIX} Security Symposium ({USENIX} Security 15)}, 2015, pp. 499--514.

\bibitem{olejnik2017smarper}
K.~Olejnik, I.~Dacosta, J.~S. Machado, K.~Huguenin, M.~E. Khan, and J.-P. Hubaux, ``Smarper: Context-aware and automatic runtime-permissions for mobile devices,'' in \emph{IEEE Symposium on Security and Privacy (SP)}.\hskip 1em plus 0.5em minus 0.4em\relax IEEE, 2017, pp. 1058--1076.

\bibitem{das2018personalized}
A.~Das, M.~Degeling, D.~Smullen, and N.~Sadeh, ``Personalized privacy assistants for the internet of things: Providing users with notice and choice,'' \emph{IEEE Pervasive Computing}, vol.~17, no.~3, pp. 35--46, 2018.

\bibitem{orekondy2017towards}
T.~Orekondy, B.~Schiele, and M.~Fritz, ``Towards a visual privacy advisor: Understanding and predicting privacy risks in images,'' in \emph{Proceedings of the IEEE International Conference on Computer Vision (ICCV '17)}.\hskip 1em plus 0.5em minus 0.4em\relax IEEE, 2017, pp. 3686--3695.

\bibitem{acs2018differentially}
G.~Acs, L.~Melis, C.~Castelluccia, and E.~De~Cristofaro, ``Differentially private mixture of generative neural networks,'' \emph{IEEE Transactions on Knowledge and Data Engineering}, vol.~31, no.~6, pp. 1109--1121, 2018.

\bibitem{denton2015deep}
E.~L. Denton, S.~Chintala, R.~Fergus \emph{et~al.}, ``Deep generative image models using a laplacian pyramid of adversarial networks,'' \emph{Advances in neural information processing systems (NIPS '15)}, vol.~28, 2015.

\bibitem{choi2017generating}
E.~Choi, S.~Biswal, B.~Malin, J.~Duke, W.~F. Stewart, and J.~Sun, ``Generating multi-label discrete patient records using generative adversarial networks,'' in \emph{Machine learning for healthcare conference}.\hskip 1em plus 0.5em minus 0.4em\relax PMLR, 2017, pp. 286--305.

\bibitem{triastcyn2018generating}
A.~Triastcyn and B.~Faltings, ``Generating artificial data for private deep learning,'' \emph{arXiv preprint arXiv:1803.03148}, 2018.

\bibitem{park2018data}
N.~Park, M.~Mohammadi, K.~Gorde, S.~Jajodia, H.~Park, and Y.~Kim, ``Data synthesis based on generative adversarial networks,'' \emph{Proceedings of the VLDB Endowment}, vol.~11, no.~10, pp. 1071--1083, 2018.

\bibitem{joon2015person}
S.~Joon~Oh, R.~Benenson, M.~Fritz, and B.~Schiele, ``{Person Recognition in Personal Photo Collections},'' in \emph{Proceedings of the IEEE international conference on computer vision (CVPR '15)}.\hskip 1em plus 0.5em minus 0.4em\relax IEEE, 2015, pp. 3862--3870.

\bibitem{sun2018face}
X.~Sun, P.~Wu, and S.~C. Hoi, ``Face detection using deep learning: An improved faster rcnn approach,'' \emph{Neurocomputing}, vol. 299, pp. 42--50, 2018.

\bibitem{mcpherson2016defeating}
R.~McPherson, R.~Shokri, and V.~Shmatikov, ``Defeating image obfuscation with deep learning,'' \emph{arXiv preprint arXiv:1609.00408}, 2016.

\bibitem{oh2016faceless}
S.~J. Oh, R.~Benenson, M.~Fritz, and B.~Schiele, ``Faceless person recognition: Privacy implications in social media,'' in \emph{Proceedings of the 14th European Conference (ECCV '16)}.\hskip 1em plus 0.5em minus 0.4em\relax Springer, 2016, pp. 19--35.

\bibitem{cheng2010you}
Z.~Cheng, J.~Caverlee, and K.~Lee, ``You are where you tweet: a content-based approach to geo-locating twitter users,'' in \emph{Proceedings of the 19th ACM international conference on Information and knowledge management (CIKM '10)}.\hskip 1em plus 0.5em minus 0.4em\relax ACM, 2010, pp. 759--768.

\bibitem{rodriguez2017age}
P.~Rodr{\'\i}guez, G.~Cucurull, J.~M. Gonfaus, F.~X. Roca, and J.~Gonzalez, ``Age and gender recognition in the wild with deep attention,'' \emph{Pattern Recognition}, vol.~72, pp. 563--571, 2017.

\bibitem{zhou2012principal}
W.~Zhou, H.~Li, Y.~Lu, and Q.~Tian, ``Principal visual word discovery for automatic license plate detection,'' \emph{IEEE transactions on image processing}, vol.~21, no.~9, pp. 4269--4279, 2012.

\bibitem{sharif2016accessorize}
M.~Sharif, S.~Bhagavatula, L.~Bauer, and M.~K. Reiter, ``Accessorize to a crime: Real and stealthy attacks on state-of-the-art face recognition,'' in \emph{Proceedings of the ACM Conference on Computer and Communications Security (CCS '16)}.\hskip 1em plus 0.5em minus 0.4em\relax ACM, 2016, pp. 1528--1540.

\bibitem{szegedy2014intriguing}
C.~Szegedy, W.~Zaremba, I.~Sutskever, J.~Bruna, D.~Erhan, I.~Goodfellow, and R.~Fergus, ``Intriguing properties of neural networks,'' in \emph{Proceedings of the 2nd International Conference on Learning Representations (ICLR '14)}, 2014.

\bibitem{papernot2016limitations}
N.~Papernot, P.~McDaniel, S.~Jha, M.~Fredrikson, Z.~B. Celik, and A.~Swami, ``The limitations of deep learning in adversarial settings,'' in \emph{IEEE European symposium on security and privacy (EuroS\&P '16)}.\hskip 1em plus 0.5em minus 0.4em\relax IEEE, 2016, pp. 372--387.

\bibitem{liu2017protecting}
Y.~Liu, W.~Zhang, and N.~Yu, ``Protecting privacy in shared photos via adversarial examples based stealth,'' \emph{Security and Communication Networks}, vol. 2017, 2017.

\bibitem{friedrich2019adversarial}
M.~Friedrich, A.~K{\"o}hn, G.~Wiedemann, and C.~Biemann, ``Adversarial learning of privacy-preserving text representations for de-identification of medical records,'' in \emph{Proceedings of the 57th Annual Meeting of the Association for Computational Linguistics (ACL '20)}, 2019, pp. 5829--5839.

\bibitem{liu2019adversaries}
B.~Liu, M.~Ding, T.~Zhu, Y.~Xiang, and W.~Zhou, ``Adversaries or allies? privacy and deep learning in big data era,'' \emph{Concurrency and Computation: Practice and Experience}, vol.~31, no.~19, p. e5102, 2019.

\bibitem{li2019anonymousnet}
T.~Li and L.~Lin, ``Anonymousnet: Natural face de-identification with measurable privacy,'' in \emph{Proceedings of the IEEE/CVF conference on computer vision and pattern recognition workshops (CVPRW '19)}, 2019.

\bibitem{su2019one}
J.~Su, D.~V. Vargas, and K.~Sakurai, ``One pixel attack for fooling deep neural networks,'' \emph{IEEE Transactions on Evolutionary Computation}, vol.~23, no.~5, pp. 828--841, 2019.

\bibitem{Zahoor2021Oct}
S.~Zahoor and R.~N. Mir, ``{Resource management in pervasive Internet of Things: A survey},'' \emph{Journal of King Saud University - Computer and Information Sciences}, vol.~33, no.~8, pp. 921--935, Oct. 2021.

\bibitem{Zeng2021Jul}
D.~Zeng, S.~Liang, X.~Hu, H.~Wang, and Z.~Xu, ``{FedLab: A Flexible Federated Learning Framework},'' \emph{arXiv}, Jul. 2021.

\bibitem{Shen2019Feb}
M.~Shen, X.~Tang, L.~Zhu, X.~Du, and M.~Guizani, ``{Privacy-Preserving Support Vector Machine Training Over Blockchain-Based Encrypted IoT Data in Smart Cities},'' \emph{IEEE Internet of Things Journal}, vol.~6, no.~5, pp. 7702--7712, Feb. 2019.

\bibitem{Wei2022Oct}
X.~Wei, Y.~Xu, Y.~Huang, H.~Lv, H.~Lan \emph{et~al.}, ``Learning extremely lightweight and robust model with differentiable constraints on sparsity and condition number,'' in \emph{{Computer Vision {\textendash} ECCV 2022}}.\hskip 1em plus 0.5em minus 0.4em\relax Cham, Switzerland: Springer, Oct. 2022, pp. 690--707.

\bibitem{Hu2021Oct}
X.~Hu, L.~Chu, J.~Pei, W.~Liu, and J.~Bian, ``Model complexity of deep learning: {A survey},'' \emph{Knowl. Inf. Syst.}, vol.~63, no.~10, pp. 2585--2619, Oct. 2021.

\bibitem{Ko2021Mar}
Y.~Ko, A.~Chadwick, D.~Bates, and R.~Mullins, ``Lane compression: {A} lightweight lossless compression method for machine learning on embedded systems,'' \emph{ACM Trans. Embedded Comput. Syst.}, vol.~20, no.~2, pp. 1--26, Mar. 2021.

\bibitem{Zhu2023Aug}
S.~Zhu, X.~Xu, J.~Zhao, and F.~Xiao, ``{LKD-STNN: A} lightweight malicious traffic detection method for internet of things based on knowledge distillation,'' \emph{IEEE Internet Things J.}, vol.~11, no.~4, pp. 6438--6453, Aug. 2023.

\bibitem{Wang2024Mar}
Y.~Wang, G.~Qin, M.~Zou, Y.~Liang, G.~Wang \emph{et~al.}, ``A lightweight intrusion detection system for internet of vehicles based on transfer learning and {MobileNetV2} with hyper-parameter optimization,'' \emph{Multimed. Tools Appl.}, vol.~83, no.~8, pp. 22\,347--22\,369, Mar. 2024.

\bibitem{Nguyen2021Apr}
D.~C. Nguyen, M.~Ding, Q.-V. Pham, P.~N. Pathirana, L.~B. Le \emph{et~al.}, ``{Federated Learning Meets Blockchain in Edge Computing: Opportunities and Challenges},'' \emph{IEEE Internet of Things Journal}, vol.~8, no.~16, pp. 12\,806--12\,825, Apr. 2021.

\bibitem{Yang2022Aug}
Z.~Yang, Y.~Shi, Y.~Zhou, Z.~Wang, and K.~Yang, ``{Trustworthy Federated Learning via Blockchain},'' \emph{IEEE Internet of Things Journal}, vol.~10, no.~1, pp. 92--109, Aug. 2022.

\bibitem{Pastaltzidis}
I.~Pastaltzidis, N.~Dimitriou, K.~Quezada-Tavarez, S.~Aidinlis, T.~Marquenie \emph{et~al.}, ``{Data augmentation for fairness-aware machine learning: Preventing algorithmic bias in law enforcement systems},'' in \emph{{FAccT '22: 2022 ACM Conference on Fairness, Accountability, and Transparency}}.\hskip 1em plus 0.5em minus 0.4em\relax Association for Computing Machinery, 2022, pp. 2302--2314.

\bibitem{Lo2022Jan}
S.~K. Lo, Y.~Liu, Q.~Lu, C.~Wang, X.~Xu \emph{et~al.}, ``{Toward Trustworthy AI: Blockchain-Based Architecture Design for Accountability and Fairness of Federated Learning Systems},'' \emph{IEEE Internet of Things Journal}, vol.~10, no.~4, pp. 3276--3284, Jan. 2022.

\bibitem{Patel2022Apr}
A.~R. Patel, J.~Chandrasekaran, Y.~Lei, R.~N. Kacker, and D.~R. Kuhn, ``{A Combinatorial Approach to Fairness Testing of Machine Learning Models},'' in \emph{{2022 IEEE International Conference on Software Testing, Verification and Validation Workshops (ICSTW)}}.\hskip 1em plus 0.5em minus 0.4em\relax IEEE, Apr. 2022, pp. 94--101.

\bibitem{Li2022Mar}
Z.~Li, Y.~Zhou, D.~Wu, T.~Tang, and R.~Wang, ``{Fairness-Aware Federated Learning With Unreliable Links in Resource-Constrained Internet of Things},'' \emph{IEEE Internet of Things Journal}, vol.~9, no.~18, pp. 17\,359--17\,371, Mar. 2022.

\bibitem{Li2022sept}
G.~Li, Q.~Zhao, D.~Zhang, M.-Y. Chen, M.~M. Hassan \emph{et~al.}, ``{GT-Chain: A Fair Blockchain for Intelligent Industrial IoT Applications},'' \emph{IEEE Trans. Network Sci. Eng.}, vol.~9, no.~5, pp. 3244--3257, Sept. 2022.

\bibitem{Liu2022Feb}
D.~Liu, C.~Huang, J.~Ni, X.~Lin, and X.~S. Shen, ``{Blockchain-Cloud Transparent Data Marketing: Consortium Management and Fairness},'' \emph{IEEE Trans. Comput.}, vol.~71, no.~12, pp. 3322--3335, Feb. 2022.

\bibitem{BibEntry2022Dec}
\BIBentryALTinterwordspacing
``{How Much Does the Internet of Things Cost? {\ifmmode---\else\textemdash\fi} ITRex},'' Dec. 2022, [Accessed 14. Apr. 2023]. [Online]. Available: \url{https://itrexgroup.com/blog/how-much-iot-cost-factors-challenges}
\BIBentrySTDinterwordspacing

\bibitem{xu2022influence}
X.~Xu, ``{The Influence of Artificial Intelligence on the Financial Industry},'' in \emph{Proceedings of the 2022 4th International Conference on Economic Management and Cultural Industry (ICEMCI '22)}.\hskip 1em plus 0.5em minus 0.4em\relax Atlantis Press, 2022, pp. 393--400.

\bibitem{Dahmen2023Apr}
\BIBentryALTinterwordspacing
S.~Dahmen-Lhuissier, ``{Internet of Things (IoT)},'' Apr. 2023, [Accessed 14. Apr. 2023]. [Online]. Available: \url{https://www.etsi.org/technologies/internet-of-things}
\BIBentrySTDinterwordspacing

\bibitem{Rabieinejad2021Dec}
E.~Rabieinejad, A.~Yazdinejad, R.~M. Parizi, and A.~Dehghantanha, ``{Generative Adversarial Networks for Cyber Threat Hunting in Ethereum Blockchain},'' \emph{Distrib. Ledger Technol.}, Dec. 2021.

\bibitem{Ferrag2023Mar}
M.~A. Ferrag, M.~Debbah, and M.~Al-Hawawreh, ``{Generative AI for Cyber Threat-Hunting in 6G-enabled IoT Networks},'' \emph{arXiv}, Mar. 2023.

\bibitem{Ferrag2023Apr}
M.~A. Ferrag, D.~Hamouda, M.~Debbah, L.~Maglaras, and A.~Lakas, ``{Generative Adversarial Networks-Driven Cyber Threat Intelligence Detection Framework for Securing Internet of Things},'' \emph{arXiv}, Apr. 2023.

\bibitem{huang2021accurate}
J.~Huang, G.~Li, J.~Tian, and S.~Li, ``Accurate interpretation of the online learning model for 6g-enabled internet of things,'' \emph{IEEE Internet of Things Journal}, vol.~8, no.~20, pp. 15\,228--15\,239, 2021.

\bibitem{krawczyk2017ensemble}
B.~Krawczyk, L.~L. Minku, J.~Gama, J.~Stefanowski, and M.~Wo{\'z}niak, ``Ensemble learning for data stream analysis: A survey,'' \emph{Information Fusion}, vol.~37, pp. 132--156, 2017.

\bibitem{qu2023learn}
Y.~Qu, X.~Yuan, M.~Ding, W.~Ni, T.~Rakotoarivelo, and D.~Smith, ``Learn to unlearn: A survey on machine unlearning,'' \emph{arXiv preprint arXiv:2305.07512}, 2023.

\bibitem{Pradhan2023Feb}
I.~P. Pradhan and P.~Saxena, ``{Reskilling Workforce for the Artificial Intelligence Age: Challenges and the Way Forward},'' in \emph{{The Adoption and Effect of Artificial Intelligence on Human Resources Management, Part B}}.\hskip 1em plus 0.5em minus 0.4em\relax Emerald Publishing Limited, Feb. 2023, pp. 181--197.

\bibitem{Albouq2022Mar}
S.~S. Albouq, A.~A.~A. Sen, N.~Almashf, M.~Yamin, A.~Alshanqiti \emph{et~al.}, ``{A Survey of Interoperability Challenges and Solutions for Dealing With Them in IoT Environment},'' \emph{IEEE Access}, vol.~10, pp. 36\,416--36\,428, Mar. 2022.

\bibitem{Chengoden2023Feb}
R.~Chengoden, N.~Victor, T.~Huynh-The, G.~Yenduri, R.~H. Jhaveri \emph{et~al.}, ``{Metaverse for Healthcare: A Survey on Potential Applications, Challenges and Future Directions},'' \emph{IEEE Access}, vol.~11, pp. 12\,765--12\,795, Feb. 2023.

\bibitem{Huang2023Mar}
Z.~Huang, C.~Xiong, H.~Ni, D.~Wang, Y.~Tao \emph{et~al.}, ``{Standard Evolution of 5G-Advanced and Future Mobile Network for Extended Reality and Metaverse},'' \emph{IEEE Internet of Things Magazine}, vol.~6, no.~1, pp. 20--25, mar 2023.

\bibitem{Lim2022Jul}
W.~Y.~B. Lim, Z.~Xiong, D.~Niyato, X.~Cao, C.~Miao \emph{et~al.}, ``{Realizing the Metaverse with Edge Intelligence: A Match Made in Heaven},'' \emph{IEEE Wireless Commun.}, jul 2022, in press.

\bibitem{Siyue2023Jan}
S.~Zhang, W.~Y.~B. Lim, W.~C. Ng, Z.~Xiong, D.~Niyato \emph{et~al.}, ``{Towards Green Metaverse Networking: Technologies, Advancements and Future Directions},'' \emph{IEEE Network}, pp. 1--10, Jan. 2023.

\bibitem{schlimmer1986incremental}
J.~C. Schlimmer and R.~H. Granger, ``Incremental learning from noisy data,'' \emph{Machine learning}, vol.~1, pp. 317--354, 1986.

\bibitem{lu2018learning}
J.~Lu, A.~Liu, F.~Dong, F.~Gu, J.~Gama, and G.~Zhang, ``Learning under concept drift: A review,'' \emph{IEEE transactions on knowledge and data engineering}, vol.~31, no.~12, pp. 2346--2363, 2018.

\bibitem{aouedi2022ensemble}
O.~Aouedi, K.~Piamrat, and B.~Parrein, ``Ensemble-based deep learning model for network traffic classification,'' \emph{IEEE Transactions on Network and Service Management}, vol.~19, no.~4, pp. 4124--4135, 2022.

\end{thebibliography}
\end{document}